\documentclass[12pt,a4paper,dvipsnames,usenames]{article}
\usepackage{amssymb}
\usepackage{amsmath}
\usepackage{amsthm}
\usepackage{latexsym}
\usepackage[dvips]{epsfig}
\usepackage{mathrsfs}
\usepackage{bm}
\usepackage{stmaryrd}
\usepackage{authblk}

\usepackage[dvipsnames]{xcolor}
\usepackage{tikz}
\usepackage{enumitem}
\usepackage{newtxtext}

\theoremstyle{plain}
\newtheorem{proposition}{Proposition}
\newtheorem{lemma}{Lemma}
\newtheorem{theorem}{Theorem}
\newtheorem{assumption}{Assumption}

\newtheorem*{main}{Theorem}
\newtheorem{definition}{Definition}
\newtheorem{remark}{Remark}

\allowdisplaybreaks

\def\bma{{\bm a}}
\def\bmb{{\bm b}}
\def\bmc{{\bm c}}
\def\bmd{{\bm d}}
\def\bme{{\bm e}}
\def\bmf{{\bm f}}
\def\bmg{{\bm g}}
\def\bmh{{\bm h}}
\def\bmi{{\bm i}}
\def\bmj{{\bm j}}
\def\bmk{{\bm k}}
\def\bml{{\bm l}}
\def\bmn{{\bm n}}

\def\bmo{{\bm o}}

\def\bms{{\bm s}}
\def\bmt{{\bm t}}
\def\bmu{{\bm u}}
\def\bmv{{\bm v}}

\def\bmx{{\bm x}}

\def\bmzero{{\bm 0}}
\def\bmone{{\bm 1}}
\def\bmtwo{{\bm 2}}
\def\bmthree{{\bm 3}}

\def\bmA{{\bm A}}
\def\bmB{{\bm B}}
\def\bmC{{\bm C}}
\def\bmD{{\bm D}}
\def\bmE{{\bm E}}
\def\bmF{{\bm F}}

\def\bmK{{\bm K}}
\def\bmL{{\bm L}}

\def\bmP{{\bm P}}
\def\bmQ{{\bm Q}}

\def\bmS{{\bm S}}

\def\bmZ{{\bm Z}}

\def\bmbeta{{\bm \beta}}

\def\bmepsilon{{\bm \epsilon}}

\def\bmiota{{\bm \iota}}
\def\bmomega{{\bm \omega}}

\def\bmpartial{{\bm \partial}}
\def\bmnabla{{\bm \nabla}}

\usepackage{mathtools}% http://ctan.org/pkg/mathtools
\usepackage{xparse}% http://ctan.org/pkg/xparse
\makeatletter
\newcommand{\raisemath}[1]{\mathpalette{\raisem@th{#1}}}
\newcommand{\raisem@th}[3]{\raisebox{#1}{$#2#3$}}
\makeatother
\NewDocumentCommand{\newrbar}{O{0pt} O{0pt}}{
  \ensuremath{\mathrlap{\raisemath{#2}{\hspace*{#1}{\mathchar'26\mkern-9mu}}}r}}

\newcounter{mnotecount}%[section]

\newcommand{\mnotex}[1]%{}
{\protect{\stepcounter{mnotecount}}$^{\mbox{\footnotesize $\bullet$\themnotecount}}$ 
\marginpar{%\color{red}%
\raggedright\tiny\em
$\!\!\!\!\!\!\,\bullet$\themnotecount: #1} }

\newcounter{mnote}

\usepackage{color}
\usepackage{xcolor}
\usepackage[normalem]{ulem}
\usepackage{soul}
\usepackage{hyperref}

\begin{document}

\title{\textbf{BMS-supertranslation charges at the critical sets of null infinity}}
 
\author[1]{Mariem Magdy Ali Mohamed  \footnote{E-mail
    address: {\tt m.m.a.mohamed@qmul.ac.uk}}}

\author[2]{Kartik Prabhu \footnote{E-mail address: {\tt kartikprabhu@rri.res.in}}}

\author[1]{Juan A. Valiente Kroon \footnote{E-mail address: {\tt j.a.valiente-kroon@qmul.ac.uk}}}

\affil[1]{School of Mathematical Sciences, Queen Mary, University of London,
Mile End Road, London E1 4NS, United Kingdom.}

\affil[2]{Raman Research Institute, Sadashivanagar, Bengaluru 560080, India.}

\maketitle

\begin{abstract}
For asymptotically flat spacetimes, a conjecture by Strominger states that asymptotic BMS-supertranslations and their associated charges at past null infinity $\mathscr{I}^{-}$ can be related to those at future null infinity $\mathscr{I}^{+}$ via an antipodal map at spatial infinity $i^{0}$. We analyse the validity of this conjecture using Friedrich’s formulation of spatial infinity, which gives rise to a regular initial value problem for the conformal field equations at spatial infinity. A central structure in this analysis is the cylinder at spatial infinity $\mathcal{I}$ representing a blow-up of the standard spatial infinity point $i^{0}$ to a 2-sphere. The cylinder $\mathcal{I}$ touches past and future null infinities $\mathscr{I}^{\pm}$ at the critical sets $\mathcal{I}^{\pm}$. We show that for a generic class of asymptotically Euclidean and regular initial data, BMS-supertranslation charges are not well-defined at $\mathcal{I}^{\pm}$ unless the initial data satisfies an extra regularity condition. We also show that given initial data that satisfy the regularity condition, BMS-supertranslation charges at $\mathcal{I}^{\pm}$ are fully determined by the initial data and that the relation between the charges at $\mathcal{I}^{-}$ and those at $\mathcal{I}^{+}$ directly follows from our regularity condition. 
\end{abstract}
%\newpage
%\tableofcontents
%\newpage
%%%%%%%%%%%%%%%%%%%%%%%%%%%%%%%%%%%%%%%%%%%%%%%%%%%%%%%%%%%%%%%%%%
\section{Introduction}
Studies of isolated systems, asymptotic structures and symmetries have received increasing interest in recent years due to their relation to black-hole physics \cite{Hawking15,HawPerryStro16,BouPor17}, the gravitational memory effect \cite{Favata10,Christodoulou91,BD92} and developments in soft theorems \cite{Weinberg65,Strominger14,HLMS15}. A common approach in these studies involves the use of conformal transformations to study the behaviour of the gravitational field ‘at infinity’ using local differential geometry by mapping points at infinity in one manifold to a region located at finite distances in another manifold, the so-called conformal boundary.

The conformal approach is inspired by R. Penrose seminal work \cite{Penrose63, Penrose64} in which the notion of asymptotically simple (AS) spacetimes was initially introduced with the aim of identifying a large class of spacetimes that admit a conformal extension similar to that of Minkowski spacetime. More precisely, a spacetime $(\Tilde{\mathcal{M}},\Tilde{\bmg})$ is asymptotically simple if there exists a smooth, oriented, time-oriented and causal spacetime $(\mathcal{M},\bmg)$ and a smooth function $\Xi$ on $\mathcal{M}$ such that (i) $\mathcal{M}$ is a manifold with boundary $\mathscr{I} \equiv \partial \mathcal{M}$; (ii) the conformal factor $\Xi$ satisfies: $\Xi > 0$ on $\mathcal{M} \setminus \mathscr{I}$ and $\Xi = 0, \bmd{\Xi} \neq 0$ on $ \mathscr{I} $; (iii) the manifolds $\Tilde{\mathcal{M}}$ and $\mathcal{M}$ are related by $\phi: \Tilde{\mathcal{M}} \rightarrow \mathcal{M}$ such that $\phi(\Tilde{\mathcal{M}}) = \mathcal{M} \setminus \mathscr{I}$ and $\phi^{*}\bmg = \Xi^2 \Tilde{\bmg}$; and finally, (iv) null geodesics of $(\Tilde{\mathcal{M}},\Tilde{\bmg})$ starts and ends on $\mathscr{I}$. In this context, we say that $\phi$ is a conformal transformation that maps $\Tilde{\mathcal{M}}$ to $\mathcal{M} \setminus \mathscr{I}$. The manifolds $\tilde{\mathcal{M}}$ and $\mathcal{M}$ are referred to as the physical and unphysical manifolds, respectively. Similarly, the metric $\tilde{\bmg}$ is the physical metric, and $\bmg$ is the unphysical metric. Generally, the boundary $\mathscr{I}$ (null infinity) can be split into two disjoint sets $\mathscr{I}^+$ (future null infinity) and $\mathscr{I}^-$ (past null infinity). 
%The notion of asymptotic simplicity is crucial in the discussion of radiation proprieties of isolated systems. On the other hand, the study of conserved quantities involves examining the behaviour of the gravitational field in the asymptotic region where future and past null infinities are joined, i.e. at spatial infinity. Therefore, most of t

The literature on the asymptotic structure of spacetimes can be divided into two categories: studies of the asymptotic structure at null infinity or at spatial infinity. In the null regime, it was expected that the Poincar\'{e} group would describe the asymptotic symmetry group for AS spacetimes, given that the background geometry of an AS spacetime is similar to that of Minkowski. However, the work of Bondi, Metzner and Sachs \cite{BMS62} revealed that the asymptotic symmetry group for AS spacetimes is given by the infinite-dimensional BMS symmetry group, which can be written as the semi-direct product of the Lorentz group with the infinite-dimensional group of angle-dependent translations (supertranslations) along null infinity. Notions of asymptotic flatness at spatial infinity also give rise to an infinite-dimensional asymptotic symmetry group at spatial infinity, known as the Spi group \cite{AshHan78,AshRom92}, with a structure similar to the BMS group ---see also \cite{HenneauxTroessaert18,HenneauxTroessaert18-2,HenneauxTroessaert18-3,HenneauxTroessaert20,PrabhuShehzad20}.

Asymptotic symmetry groups define corresponding conserved quantities or asymptotic charges. At spatial infinity, notions of conserved mass, momentum and angular momentum can be derived using a Hamiltonian formulation \cite{ADM59,RegTeit74}. More recently, it was shown that charges associated with supertranslations at spatial infinity are generally non-vanishing \cite{HenneauxTroessaert18-2,PrabhuShehzad20}. At null infinity, the challenge is that BMS charges can not be defined using a standard Hamiltonian formulation as, generically, there exists no Hamiltonian that generates BMS transformations at null infinity. This observation can be linked to the fact that the symplectic current can be radiated away at null infinity, i.e., BMS charges are not exactly conserved. In fact, BMS charges have non-vanishing fluxes through null infinity. Nevertheless, the discussion in \cite{WaldZoupas00} provides a general definition of ‘conserved quantities’ associated with asymptotic symmetries, even in scenarios in which the Hamiltonian does not exist.

A conjecture by Strominger \cite{Strominger14} states that a priori independent asymptotic symmetry groups at past and future null infinities $\mathscr{I}^{\pm}$, denoted by $\text{BMS}^{+}$ and $\text{BMS}^{-}$, respectively, can be related via an antipodal reflection map near spatial infinity. The verification of this conjecture, referred to as the matching problem, would imply a global diagonal asymptotic symmetry group $\text{BMS}^{+} \times \text{BMS}^{-}$. In other words, the incoming fluxes associated with $\text{BMS}^{+}$ would be equal to the outgoing fluxes associated with $\text{BMS}^{-}$. The matching problem is also a crucial ingredient in the conjectured equivalence relation between asymptotic symmetries, the soft graviton theorem and the gravitational memory effect \cite{Strominger17,BarTroe16}.

The strategy in validating the matching of $\text{BMS}^{+}$ and $\text{BMS}^{-}$ and their associated charges involves expanding the fields in suitable coordinates around null and spatial infinity. On Minkowski spacetime, the matching of asymptotic charges at past and future null infinities has been verified for the spin-1 and spin-2 fields \cite{CampEyh17,Troessaert18,MohamedKroon22}. For more general spacetimes, the analysis is complicated due to the singular conformal structure at spatial infinity for spacetimes with non-vanishing Arnowitt-Deser-Misner (ADM) mass, referred to as the problem of spatial infinity ---see e.g. Chapter 20 in \cite{kroon16}. Another challenge is that one requires a transformation between adapted coordinates at null and spatial infinity, which can be explicitly computed on Minkowski spacetime but is not generally known for general spacetimes.  Nevertheless, the covariant formulation of Ashtekar and Hansen \cite{AshHan78} was used in \cite{Prabhu18,Prabhu19} to prove the matching of asymptotic charges for the spin-1 and gravitational fields on spacetimes that satisfy Ashtekar-Hansen's notion of asymptotic flatness. Similar techniques were used in \cite{PrabhuShehzad22} to investigate the matching of Lorentz charges for the gravitational field on Ashtekar-Hansen asymptotically flat spacetimes ---see also \cite{CaponeNguyenParisini23}. 

\medskip
The purpose of this article is to verify the matching of BMS asymptotic charges in a full GR setting using an initial value formulation of the field equations. The argument made in \cite{Strominger14} is that the matching of BMS-supertranlation charges should hold for Christodoulou-Klainerman class of spacetimes \cite{Christodoulou1993}. However, as it follows from the analysis presented in this article, the Christodoulou-Klainerman class of spacetimes are not general enough to obtain non-trivial asymptotic charges near spatial infinity since they lead to vanishing BMS-supertranslation charges at spatial infinity ---this statement is further elaborated in Section \ref{Subsection:MainResult} ----see Remark \ref{Remark:C-K}. Therefore, the aim of this article is to identify a generic class of initial data and conditions on the initial data that implies non-trivial, well-defined asymptotic charges at spatial infinity.

In the context of the initial value problem, Einstein's field equations are split into constraint equations and evolution equations; the constraint equations are satisfied by an initial data set $(\Tilde{\mathcal{S}},\Tilde{\bmh},\Tilde{\bmK})$ prescribed on an initial Cauchy hypersurface $\Tilde{\mathcal{S}}$, where $\Tilde{\bmh}$ is the intrinsic metric on $\Tilde{\mathcal{S}}$ and $\Tilde{\bmK}$ is the extrinsic curvature. The well-posedness of the Cauchy problem ensures that there exists a vacuum spacetime $(\tilde{\mathcal{M}},\tilde{\bmg})$, referred to as the development of the initial data $(\Tilde{\mathcal{S}},\Tilde{\bmh},\Tilde{\bmK})$, such that $\tilde{\mathcal{S}}$ is a spacelike hypersurface in $\tilde{\mathcal{M}}$ with an intrinsic metric $\tilde{\bmh}$ induced on $\tilde{\mathcal{S}}$ by $\tilde{\bmg}$ with an associated extrinsic curvature $\Tilde{\bmK}$. It is worth noting that not every spacetime can be globally constructed from an initial value problem. A spacetime is said to be globally hyperbolic if it can be constructed from an initial value problem of Einstein's field equations. A special class of initial data relevant to this article are the so-called asymptotically Euclidean and regular initial data \cite{Geroch72} defined by:
\begin{definition}[asymptotically Euclidean and regular]
A three-dimensional Riemannian manifold $(\Tilde{\mathcal{S}},\Tilde{\bmh})$ is asymptotically Euclidean and regular if there exists a three-dimensional, orientable, compact manifold $(\mathcal{S},\bmh)$ with points $i_k \in \mathcal{S}, k = 1, \dotso, N$ with $N$ some integer, a function $\Omega \in C^2$ and a diffeomorphism $\varphi: \mathcal{S} \setminus \{i_1, \dotso,i_N \} \rightarrow \Tilde{\mathcal{S}}$ such that
\begin{enumerate}
    \item $\Omega(i_k)=0, \bmd \Omega(i_k) =0$ and $\textbf{Hess } \Omega(i_k) = -2 \bmh(i_k)$, for all $i_k \in \{i_1, \dotso,i_N \}$,
    \item $\Omega > 0$ on $\mathcal{S} \setminus \{i_1, \dotso,i_N \}$, and
    \item $\bmh = \Omega^2 \varphi^{*} \Tilde{\bmh}$ on $\mathcal{S} \setminus \{i_1, \dotso,i_N \}$ with $\bmh \in C^2(\mathcal{S}) \cap C^{\infty}(\mathcal{S} \setminus \{i_1, \dotso,i_N \})$. 
\end{enumerate}
\label{Definition:AsympEuclideanAndRegular}
\end{definition}
In the above, neighbourhoods of the points $i_k$ can be mapped to the asymptotic ends of $\Tilde{\mathcal{S}}$ and thus, each of these points represents spacelike infinity. Compared to the standard definition of asymptotically Euclidean manifolds which describes the asymptotic expansion of the intrinsic fields on $\Tilde{\mathcal{S}}$ near the asymptotic ends, Definition \ref{Definition:AsympEuclideanAndRegular} is more geometric in nature and it imposes extra conditions on the smoothness of the initial data, which in turn affects the asymptotic behaviour of their evolution in time \cite{Geroch72}. In this article, the aim is to obtain non-trivial BMS asymptotic charges associated with the development of some initial data that satisfy the constraint equations with prescribed behaviour in the asymptotic region. These initial data were first considered in \cite{Huang10} and are obtained by means of a gluing construction. This class of initial data includes, as a particular case, boosted solutions to the constraints and allows for a term with arbitrary multipolar structure in the initial metric, which appears at the same order as the mass. As will become evident from our analysis, this arbitrary multipolar structure will be responsible for the existence of non-trivial BMS asymptotic charges.

It should be noted that the transformation of Einstein's field equations from the physical manifold $(\tilde{\mathcal{M}},\tilde{\bmg})$ to the unphysical manifold $(\mathcal{M}, \bmg)$ implies singular equations at the conformal boundary $\Xi =0$. An alternative set of field equations, the so-called metric conformal field equations, can be constructed following the discussion in \cite{Friedrich81a,Friedrich81b,Friedrich83}. These equations are regular at $\Xi=0$, and they imply solutions to Einstein's field equations at the points where $\Xi \neq 0$ ---see \cite{kroon16}, Chapter 8. In this article, we will make use of the extended conformal field equations (ECFEs). These equations are formulated in terms of a Weyl connection $\hat{\bmnabla}$ and exhibit additional gauge freedom in contrast to the metric conformal field equations, which are formulated in terms of the Levi-Civita connection $\bmnabla$ associated with $\bmg$. As such, the main goal of this project will be to use the ECFEs to evaluate BMS asymptotic charges near spatial infinity.

To address the singular conformal structure at spatial infinity, we make use of Friedrich's formulation of spatial infinity originally introduced in \cite{Friedrich98} with the aim of obtaining a regular initial value problem for the conformal field equations at spatial infinity. A central structure in this formulation is the cylinder at spatial infinity $\mathcal{I}$ corresponding to a blow-up of the spatial infinity point to a 2-sphere. The cylinder $\mathcal{I}$ touches the endpoints of past and future null infinities $\mathscr{I}^{\pm}$ at the critical sets $\mathcal{I}^{\pm}$. Associated with this formulation is a particular choice of gauge, referred to as the F-gauge, in which the coordinates and frames on an initial hypersurface are propagated along conformal geodesics. One of the remarkable properties of conformal geodesics is that they introduce a canonical conformal factor that depends on the proper time along the curves and the initial data. In other words, the F-gauge is constructed so that the location of the conformal boundary is known a priori. Moreover, in this particular choice of gauge, the cylinder $\mathcal{I}$ is a total characteristic of the ECFEs, i.e., the associated evolution equations can be written as a system of transport equations on $\mathcal{I}$. A significant advantage of Friedrich's formulation is that it allows us to link quantities at the critical sets $\mathcal{I}^{\pm}$ with the initial data prescribed on an initial hypersurface. This approach was used in \cite{FriedrichKannar00,GasperinKroon20} to express the Newman-Penrose (NP) constants in terms of initial data, and it illustrates our strategy in this work where the goal is to express BMS asymptotic charges at $\mathcal{I}^{\pm}$ in terms of our initial data. 
%Conformal geodesics are a pair of a curve and a covector satisfying some conformal invariant equations. 

As mentioned earlier, one of the challenges in verifying the matching of asymptotic charges is that a transformation between adapted coordinates at null and spatial infinity is required. In this article, BMS asymptotic charges are expressed in the NP-gauge, comprised of certain conformal gauge conditions, certain coordinates and an orthonormal frame field satisfying certain frame gauge conditions. The main difference between the F-gauge and the NP-gauge is that the former is adapted to Cauchy hypersurfaces while the latter is adapted to null infinity $\mathscr{I}^{\pm}$. The discussion in \cite{FriedrichKannar00} provides a prescription of the transformation between the NP-gauge and the F-gauge. Given a solution to the ECFEs, an explicit transformation can be obtained, allowing us to express the BMS asymptotic charges in terms of the F-gauge. In turn, the BMS-asymptotic charges can be evaluated at $\mathcal{I}^{\pm}$ given the solution to the ECFEs.

\subsection*{Main result}
The main results of this article can be summarised in the following:
\begin{main}
    For the generic initial data in \cite{Huang10}, asymptotic BMS-supertranslation charges are not well-defined at the critical sets $\mathcal{I}^{\pm}$ unless the conformal initial data satisfy the regularity condition given in Lemma \ref{Regularity-conditions-GR}. If the initial data are chosen to satisfy the extra regularity condition, the BMS-supertranslation charges at $\mathcal{I}^{\pm}$ are fully determined by the initial data and the matching between charges at $\mathcal{I}^{+}$ and $\mathcal{I}^{-}$ follows directly from the regularity condition.
\end{main}

As it will be discussed in the main body of the article, the piece of the freely specifiable initial data from which the value of the BMS charges at $\mathcal{I}^\pm$ arise correspond to a function $\xi\in C^2(\mathbb{S})$. The regularity condition in Lemma \ref{Regularity-conditions-GR} ensuring that the charges are well-defined is a statement about the parity of the function $\xi$. More precisely, all odd parity harmonics, except for the one with $l=1$, are required to vanish. This result provides evidence that Strominger's antipodal matching condition is, in fact, a regularity condition on spatial infinity. A full proof of this statement would require a  clarification of the relation between the asymptotic expansions used in our analysis and full solutions to the conformal Einstein field equations. This, in turn, requires the construction of detailed estimates for the remainders of the asymptotic expansions along the lines of what was done in \cite{Friedrich03-2} for the spin-2 field ---the latter is, however, beyond the scope of this article. 

As pointed in \cite{Hen23} the type of parity condition arising from our analysis (and an analogous one for the extrinsic curvature, which is not required in the present analysis) is required to make all Poincar\'e charges at spatial infinity well-defined. This condition can be traced back to the seminal article \cite{RegTeit74}. Our analysis points out a deep connection between the regularity of null infinity and the physical requirement that asymptotic charges are well-defined. Again, a systematic analysis of these ideas goes beyond the scope of this article.

\subsection*{Outline of the article}
In Section \ref{Section: Conformal tools}, we start by introducing some of the basic conformal tools used throughout this article ---e.g. the conformal field equations and conformal geodesics. Friedrich's formulation of spatial infinity and the F-gauge are introduced in Section \ref{Section:Friedrich formulation of spatial infinity}. In Section \ref{Section: The NP-gauge}, we discuss the NP-gauge conditions and the relation between the NP-gauge and the F-gauge. The expressions for BMS-supertranslation charges are introduced in Section \ref{Section: BMS-supertranslations charges} along with their translation to the F-gauge. In Section \ref{Section: Initial data}, we present the initial data utilised in our analysis of the conformal field equations. We conclude the analysis of this article in Section \ref{Section:Evaluating-BMS-charges} by obtaining the zero-order solution of the conformal field equations and evaluating the BMS-supertranslation charges at the critical sets $\mathcal{I}^{\pm}$.

\subsection*{Notations and conventions}
In this article, Latin letters $a,\,b,\,c,\ldots$ will denote spacetime abstract tensorial indices while $i,\,j,\,k,\ldots$ will denote spatial abstract tensorial indices. Capital Latin letters $A,\,B,\,C,\ldots$ will denote abstract spinorial indices.  

To discuss the components of tensors with respect to a coordinate basis, the Greek letters $\mu,\,\nu,\dots$ will be used as spacetime coordinate indices while $\alpha,\, \beta,\ldots$ will be used as spatial coordinate indices. Then, the components of a generic spacetime tensor $T_{ab}$ with respect to an arbitrary coordinate system $(x^{\mu})$ will be written as $T_{\mu \nu} = T_{ab} (\bmpartial/ \bmpartial x^{\mu})^{a} (\bmpartial/ \bmpartial x^{\nu})^{b}$. Similarly, the components of a generic spatial tensor $l_{ij}$ with respect to an arbitrary coordinate system $(x^{\alpha})$ will be written as $l_{\alpha \beta} = l_{ij} (\bmpartial_{\alpha})^{i} (\bmpartial_{\beta})^{j}$, where $(\bmpartial_{\alpha})^{i} \equiv (\bmpartial/ \bmpartial x^{\alpha})^{i}$.

To discuss the components of tensors and spinors with respect to a frame basis, let $\bma, \bmb, \bmc, \ldots$ denote tensorial frame indices and $\bmA, \bmB, \bmC, \ldots$ denote spinorial frame indices. Then, the components of a generic tensor $T_{ab}$ with respect to an arbitrary basis $\bme_{a} \equiv \{ \bme_{\bma} \}$ will be written as $T_{\bma \bmb} = T_{ab} \bme_{\bma}{}^a{} \bme_{\bmb}{}^b{}$ with $\bma, \bmb \in \{ \bmzero, \bmone, \bmtwo, \bmthree \}$. Moreover, if $\{ \bmo, \bmiota \}$ denote a spin bases satisfying $\llbracket \bmo, \bmiota \rrbracket =1$, where $\llbracket .,.\rrbracket$ is the antisymmetric product, then the components of a generic spinor $\zeta_{A}$ can be written as $\zeta_{\bmA} = \zeta_{A} \bmepsilon_{\bmA}{}^A{}$, where 
\begin{eqnarray}
&& \bmo^A = \bmepsilon_{\bmzero}{}^A{}, \qquad  \bmiota^A = \bmepsilon_{\bmone}{}^A{}, \nonumber \\
&& \bmo_A = \bmepsilon^{\bmone}{}_A{} , \qquad \bmiota_A = - \bmepsilon^{\bmzero}{}_A{}.
\label{Spin-frames-definition}
\end{eqnarray}
The antisymmetric product $\llbracket .,. \rrbracket$ of two generic spinors $\zeta$ and $\lambda$ can be expressed as 
$$
\llbracket \zeta, \lambda \rrbracket = \zeta_B \lambda^B = \epsilon_{AB} \zeta^A \lambda^B,
$$
where $\epsilon_{AB}$ is the $\epsilon$-spinor that can be regarded as an index raising/lowering object for spinors. Throughout, we express spacetime frames $\{ \bme_{\bma} \}$ in spinorial notation. The spinorial counterpart of $\{ \bme_{\bma} \}$ is given by
$$
\bme_{\bmA \bmA'} = \sigma^{\bma}{}_{\bmA \bmA'} \bme_{\bma}
$$
where $ \sigma^{\bma}{}_{\bmA \bmA'}$ are the Infeld-van der Waerden symbols. 
Finally, the signature convention for spacetime metrics used in this article is
$(+,-,-,-)$. Throughout, we mostly follow the notation and conventions
of Penrose and Rindler \cite{PenRind86} ---see also \cite{kroon16}.

\section{Conformal geometry tools}
\label{Section: Conformal tools}
The purpose of this section is to provide a brief introduction of the conformal tools utilised throughout this article. In the following, let $\tilde{\mathcal{M}}$ denote a four-dimensional Lorentzian manifold. The metrics $\tilde{\bmg}$ and $\hat{\bmg}$ on $\tilde{\mathcal{M}}$ are \emph{conformally related} if there exists a positive function $\Omega$ such that
\begin{equation*}
    \hat{\bmg} = \Omega^{2} \tilde{\bmg},
\end{equation*}
where $\Omega$ is known as the \emph{conformal factor}. On the other hand, if $\mathcal{M}$ is a four-dimensional manifold with metric $\bmg$, then one defines a \emph{conformal transformation} $\phi$ as the diffeomorphic map $\phi: \tilde{\mathcal{M}} \to \mathcal{M}$ such that
\begin{equation*}
    \phi^{*} \bmg = \Xi^{2} \tilde{\bmg}.
\end{equation*}
where $\Xi$ denotes the conformal factor which is a positive function on $\tilde{\mathcal{M}}$ and $\phi^{*} \bmg$ is the pull-back of $\bmg$ to $\tilde{\mathcal{M}}$. Given the above, one can define:
\begin{definition}[conformal compactification]{\em
    Let $\mathcal{U}$ denote a compact, connected and open subset of $\mathcal{M}$, then the diffeomorphic map $\phi: \tilde{\mathcal{M}} \to \mathcal{U}$ defines a conformal compactification of $\tilde{\mathcal{M}}$ if there exists a positive function $\Xi$ satisfying 
    \begin{enumerate}
        \item[i.] $\Xi > 0$ in $\mathcal{U}$,
        \item[ii.] $\Xi =0$ on the boundary of the open set $\mathcal{U}$, denoted $\partial \mathcal{U}$,
    \end{enumerate}
    and if $\bmg$ is related to $\tilde{\bmg}$ by 
    \begin{equation}
        \bmg = (\phi^{*})^{-1} (\Xi^{2} \tilde{\bmg}) \qquad \text{in } \quad \mathcal{U}.
        \label{Conformal-compactification}
    \end{equation}
    In this context, $\partial \mathcal{U}$ is known as the conformal boundary of $\tilde{\mathcal{M}}$.}
\end{definition}
\begin{remark}
   {\em Throughout, we omit $\phi^{*}$ and $(\phi^{*})^{-1} $ when discussing conformal transformations and compactifications ---e.g., eq. \eqref{Conformal-compactification} will be written as $\bmg = \Xi^{2} \tilde{\bmg}$.}
\end{remark}
\subsection{The metric conformal field equations}
The conformal transformation given by eq. \eqref{Conformal-compactification} implies transformation laws for the Levi-Civita connections associated with $\tilde{\bmg}$ and $\bmg$ (denoted by $\tilde{\bmnabla}$ and $\bmnabla$) and other related fields ---e.g. the Riemann curvature tensors $\tilde{R}^{a}{}_{bcd}$ and $R^{a}{}_{bcd}$, and the Ricci tensors $\tilde{R}_{ab}$ and $R_{ab}$ etc. The derivation of these formulae is discussed in \cite{kroon16} but will not be necessary for our discussion.

Now, assume that $(\tilde{\mathcal{M}},\tilde{\bmg})$ satisfy the vacuum Einstein field equations, so that
\begin{equation}
    \tilde{R}_{ab} =0.
    \label{vacuum-Einstein-field-equations}
\end{equation}
The transformation law of the Ricci tensor implied by eq. \eqref{Conformal-compactification} yields a singular expression for $R_{ab}$ at $\Xi=0$. However, the prescription in \cite{Friedrich81a,Friedrich81b,Friedrich83,kroon16} introduces a set of field equations on $\mathcal{M}$ that are well-defined at the conformal boundary. We will refer to these equations as the \emph{metric conformal field equations}, and they are given by
\begin{subequations}
    \begin{eqnarray}
        && \nabla_{a} \nabla_{b} \Xi = - \Xi L_{ab} + s g_{ab}, \label{vacuum-metric-conformal-field-equations1} \\
        && \nabla_{a} s = - L_{ac} \nabla^{c} \Xi, \label{vacuum-metric-conformal-field-equations2} \\
        && \nabla_{c} L_{db} - \nabla_{d} L_{cb} = \nabla_{a} \Xi d^{a}{}_{bcd}, \label{vacuum-metric-conformal-field-equations3} \\
        && \nabla_{a}d^{a}{}_{bcd} =0,
        \label{vacuum-metric-conformal-field-equations4} \\
        && 6 \Xi s - 3 \nabla_{c} \Xi \nabla^{c} \Xi =0, \label{vacuum-metric-conformal-field-equations5}
    \end{eqnarray}
    \label{vacuum-metric-conformal-field-equations}
\end{subequations}
where $\bmnabla$ is the $\bmg$-Levi-Civita connection, $L_{ab}$ is the Schouten tensor associated with $\bmnabla$, $d^{a}{}_{bcd}$ is the rescaled Weyl tensor, defined in terms of the Weyl tensor $C^{a}{}_{bcd}$ as
\begin{equation}
    d^{a}{}_{bcd} = \Xi^{-1} C^{a}{}_{bcd}.
    \label{Definition-rescaled-Weyl-tensor}
\end{equation}
Finally, $s$ denotes Friedrich's scalar, given by
\begin{equation*}
    s \equiv \frac{1}{4} \nabla^{c} \nabla_{c} \Xi + \frac{1}{24} R \Xi,
\end{equation*}
where $R$ is the Ricci scalar associated with $\bmnabla$. 

Note that eqs. \eqref{vacuum-metric-conformal-field-equations} exhibit conformal gauge freedom manifested by the fact that a solution to eq. \eqref{vacuum-Einstein-field-equations} will correspond to an infinite number of solutions to eq. \eqref{vacuum-metric-conformal-field-equations}. In our analysis, our focus will be on an equivalent set of conformal field equations exhibiting additional gauge freedom compared to eqs. \eqref{vacuum-metric-conformal-field-equations}. 
\subsection{The extended conformal field equations}
\label{Section: The extended conformal field equations}
This section briefly overviews the conformal formulation of the Einstein field equations introduced by Friedrich in \cite{Friedrich98}. The result of this formulation is a set of equations known as the extended conformal field equations (ECFEs).

Following the discussion in \cite{kroon16}, let $(\Tilde{\mathcal{M}}, \Tilde{\bmg})$ denote the physical spacetime satisfying eqs. \eqref{vacuum-Einstein-field-equations}, and let $(\mathcal{M},\bmg)$ denote the unphysical spacetime with
\begin{equation}
    g_{ab} = \Xi^2 \Tilde{g}_{ab},
    \label{metric-conformal-transformation}
\end{equation}
Given a $\bmg$-orthonormal frame $\{ \bme_{\bma} \}$ and a dual frame $\{ \bmomega^{\bma} \}$, one can write
\begin{subequations}
    \begin{eqnarray}
        && g_{\bma \bmb} = \Xi^2 \Tilde{g}_{\bma \bmb}, \label{metric-frames} \\
        && g^{\bma \bmb} = \Xi^{-2} \Tilde{g}^{\bma \bmb}.
        \label{co-metric-frames}
    \end{eqnarray}
\end{subequations}
To introduce the ECFEs, introduce the \emph{Weyl connection} as a torsion-free connection $\hat{\bmnabla}$ satisfying
\begin{equation}
    \hat{\nabla}_{\bma} g_{\bmb \bmc} = - 2 f_{\bma} g_{\bmb \bmc},
    \label{Weyl-connection}
\end{equation}
where $\hat{\nabla}_{\bma} = \bme_{\bma}{}^{a} \hat{\nabla}_a$. The relation between $\hat{\bmnabla}$, $\bmnabla$ and $\tilde{\bmnabla}$ is then given by
\begin{subequations}
    \begin{eqnarray}
    && \hat{\bm{ \nabla}} - \bmnabla = \bmS(\bmf), \\
    && \hat{\bmnabla} - \Tilde{\bmnabla} = \bmS(\bmbeta),
\end{eqnarray}
\label{Weyl-Unphysical-physical-Connections-relation}
\end{subequations}
where $\bmbeta \equiv \bmf + \Xi^{-1} \bmd{\Xi}$ and $\bmS(\bmf)$ can be written as
\begin{equation*}
    \bmS(\bmf) = S_{a b}{}^{c d} f_{d},
\end{equation*}
with
\begin{equation}
    S_{a b}{}^{c d} = \delta_{a}{}^{c} \delta_{b}{}^{d} + \delta_{a}{}^{d} \delta_{b}{}^{c}-g_{a b} g^{c d},
    \label{The-S-tensor}
\end{equation}
where $\delta_{a}{}^{c}$ is the Kronecker delta. Given the above, the extended conformal field equations can be written explicitly in terms of the zero quantities $\hat{\Sigma}_{\bma \bmb}, \hat{\Xi}^{\bmc}{}_{\bmd \bma \bmb}, \hat{\Delta}_{\bmc \bmd \bmb} \text{ and } \hat{\Lambda}_{\bmb \bmc \bmd}$ as 
\begin{equation}
    \hat{\Sigma}_{\bma \bmb}=0, \quad \hat{\Xi}^{\bmc}{}_{\bmd \bma \bmb}=0, \quad \hat{\Delta}_{\bmc \bmd \bmb}=0, \quad \hat{\Lambda}_{\bmb \bmc \bmd}=0,
    \label{ECFE}
\end{equation}
where $(\hat{\Sigma}_{\bma \bmb}, \hat{\Xi}^{\bmc}{}_{\bmd \bma \bmb}, \hat{\Delta}_{\bmc \bmd \bmb}, \hat{\Lambda}_{\bmb \bmc \bmd})$ are defined as
\begin{subequations}
    \begin{eqnarray*}
        && \hat{\Sigma}_{\bma \bmb} \equiv [\bme_{\bmb}, \bme_{\bma}] - (\hat{\Gamma}_{\bma}{}^{\bmc}{}_{\bmb} - \hat{\Gamma}_{\bmb}{}^{\bmc}{}_{\bma}) \bme_{\bmc}, \\
        && \hat{\Xi}^{\bmc}{}_{\bmd \bma \bmb} \equiv \hat{P}^{\bmc}{}_{\bmd \bma \bmb} - \hat{\rho}^{\bmc}{}_{\bmd \bma \bmb}, \\
        && \hat{\Delta}_{\bmc \bmd \bmb} \equiv \hat{\nabla}_{\bmc} \hat{L}_{\bmd \bmb} - \hat{\nabla}_{\bmd} \hat{L}_{\bmc \bmb} - d_{\bma} d^{\bma}{}_{\bmb \bmc \bmd}, \\
        && \hat{\Lambda}_{\bmb \bmc \bmd} \equiv \hat{\nabla}_{\bma}d^{\bma}{}_{\bmb \bmc \bmd} - f_{\bma} d^{\bma}{}_{\bmb \bmc \bmd}.
    \end{eqnarray*}
    \label{The-ECFEs-zero-quantities}
\end{subequations}
Here, $[\bme_{\bmb}, \bme_{\bma}]$ is the commutator defined as $[\bme_{\bmb}, \bme_{\bma}] \equiv \bme_{\bmb}(\bme_{\bma}(f))- \bme_{\bma}(\bme_{\bmb}(f))$ for any function $f$ on $\mathcal{M}$, $\hat{L}_{\bma \bmb}$ are the components of the Schouten tensor $\hat{\bmL}$ associated with $\hat{\bmnabla}$, $\hat{\Gamma}_{\bma}{}^{\bmb}{}_{\bmc}$ are the $\hat{\bmnabla}$-connection coefficients, and $d_a$ is a 1-form related to $f_{a}$ by
\begin{equation*}
    d_a = \Xi f_a + (\bmd{\Xi})_a.
\end{equation*}
Finally, $\hat{P}^{\bmc}{}_{\bmd \bma \bmb}$ and $\hat{\rho}^{\bmc}{}_{\bmd \bma \bmb}$ are the components of the so-called geometric and algebraic curvature with respect to $\{ \bme_{\bma} \}$, defined by
\begin{subequations}
    \begin{eqnarray*}
        && \hat{P}^{\bmc}{}_{\bmd \bma \bmb} \equiv \bme_{\bma}(\hat{\Gamma}_{\bmb}{}^{\bmc}{}_{\bmd}) - \bme_{\bmb}(\hat{\Gamma}_{\bma}{}^{\bmc}{}_{\bmd}) + \hat{\Gamma}_{\bmf}{}^{\bmc}{}_{\bmd} (\hat{\Gamma}_{\bmb}{}^{\bmf}{}_{\bma}-\hat{\Gamma}_{\bma}{}^{\bmf}{}_{\bmb}) + \hat{\Gamma}_{\bmb}{}^{\bmf}{}_{\bmd} \hat{\Gamma}_{\bma}{}^{\bmc}{}_{\bmf} - \hat{\Gamma}_{\bma}{}^{\bmf}{}_{\bmd} \hat{\Gamma}_{\bmb}{}^{\bmc}{}_{\bmf}, \\
        && \hat{\rho}^{\bmc}{}_{\bmd \bma \bmb} \equiv \Xi d^{\bmc}{}_{\bmd \bma \bmb} + 2 S_{\bmd [\bma}{}^{\bmc \bmf} \hat{L}_{\bmb] \bmf}.
    \end{eqnarray*}
\end{subequations}
The ECFEs yield partial differential equations to be solved for the unknowns
\begin{equation*}
    (\bme_{\bma}, \hat{\Gamma}_{\bma}{}^{\bmb}{}_{\bmc}, \hat{L}_{\bma \bmb}, d^{\bma}{}_{\bmb \bmd \bmc}).
\end{equation*}
In addition to eqs. \eqref{ECFE}, introduce the zero quantities $\delta_{\bma}, \gamma_{\bma \bmb}$ and $\varsigma_{\bma \bmb}$ satisfying
\begin{equation}
    \delta_{\bma} =0, \quad \gamma_{\bma \bmb} =0, \quad \varsigma_{\bma \bmb} =0,
    \label{Supplementary-ECFE}
\end{equation}
where 
\begin{subequations}
    \begin{eqnarray*}
        && \delta_{\bma} \equiv d_{\bma} - \Xi f_{\bma} - \hat{\nabla}_{\bma} \Xi, \\
        && \gamma_{\bma \bmb} = \hat{L}_{\bma \bmb} - \hat{\nabla}_{\bma} \beta_{\bmb} - \frac{1}{2} S_{\bma \bmb}{}^{\bmc \bmd} \beta_{\bmc} \beta_{\bmd}, \\
        && \varsigma_{\bma \bmb} = \hat{L}_{[\bma \bmb]} - \hat{\nabla}_{[\bma} f_{\bmb]}.
    \end{eqnarray*}
\end{subequations}
The supplementary eqs. \eqref{Supplementary-ECFE} relate the solutions of the ECFEs to Einstein's field equations. In particular, given a solution $(\bme_{\bma}, \hat{\Gamma}_{\bma}{}^{\bmb}{}_{\bmc}, \hat{L}_{\bma \bmb}, d^{\bma}{}_{\bmb \bmd \bmc})$ to the ECFEs with a choice of $\Xi$ and $d_{\bma}$ that satisfies the supplementary eqs. \eqref{Supplementary-ECFE}, then if $\Xi \neq 0$ and $\text{det}(\eta^{\bma \bmb} \bme_{\bma} \otimes \bme_{\bmb}) \neq 0$ on some open set $\mathcal{U}$ of $\mathcal{M}$, the metric 
$$
\Tilde{g}_{\bma \bmb} = \Xi^{-2} \eta_{\bma \bmb} \bmomega^{\bma} \otimes \bmomega^{\bmb}
$$
is a solution to eq. \eqref{vacuum-Einstein-field-equations} on $\mathcal{U}$. Here $\eta_{\bma \bmb}$ is used to denote the components of the Minkowski metric with respect to a Cartesian coordinate frame field, i.e., $\eta_{\bma \bmb} = \text{diag}(1,-1,-1,-1)$.

Similar to the metric conformal field equations introduced in the previous section, the ECFEs exhibit conformal gauge freedom. To demonstrate this, define the conformal metric $\grave{\bmg}$ such that
\begin{subequations}
    \begin{eqnarray*}
        && \grave{g}_{\bma \bmb} = \grave{\Xi}^2 \Tilde{g}_{\bma \bmb}, \\
        && \grave{g}_{\bma \bmb} = \kappa^2 g_{\bma \bmb}.
    \end{eqnarray*}
\end{subequations}
Thus, we have $\grave{\Xi} = \kappa^{-1} \Xi$. Then, introduce the Levi-Civita connection $\grave{\bmnabla}$ associated with $\grave{\bmg}$ and define the Weyl connection $\check{\bmnabla}$ as 
$$
\check{\nabla}_{\bma}g_{\bmb \bmc} = - 2 \grave{f}_{\bma} g_{\bmb \bmc},
$$
so that the relation between $\grave{\bmnabla}$, $\check{\bmnabla}$ and $\hat{\bmnabla}$ is given by
\begin{subequations}
    \begin{eqnarray*}
        && \grave{\bmnabla} - \check{\bmnabla} = \bmS(\grave{\bmf}), \\
        && \hat{\bmnabla} - \check{\bmnabla} = \bmS(\bmk + \kappa^{-1} \bmd{\kappa}), 
    \end{eqnarray*}
\end{subequations}
where $\bmk = \bmf - \grave{\bmf}$. In terms of the above, the conformal covariance of the ECFEs can now be expressed as follows: if $(\bme_{\bma}, \hat{\Gamma}_{\bma}{}^{\bmb}{}_{\bmc}, \hat{L}_{\bma \bmb}, d^{\bma}{}_{\bmb \bmd \bmc})$ is a solution to eqs. \eqref{ECFE}, the collection $(\grave{\bme}_{\bma}, \check{\Gamma}_{\bma}{}^{\bmb}{}_{\bmc}, \check{L}_{\bma \bmb}, \grave{d}^{\bma}{}_{\bmb \bmd \bmc})$ with 
\begin{subequations}
    \begin{eqnarray*}
        && \grave{\bme}_{\bma} = \kappa \bme_{\bma}, \\
        && \check{\Gamma}_{\bma}{}^{\bmb}{}_{\bmc} = \kappa \hat{\Gamma}_{\bma}{}^{\bmb}{}_{\bmc} + \delta_{\bmc}{}^{\bmb} \hat{\nabla}_{\bma}\kappa - \kappa S_{\bma \bmc}{}^{\bmb \bmd} (k_{\bmd}+ \kappa^{-1} \hat{\nabla}_{\bmd}\kappa), \\ 
        && \check{L}_{\bma \bmb} = \kappa^2 \hat{L}_{\bma \bmb} - \kappa^2 \hat{\nabla}_{\bma}(k_{\bmb}+ \kappa^{-1} \hat{\nabla}_{\bmb}\kappa) - \frac{1}{2} \kappa^2 S_{\bma \bmb}{}^{\bmc \bmd} (k_{\bmc}+ \kappa^{-1} \hat{\nabla}_{\bmc}\kappa) (k_{\bmd}+ \kappa^{-1} \hat{\nabla}_{\bmd}\kappa), \\
        && \grave{d}^{\bma}{}_{\bmb \bmc \bmd} = \kappa^3 d^{\bma}{}_{\bmb \bmc \bmd},
    \end{eqnarray*}
\end{subequations}
is also a solution to the ECFEs. However, note that the ECFEs exhibit additional gauge freedom, corresponding to the freedom in the choice of $\hat{\bmnabla}$, compared to the metric conformal field equations. Therefore, to obtain a solution to eqs. \eqref{ECFE}, our analysis of the ECFEs will require a gauge choice. In Section \ref{Section:The conformal field equations in the Conformal Gaussian gauge}, the final form of the ECFEs will be obtained by introducing the so-called conformal Gaussian gauge that allows us to fix the conformal factor $\Xi$ and $\hat{\bmnabla}$.

\begin{remark}
{\em  As our main focus will be on obtaining solutions to the ECFEs, rather than the metric conformal field equations, the term conformal field equations will be used interchangeably with ECFEs in later discussions.}
\end{remark}

\subsubsection*{The extended conformal field equations in spinor formulation}
The analysis of the ECFEs will be carried out using spinors. In the following, let $\{ \bme_{\bma} \}$ denote the $\bmg$-orthonormal frame introduced in the previous section and let $\{ \bme_{\bmA \bmA'} \}$ refer to the spinorial version of the frame satisfying 
\begin{equation}
    \bmg(\bme_{\bmA \bmA'},\bme_{\bmB \bmB'}) = \epsilon_{\bmA \bmB} \bar{\epsilon}_{\bmA' \bmB'}.
\end{equation}
Then, introduce the spinorial counterparts of $\hat{P}^{\bmc}{}_{\bmd \bma \bmb}, \hat{\rho}^{\bmc}{}_{\bmd \bma \bmb}, \hat{L}_{\bma \bmb}, d_{\bma}, d^{\bma}{}_{\bmb \bmc \bmd}$ and $f_{\bma}$ as
$$
\hat{P}^{\bmC \bmC'}{}_{\bmD \bmD' \bmA \bmA' \bmB \bmB'}, \hat{\rho}^{\bmC \bmC'}{}_{\bmD \bmD' \bmA \bmA' \bmB \bmB'}, \hat{L}_{\bmA \bmA' \bmB \bmB'}, d_{\bmA \bmA'}, d^{\bmA \bmA'}{}_{\bmB \bmB' \bmC \bmC' \bmD \bmD'}, f_{\bmA \bmA'}
$$
Since the definition of geometric and algebraic curvature is given in terms of the connection coefficients $\hat{\Gamma}_{\bma}{}^{\bmb}{}_{\bmc}$, introduce the spinorial connection coefficients $\hat{\Gamma}_{\bmA \bmA'}{}^{\bmB \bmB'}{}_{\bmC \bmC'}$ which can be decomposed as
\begin{equation}
    \hat{\Gamma}_{\bmA \bmA'}{}^{\bmB \bmB'}{}_{\bmC \bmC'} = \hat{\Gamma}_{\bmA \bmA'}{}^{\bmB}{}_{\bmC} \delta_{\bmC'}{}^{\bmB'} + \bar{\hat{\Gamma}}_{\bmA \bmA'}{}^{\bmB'}{}_{\bmC'} \delta_{\bmC}{}^{\bmB},
    \label{Decomposition-Connection-Coefficients-2}
\end{equation}
where $\hat{\Gamma}_{\bmA \bmA'}{}^{\bmB}{}_{\bmC}$ are known as the spin connection coefficients. Given the above, the relation between the Weyl connection $\hat{\bmnabla}$ and the unphysical connection $\bmnabla$ given by eq. \eqref{Weyl-Unphysical-physical-Connections-relation} implies that
$$
\hat{\Gamma}_{\bmA \bmA'}{}^{\bmB}{}_{\bmC} = \Gamma_{\bmA \bmA'}{}^{\bmB}{}_{\bmC} + \delta_{\bmA}{}^{\bmB} f_{\bmC \bmA'}, \qquad \hat{\Gamma}_{\bmA \bmA'}{}^{\bmQ}{}_{\bmQ} = f_{\bmA \bmA'}.
$$
Then, $\hat{P}^{\bmC \bmC'}{}_{\bmD \bmD' \bmA \bmA' \bmB \bmB'}$ and $\hat{\rho}^{\bmC \bmC'}{}_{\bmD \bmD' \bmA \bmA' \bmB \bmB'}$ can be decomposed as 
\begin{subequations}
    \begin{eqnarray*}
        && \hat{P}^{\bmC \bmC'}{}_{\bmD \bmD' \bmA \bmA' \bmB \bmB'} = \hat{P}^{\bmC}{}_{\bmD \bmA \bmA' \bmB \bmB'} \delta_{\bmD'}{}^{\bmC'} + \bar{\hat{P}}^{\bmC'}{}_{\bmD' \bmA \bmA' \bmB \bmB'} \delta_{\bmD}{}^{\bmC}, \\
        && \hat{\rho}^{\bmC \bmC'}{}_{\bmD \bmD' \bmA \bmA' \bmB \bmB'} = \hat{\rho}^{\bmC}{}_{\bmD \bmA \bmA' \bmB \bmB'} \delta_{\bmD'}{}^{\bmC'} + \bar{\hat{\rho}}^{\bmC'}{}_{\bmD' \bmA \bmA' \bmB \bmB'} \delta_{\bmD}{}^{\bmC},
    \end{eqnarray*}
\end{subequations}
where 
\begin{subequations}
    \begin{eqnarray*}
        && \hat{P}_{\bmA \bmB \bmC \bmC' \bmD \bmD'} = \hat{P}_{(\bmA \bmB) \bmC \bmC' \bmD \bmD'} + \frac{1}{2} \epsilon_{\bmA \bmB} \left(\hat{\nabla}_{\bmC \bmC'} f_{\bmD \bmD'} - \hat{\nabla}_{\bmD \bmD'} f_{\bmC \bmC'} \right), \\
        && \hat{\rho}_{\bmA \bmB \bmC \bmC' \bmD \bmD'} = \hat{\rho}_{(\bmA \bmB) \bmC \bmC' \bmD \bmD'} + \frac{1}{2} \epsilon_{\bmA \bmB} \left( \hat{L}_{\bmC \bmC' \bmD \bmD'} -  \hat{L}_{\bmD \bmD' \bmC \bmC'} \right). 
    \end{eqnarray*}
\end{subequations}
Given these definitions, the extended conformal field equations can be written in terms of the zero quantities $(\hat{\Sigma}_{\bmA \bmA' \bmB \bmB'}, \hat{\Xi}^{\bmC}{}_{\bmD \bmA \bmA' \bmB \bmB'} , \hat{\Delta}_{\bmC \bmC' \bmD \bmD' \bmB \bmB'}, \hat{\Lambda}_{\bmB \bmB' \bmC \bmC' \bmD \bmD'} )$ as
\begin{equation}
    \hat{\Sigma}_{\bmA \bmA' \bmB \bmB'} = 0, \qquad \hat{\Xi}^{\bmC}{}_{\bmD \bmA \bmA' \bmB \bmB'} =0, \qquad \hat{\Delta}_{\bmC \bmC' \bmD \bmD' \bmB \bmB'} =0, \qquad \hat{\Lambda}_{\bmB \bmB' \bmC \bmC' \bmD \bmD'} =0,
\end{equation}
where 
\begin{subequations}
    \begin{eqnarray*}
        && \hat{\Sigma}_{\bmA \bmA' \bmB \bmB'} \equiv [\bme_{\bmA \bmA'}, \bme_{\bmB \bmB'}] - \left( \hat{\Gamma}_{\bmA \bmA'}{}^{\bmC \bmC'}{}_{\bmB \bmB'} - \hat{\Gamma}_{\bmB \bmB'}{}^{\bmC \bmC'}{}_{\bmA \bmA'} \right) \bme_{\bmC \bmC'}, \\
        && \hat{\Xi}^{\bmC}{}_{\bmD \bmA \bmA' \bmB \bmB'} \equiv \hat{P}^{\bmC}{}_{\bmD \bmA \bmA' \bmB \bmB'} - \hat{\rho}^{\bmC}{}_{\bmD \bmA \bmA' \bmB \bmB'}, \\
        && \hat{\Delta}_{\bmC \bmC' \bmD \bmD' \bmB \bmB'} \equiv \hat{\nabla}_{\bmC \bmC'} \hat{L}_{\bmD \bmD' \bmB \bmB'} - \hat{\nabla}_{\bmD \bmD'} \hat{L}_{\bmC \bmC' \bmB \bmB'} - d_{\bmA \bmA'} d^{\bmA \bmA'}{}_{\bmB \bmB' \bmC \bmC' \bmD \bmD'}, \\
        && \hat{\Lambda}_{\bmB \bmB' \bmC \bmC' \bmD \bmD'} \equiv \hat{\nabla}_{\bmA \bmA'} d^{\bmA \bmA'}{}_{\bmB \bmB' \bmC \bmC' \bmD \bmD'} - f_{\bmA \bmA'} d^{\bmA \bmA'}{}_{\bmB \bmB' \bmC \bmC' \bmD \bmD'}.
    \end{eqnarray*}
\end{subequations}
One can also define the spinorial counterparts of the zero quantities $\delta_{\bma}, \gamma_{\bma \bmb}$ and $\varsigma_{\bma \bmb}$ as 
$$
\delta_{\bmA \bmA'}, \qquad \gamma_{\bmA \bmA' \bmB \bmB'}, \qquad \varsigma_{\bmA \bmA' \bmB \bmB'},
$$
that can be expressed in terms of the spinorial counterparts of $d_{\bma}, f_{\bma}, \hat{L}_{\bma \bmb}, \beta_{\bma}$. Then, eqs. \eqref{Supplementary-ECFE} can be written as 
\begin{equation}
    \delta_{\bmA \bmA'} = 0, \qquad \gamma_{\bmA \bmA' \bmB \bmB'} =0, \qquad \varsigma_{\bmA \bmA' \bmB \bmB'} =0.
\end{equation}

\subsection{Conformal Geodesics}
\label{Section:Conformal geodesics}
As will become evident, Friedrich's formulation of spatial infinity relies on a class of conformal curves known as conformal geodesics. The aim of this section is to briefly introduce the notion of conformal geodesics utilised in this work.
\begin{definition}[conformal geodesics] {\em
    Let $I \subseteq \mathbb{R}$ and $\tau \in I$, then the curve $x(\tau)$ is a conformal geodesic on the physical spacetime $(\Tilde{\mathcal{M}},\Tilde{\bmg})$ if there exists a 1-form $\bmbeta(\tau)$ along $x(\tau)$ such that
\begin{subequations}
    \begin{eqnarray}
        && \Tilde{\nabla}_{\dot{\bmx}} \dot{\bmx} = - 2 \langle \bmbeta,\dot{\bmx}  \rangle \dot{\bmx} + \Tilde{\bmg}(\dot{\bmx},\dot{\bmx}) \bmbeta^{\sharp}, \\
        && \Tilde{\nabla}_{\dot{\bmx}} \bmbeta = \langle \bmbeta , \dot{\bmx} \rangle - \frac{1}{2} \Tilde{\bmg}^{\sharp}(\bmbeta, \bmbeta) \dot{\bmx}^{\flat} + \Tilde{\bmL}(\dot{\bmx},.),
    \end{eqnarray}
    \label{conformal-geodesic-equations}
\end{subequations}
where $\dot{\bmx}$ denotes the tangent vector and $\tilde{\bmL}$ is the Schouten tensor associated with $\tilde{\bmnabla}$.
}
\end{definition}
In the following, let $\tilde{\mathcal{S}}$ denote a spacelike submanifold of $(\tilde{\mathcal{M}},\tilde{\bmg})$ and consider the smooth initial data on $\tilde{\mathcal{S}}$ as
$$
x_{\star} \in \tilde{\mathcal{S}}, \qquad \dot{\bmx}_{\star} \in T|_{x_{\star}}(\tilde{\mathcal{S}}), \qquad \bmbeta_{\star} \in T^{*}|_{x_{\star}}(\tilde{\mathcal{S}}),
$$
where $T|_{x_{\star}}(\tilde{\mathcal{S}})$ and $T^{*}|_{x_{\star}}(\tilde{\mathcal{S}})$ denote the tangent and dual tangent space at $x_{\star} \in \tilde{\mathcal{S}}$, respectively. Given these initial data, there exists a unique conformal geodesic $(x(\tau), \bmbeta(\tau))$ passing through each $x_{\star} \in \tilde{\mathcal{S}}$ such that
\begin{equation}
    x(0) \equiv x_{\star}, \qquad \dot{\bmx}(0) \equiv \dot{\bmx}_{\star}, \qquad \bmbeta(0) \equiv \bmbeta_{\star}.
    \label{conformal-geodesics-initial-data}
\end{equation}
To illustrate some of the useful properties of conformal geodesics, introduce the Weyl connection $\hat{\bmnabla}$ satisfying
\begin{equation}
     \hat{\nabla}_{a} \tilde{g}_{bc} = -2 \beta_{a} \tilde{g}_{bc},
     \label{Conformal-geodesics-Weyl-connection}
\end{equation}
where $\beta_{a}$ is the 1-form satisfying the conformal geodesics eqs. \eqref{conformal-geodesic-equations}. Given this definition, the relation between $\tilde{\bmnabla}$ and $\hat{\bmnabla}$ is given by
\begin{equation*}
    \hat{\bmnabla} - \tilde{\bmnabla} = \bmS(\bmbeta),
\end{equation*}
where $\bmS(\bmbeta)$ can be written as
$$
\bmS(\bmbeta) = S_{a b}{}^{c d} \beta_{d},
$$
with
\begin{equation}
    S_{a b}{}^{c d} = \delta_{a}{}^{c} \delta_{b}{}^{d} + \delta_{a}{}^{d} \delta_{b}{}^{c}-\tilde{g}_{a b}\tilde{g}^{c d}.
    \label{The-S-tensor-tilde-g}
\end{equation}
Then, eqs. \eqref{conformal-geodesic-equations} implies 
\begin{equation}
    \hat{\bmnabla}_{\dot{\bmx}} \dot{\bmx} =0, \qquad \hat{\bmL}(\dot{\bmx}, .) =0.
\end{equation}
where $\hat{\bmL}$ is the Schouten tensor associated with $\hat{\bmnabla}$. Furthermore, one can introduce a Weyl-propagated frame $\{ \bme_{\bma} \}$ as 
\begin{equation}
    \hat{\nabla}_{\dot{\bmx}} \bme_{\bma} = 0.
    \label{Conformal-geodesics-Weyl-propagated-frame}
\end{equation}
Hence, a congruence of conformal geodesics satisfying eqs. \eqref{conformal-geodesic-equations} given initial data eq. \eqref{conformal-geodesics-initial-data} singles out a Weyl connection $\hat{\bmnabla}$ and a Weyl-propagated frame $\{ \bme_{\bma} \}$ as suggested by eq. \eqref{Conformal-geodesics-Weyl-connection} and eq. \eqref{Conformal-geodesics-Weyl-propagated-frame}. 

Another essential feature of conformal geodesics is that a non-intersecting congruence of conformal geodesics singles out a canonical conformal factor $\Xi$ and a metric $\bmg = \Xi^2 \Tilde{\bmg}$ such that
\begin{equation}
    \bmg(\dot{\bmx},\dot{\bmx}) = 1.
    \label{conformal-geodesics-g-condition}
\end{equation}
In other words, the metric $\bmg$ is singled out by enforcing that the parameter $\tau$ of the conformal geodesics corresponds to the $\bmg$-proper time. In particular, we have 
\begin{proposition}
\emph{
\label{canconical-confromalFactor}
    Let $(\mathcal{\tilde{M}}, \tilde{\bmg})$ denote a vacuum spacetime satisfying eq. \eqref{vacuum-Einstein-field-equations} and let  $(x(\tau), \bmbeta(\tau))$ denote a solution to the conformal geodesics eqs. \eqref{conformal-geodesic-equations} with initial data eq. \eqref{conformal-geodesics-initial-data}. Then, if $\bmg = \Xi^2 \tilde{\bmg}$ is defined such that $\bmg(\dot{\bmx},\dot{\bmx})=1$, the conformal factor $\Xi$ can be written as a quadratic polynomial in terms of $\tau$, i.e.
    \begin{eqnarray}
    && \Xi(\tau) = \Xi_{\star} + \dot{\Xi}_{\star} \tau + \frac{1}{2} \ddot{\Xi}_{\star} \tau^2 
    \label{canconical-conformalFactor1}
    \end{eqnarray}
    with 
    \begin{eqnarray}
    \dot{\Xi}_{\star} = \langle \bmbeta_{\star},\dot{\bmx}_{\star} \rangle \Xi_{\star}, \qquad \qquad \Xi_{\star} \ddot{\Xi}_{\star} = \frac{1}{2} \tilde{\bmg}^{\sharp}(\bmbeta_{\star},\bmbeta_{\star}).
    \label{canconical-conformalFactor2}
\end{eqnarray}
If $\{ \bme_{\bma} \}$ is a $\bmg$-orthonormal Weyl-propagated frame satisfying eq. \eqref{Conformal-geodesics-Weyl-propagated-frame}, one can show that the components of $\bmbeta$ with respect to $\{ \bme_{\bma} \}$ satisfy
\begin{equation*}
    \Xi \beta_{\bmzero} = \dot{\Xi}, \qquad \Xi \beta_{\bmi}= \Xi_{\star} \beta_{\bmi \star},
\end{equation*}
where $\beta_{\bma} \equiv \langle \bmbeta, \bme_{\bma} \rangle$.}
\end{proposition}
In the next section, we will use the unique properties of conformal geodesics to introduce Friedrich's formulation of spatial infinity.

\section{Friedrich's formulation of spatial infinity}
\label{Section:Friedrich formulation of spatial infinity}
As discussed in the introduction, the strategy of this work is to make use of Friedrich's regular initial value problem of the conformal field equations \cite{Friedrich98} to analyse the behaviour of asymptotic charges near spatial infinity. Following the discussion in \cite{FriedrichKannar00}, let $\Tilde{\mathcal{S}}$ denote a spacelike submanifold of $(\Tilde{\mathcal{M}},\Tilde{\bmg})$ which is asymptotically Euclidean and regular (Definition \ref{Definition:AsympEuclideanAndRegular}) with one asymptotic end, and let $\Tilde{\bmh}$ denote the intrinsic metric induced by $\Tilde{\bmg}$ on $\Tilde{\mathcal{S}}$. Let $\mathcal{S}$ denote the hypersurface mentioned in Definition \ref{Definition:AsympEuclideanAndRegular} with the asymptotic point $i$ and denote by $SU(\mathcal{S})$ the bundle of normalised spin frames over $\mathcal{S}$ with the structure group $SU(2,\mathbb{C})$. Let $\{ \bmepsilon_{\bmA}{}^{A} \}$ with $\bmA \in \{ \bmzero, \bmone \}$ denote the spin frame chosen so that the components of the $\bmepsilon$-spinor are given by
$$
\epsilon_{\bmzero \bmone} = 1, \quad \epsilon^{\bmzero \bmone} = 1.
$$
Also, let $\tau^{AA'}$ denote the spinorial counterpart of the future directed normal vector of $\Tilde{\mathcal{S}}$ satisfying $\tau_{A A'} \tau^{AA'} =2$. Then, $\tau^{AA'}$ can be written as 
\begin{equation}
    \tau^{AA'} = \epsilon_{\bmzero}{}^{A} \bar{\epsilon}_{\bmzero'}{}^{A'} + \epsilon_{\bmone}{}^{A} \bar{\epsilon}_{\bmone'}{}^{A'}.
    \label{Tau-adapted-frame}
\end{equation}
Given $\bmt = \{ t^{\bmA}{}_{\bmB} \} \in SU(2,\mathbb{C})$, one can show 
\begin{subequations}
    \begin{eqnarray*}
        && \epsilon_{\bmA \bmB} t^{\bmA}{}_{\bmC} t^{\bmB}{}_{\bmD} = \epsilon_{\bmC \bmD},\\
        && \tau_{\bmA \bmA'} t^{\bmA}{}_{\bmB} \bar{t}^{\bmA'}{}_{\bmB'} = \tau_{\bmB \bmB'}.
    \end{eqnarray*}
\end{subequations}
Then, given a fixed spin frame $\{ \bmepsilon_{\bmA}{}^{A} \}$ at $i$ and $\bmt \in SU(2,\mathbb{C})$, introduce the transformed frame $\bmepsilon_{\bmA}{}^{A} (\bmt)$ as 
$$
\bmepsilon_{\bmA}{}^{A} (\bmt) = t_{\bmA}{}^{\bmB} \bmepsilon_{\bmB}{}^{A}.
$$
To introduce spin frames in a neighbourhood of the asymptotic point $i \in \mathcal{S}$, consider the metric ball $\mathcal{B}_{a}(i)$ in $\mathcal{S}$ centered at $i$ and choose $a>0$ such that $\mathcal{B}_{a}(i)$ is geodesically convex and the metric $\bmh$ is smooth on $\mathcal{B}_{a}(i)$. Then, construct the $\bmh$-geodesic starting at $i$ and let $\rho$ denote the affine parameter along the geodesic so that $\rho(i)=0$. By fixing $\bmt$, the spin frames $\bmepsilon_{\bmA}{}^{A} (\bmt)$ can be parallelly transported along the geodesic and the frames obtained are denoted by $\bmepsilon_{\bmA}(\rho,\bmt)$, where the upper index for the spin frames will be removed in the subsequent discussion for convenience. Then, let $\mathcal{C}_a$ denote the subset of $SU(\mathcal{S})$ defined as
\begin{equation}
    \mathcal{C}_a = \{ \bmepsilon_{\bmA}(\rho, \bmt) \in SU(\mathcal{S}) | -a < \rho < a, \bmt \in SU(2,\mathbb{C}) \}.
    \label{Fiber-Manifold-Ca}
\end{equation}
From the above, $\mathcal{C}_a$ is diffeomorphic to $\bar{\mathcal{S}}_a$ given by
\begin{equation*}
    \bar{\mathcal{S}}_a = \{ (\rho, \bmt) \in \mathbb{R} \times SU(2,\mathbb{C}) | -a < \rho < a \},
\end{equation*}
To relate the structures on $\mathcal{C}_a$ with those on $\mathcal{S}$, let $\pi$ denote the projection from $SU(\mathcal{S})$ to $\mathcal{S}$ and then denote by $\pi'$ the restriction to $\mathcal{C}_a$ so that $\pi'$ is a projection map from $\mathcal{C}_a$ to $\mathcal{B}_a(i)$. The action of $U(1)$ on $SU(\mathcal{S})$ will imply an action of $U(1)$ on $\mathcal{C}_a$. The quotient under this action $\mathcal{C}'_a = \mathcal{C}_a / U(1)$ is diffeomorphic to $( -a, a  ) \times \mathbb{S}^2$. Subsequently, if $\bmt \in SU(2,\mathbb{C})$ and $s' \in U(1)$, then $\bmepsilon_{\bmA}(\bmt)$ and $\bmepsilon_{\bmA}(s' \bmt)$ can be parallelly transported along the same geodesic and will be given by $\bmepsilon_{\bmA}(\rho,\bmt)$ and $\bmepsilon_{\bmA}(\rho,s' \bmt)$, respectively. Since the function $\rho$ is invariant under the action of $U(1)$, we have
$$
\pi'(\bmepsilon_{\bmA}(\rho,\bmt)) = \pi'(\bmepsilon_{\bmA}(\rho,s' \bmt)),
$$
and the map $\pi'$ can be factored as 
$$
\mathcal{C}_a\xrightarrow{\pi'_{1}}\mathcal{C}'_{a}\xrightarrow{\pi'_{2}} \mathcal{B}_a(i),
$$
where $\pi'_{1}$ is the Hopf fibration and $\pi'_{2}$ is the exponential map. Note that the set $\mathcal{C}'_a$ can be split into two components: $\mathcal{C}'_{a}{}^+$ on which $\rho>0$ and $\mathcal{C}'_{a}{}^-$ on which $\rho<0$. Each of these components can be mapped into the punctured disk $\mathcal{B}_{a}(i) \setminus \{ i \}$ using $\pi'_2$. Additionally, given these projections, the point $i$ can be replaced by the set $\pi'_{2}{}^{-1}(i)$ which is diffeomorphic to $\mathbb{S}^2$. Then if $I_{0} = \{ \rho =0 \} \subset \mathcal{C}_a$, we can identify $\pi'_1(I_0) = \pi'_{2}{}^{-1}(i)$ and $\pi'_{2}{}^{-1}(i)$ glues together the components $\mathcal{C}'_{a}{}^+$ and $\mathcal{C}'_{a}{}^-$.

In the following, it will be assumed that $(\tilde{\mathcal{M}}, \tilde{\bmg})$ is the development of the initial data $(\tilde{\mathcal{S}}, \tilde{\bmh})$ satisfying Definition \ref{Definition:AsympEuclideanAndRegular} and that $(\mathcal{M},\bmg, \Theta)$ is the smooth conformal extension of $(\tilde{\mathcal{M}}, \tilde{\bmg})$ such that $\mathcal{M} = \tilde{\mathcal{M}} \cup \mathscr{I}^+ \cup \mathscr{I}^-$, where $\mathscr{I}^{\pm}$ denotes future and past null infinity, respectively and $\Theta$ is the conformal factor satisfying i) $\Theta > 0$ and $\bmg = \Theta^2 \tilde{\bmg}$ on $\tilde{\mathcal{M}}$, ii) $\Theta = 0$ and $d \Theta \neq 0$ on $\mathscr{I}^{\pm}$. Then, introduce the manifold $\mathcal{M}_{a,\kappa}$ by
$$
\mathcal{M}_{a,\kappa} = \bigg\{ (\tau, \rho, \bmt) \in \mathbb{R} \times \mathbb{R} \times SU(2,\mathbb{C})| \: 0 \leq \rho < a, - \frac{\omega}{\kappa} \leq \tau \leq \frac{\omega}{\kappa} \bigg\},
$$
where $\kappa$ is an arbitrary function such that $\kappa = \rho \kappa'$ with $\kappa'$ smooth and $\kappa'(i) =1$, $\omega \equiv \omega(\rho, \bmt)$ is a smooth non-negative function such that $\omega / \kappa \to 1$ as $\rho \to 0$. Note that a coordinate system can be defined on $\mathcal{M}_{a,\kappa}$ by introducing a coordinate system on $SU(2,\mathbb{C})$ together with the functions $\tau$ and $\rho$.

Similar to previous discussion, the action of $U(1)$ on $SU(2,\mathbb{C})$ implies an action of $U(1)$ on $\mathcal{M}_{a,\kappa}$. The quotient under this action $\mathcal{M}_{a,\kappa} / U(1)$ will be denoted by $\mathcal{M}'_{a,\kappa}$ and the projection from $\mathcal{M}_{a,\kappa}$ onto $\mathcal{N} \subset \mathcal{M}$ by $\bar{\pi}'$, where $\mathcal{N}$ is the domain of influence of $\mathcal{B}_a(i) \setminus \{ i \}$. The map $\bar{\pi}'$ can be factored as 
$$
\mathcal{M}_{a,\kappa} \xrightarrow{\bar{\pi}'_{1}} \mathcal{M}'_{a,\kappa} \xrightarrow{\bar{\pi}'_{2}} \mathcal{N}.
$$
Given the mentioned construction, define the following subsets of $\mathcal{M}_{a,\kappa}$
\begin{subequations}
    \begin{eqnarray}
        && \mathscr{I}^{\pm}_a = \{ (\tau, \rho, \bmt) \in \mathcal{M}_{a,\kappa} | \: 0 <\rho < a, \tau = \pm \frac{\omega}{\kappa} \}, \label{Sets-null-infinity} \\
        && \mathcal{I} = \{ (\tau, \rho, \bmt) \in \mathcal{M}_{a,\kappa} | \: \rho =0, -1 <\tau <1 \}, \\
        && \mathcal{I}^{\pm} = \{ (\tau, \rho, \bmt) \in \mathcal{M}_{a,\kappa} | \: \rho =0, \tau = \pm 1 \}, \\
        && \mathcal{I}_0 = \{ (\tau, \rho, \bmt) \in \mathcal{M}_{a,\kappa} | \: \rho =0, \tau = 0 \},
    \end{eqnarray}
\end{subequations}
where the sets $\mathscr{I}^{\pm}_a$ represent past/future null infinity, $\mathcal{I}$ is the cylinder at spacelike infinity and $\mathcal{I}^{\pm}$ are the sets at which $\mathscr{I}^{\pm}_a$ touches $\mathcal{I}$, known as the \emph{critical sets}. 

\begin{remark}
{\em $\phantom{}$
\begin{enumerate}
    \item[i.] The subscript ${}_a$ in $\mathscr{I}^{\pm}_a$ is used to indicate that the sets $\mathscr{I}^{\pm}_a$ do not represent the entirety of past/future null infinity, rather they map to a part of null infinity close to spacelike infinity. In subsequent discussions, we will drop the subscript for convenience. 
    \item[ii.] Given that $\pi'_1(I_0) = \pi'_{2}{}^{-1}(i)$ and that $\pi'_{2}{}^{-1}(i)$ is diffeomorphic to $\mathbb{S}^2$, it is clear to see that $\bar{\pi}'_{1}(\mathcal{I})$ is diffeomorphic to $\mathbb{R} \times \mathbb{S}^2$, hence the use of the term \emph{cylinder at spatial infinity} to refer to $\mathcal{I}$.
\end{enumerate}}
\end{remark}
So far, the construction $\mathcal{C}_a$ has not been extended to the spacetime $\mathcal{M}_{a,\kappa}$. To do so, assume $\mathcal{C}_a$ defined as previously and define $\mathcal{S}_a = \{ \rho > 0 \} \subset \bar{\mathcal{S}}_a$, then the spin frames $\bmepsilon_{\bmA}(\rho, \bmt) \in SU(\mathcal{S})$ will be transported off $\mathcal{S}_a$ into the spacetime by a certain propagation law along conformal geodesics which are orthogonal to $\mathcal{S}_a$. This allows us to determine the spin frames $\bmepsilon_{\bmA}(\tau, \rho, \bmt)$ at points of $\mathcal{M}_{a,\kappa} \setminus (\mathcal{I} \cup \mathcal{I}^+ \cup \mathcal{I}^-)$ up to multiplication by a phase parameter that corresponds to the action of $U(1)$ on $SU(\mathcal{M})$. 
\begin{remark}
    {\em Subsequent calculations will be carried out on $\mathcal{M}_{a,\kappa}$. We will use the same notation to refer to fields on $\mathcal{M}$ and their pull-back to $\mathcal{M}_{a,\kappa}$ via $\bar{\pi}'$ e.g. use $\Theta$ for $\bar{\pi}'^{*} \Theta$.}
\end{remark}

Following the discussion in \cite{FriedrichKannar00}, let $Z_{u_i}$ denote the vector fields generated by $u_i$, the basis of the Lie Algebra of $SU(2,\mathbb{C})$, given by
\begin{equation}
u_1 = \frac{1}{2} \begin{pmatrix} 
0 & i \\
i & 0
\end{pmatrix}, \qquad \; u_2 = \frac{1}{2} \begin{pmatrix} 
0 & -1 \\
1 & 0
\end{pmatrix}, \qquad \; u_3 = \frac{1}{2} \begin{pmatrix} 
i & 0 \\
0 & -i
\end{pmatrix}.
\label{Basis-Lie-Algebra-SU2C}
\end{equation}
Then, introduce the complex vector fields 
\begin{equation*}
    X_+ = - (Z_{u_2} + i Z_{u_1}), \qquad X_- = - (Z_{u_2} - i Z_{u_1}), \qquad X = - 2 i Z_{u_3},
\end{equation*}
satisfying 
\begin{equation*}
    [X,X_+] = 2X_+, \qquad [X, X_- ] = -2X_-, \qquad [X_+,X_-] = - X.
\end{equation*}
Given the above, consider the smooth vector field
\begin{equation}
    \bme_{\bmA \bmA'} = \bme^{0}{}_{\bmA \bmA'} \partial_{\tau} + \bme^{1}{}_{\bmA \bmA'} \partial_{\rho} + \bme^{+}{}_{\bmA \bmA'} X_+ - \bme^{-}{}_{\bmA \bmA'} X_-,
    \label{F-gauge-frame-fields}
\end{equation}
where $\bme^{\mu}{}_{\bmA \bmA'}$ denotes the components of $\bme_{\bmA \bmA'}$ with respect to the local coordinate system $( \tau, \rho, \bmt )$. If $\bmomega^{\bmA \bmA'}$ denotes the dual frame, then $\langle \bmomega^{\bmA \bmA'}, \bme_{\bmB \bmB'} \rangle = \epsilon_{\bmB}{}^{\bmA} \bar{\epsilon}_{\bmB'}{}^{\bmA'}$ on $\mathcal{M}_{a,\kappa} \setminus \mathcal{I}$. 

Given that Friedrich's formulation is an initial value problem formulation of the conformal field equation, we will be interested in reintroducing the $1+3$ decomposition of the field equations in terms of so-called space-spinors ---see \cite{PenRind84, PenRind86, kroon16} for details. In the space-spinor formulation, the frame fields $\bme_{\bmA \bmA'}$ can be decomposed as follows
$$
\bme_{\bmA \bmA'} = \frac{1}{\sqrt{2}} \tau_{\bmA \bmA'} \partial_{\tau} - \tau^{\bmB}{}_{\bmA'} \bme_{\bmA \bmB},
$$
where $\tau_{\bmA \bmA'}$ denotes the components of the future directed normal of $\mathcal{S}$ and $\bme_{\bmA \bmB}$ is defined by
\begin{equation}
    \bme_{\bmA \bmB} \equiv \tau_{(\bmA}{}^{\bmB'} \bme_{\bmB) \bmB'} = \bme^{0}_{\bmA \bmB} \partial_{\tau} + \bme^{1}_{\bmA \bmB} \partial_{\rho} +\bme^{+}_{\bmA \bmB} X_+ + \bme^{-}_{\bmA \bmB} X_-. 
    \label{Space-spinor-frames}
\end{equation}
Since the ECFEs are written in terms of a Weyl connection, let $\hat{\bmnabla}$ denote a Weyl connection such that 
\begin{equation}
    \hat{\nabla}_{\bma} g_{\bmb \bmc} = - 2 f_{\bma} g_{\bmb \bmc},
\end{equation}
where $\{ \bme_{\bma} \}$ is the tensorial counterpart of the frame fields defined by eq. \eqref{F-gauge-frame-fields}, $g_{\bma \bmb} \equiv \bmg (\bme_{\bma}, \bme_{\bmb})$, and $f_{\bma}$ is an arbitrary 1-form which will be fixed in later discussions.

Now, let $\chi_{\bmA \bmB \bmC \bmD}$ and $\xi_{\bmA \bmB \bmC \bmD}$ denote the real and imaginary parts of $\hat{\Gamma}_{\bmA \bmB \bmC \bmD}$ of the spin connection coefficients $\hat{\Gamma}_{\bmA \bmB \bmC \bmD} \equiv \tau_{(\bmB}{}^{\bmA'} \hat{\Gamma}_{\bmA) \bmA' \bmC \bmD}$ associated with $\hat{\bmnabla}$, defined by
\begin{subequations}
    \begin{eqnarray*}
        && \chi_{\bmA \bmB \bmC \bmD} \equiv - \frac{1}{\sqrt{2}} (\hat{\Gamma}_{\bmA \bmB \bmC \bmD} + \hat{\Gamma}^{+}_{\bmA \bmB \bmC \bmD}), \\
        && \xi_{\bmA \bmB \bmC \bmD} \equiv \frac{1}{\sqrt{2}} (\hat{\Gamma}_{\bmA \bmB \bmC \bmD} - \hat{\Gamma}^{+}_{\bmA \bmB \bmC \bmD}),
    \end{eqnarray*}
\end{subequations}
where $\hat{\Gamma}^{+}_{\bmA \bmB \bmC \bmD}$ is the Hermitian conjugate of $\hat{\Gamma}_{\bmA \bmB \bmC \bmD}$. Then, $\hat{\Gamma}_{\bmA \bmB \bmC \bmD}$ can be written as 
\begin{equation}
    \hat{\Gamma}_{\bmA \bmB \bmC \bmD} = \frac{1}{\sqrt{2}} (\xi_{\bmA \bmB \bmC \bmD} - \chi_{\bmA \bmB \bmC \bmD}) = \frac{1}{\sqrt{2}} (\xi_{\bmA \bmB \bmC \bmD} - \chi_{(\bmA \bmB) \bmC \bmD}) - \frac{1}{2} \epsilon_{\bmA \bmB} f_{\bmC \bmD}, 
    \label{Decomposition-Connection-Coefficients}
\end{equation}
where $f_{\bmA \bmB} \equiv \tau_{(\bmB}{}^{\bmA'} f_{\bmA) \bmA'} = \tau_{(\bmB}{}^{\bmA'} \sigma^{\bma}{}_{\bmA) \bmA'} f_{\bma}$. 

Given that the curvature tensor $\hat{R}^{a}{}_{bcd}$ associated with $\hat{\bmnabla}$ can be written in terms of the rescaled Weyl tensor $d^{a}{}_{bcd}$ and the Schouten tensor $\hat{L}_{ab}$, it is straightforward to see that the spinorial counterpart of $\hat{R}^{a}{}_{bcd}$ is fully determined by the rescaled Weyl spinor $\phi_{\bmA \bmB \bmC \bmD} \equiv \phi_{(\bmA \bmB \bmC \bmD)}$ and the space-spinor counterpart of the Schouten tensor $\Theta_{\bmA \bmB \bmC \bmD}$, where $\phi_{\bmA \bmB \bmC \bmD}$ and $\Theta_{\bmA \bmB \bmC \bmD}$ are defined by
\begin{subequations}
    \begin{eqnarray*}
        && \phi_{\bmA \bmB \bmC \bmD} \equiv - \frac{1}{4} \epsilon^{\bmA' \bmB'} \epsilon^{\bmC' \bmD'} d_{\bmA \bmA' \bmB \bmB' \bmC \bmC' \bmD \bmD'} \\
        && \Theta_{\bmA \bmB \bmC \bmD} \equiv \Theta_{\bmA \bmB (\bmC \bmD)} \equiv \tau_{\bmB}{}^{\bmA'} \tau_{\bmD}{}^{\bmC'} \hat{L}_{\bmA \bmA' \bmC \bmC'}.
    \end{eqnarray*}
\end{subequations}
A further decomposition of $\Theta_{\bmA \bmB \bmC \bmD} \equiv \Theta_{\bmA \bmB (\bmC \bmD)}$ yields
$$
\Theta_{\bmA \bmB \bmC \bmD} = \Theta_{(\bmA \bmB) \bmC \bmD} - \frac{1}{2} \epsilon_{\bmA \bmB} \Theta_{\bmQ}{}^{\bmQ}{}_{\bmC \bmD}.
$$
As shown in \cite{Friedrich95}, the gauge choice based on conformal geodesics implies that the conformal factor $\Theta$ can be expressed in terms of initial data as follows:
\begin{equation}
    \Theta = \kappa^{-1} \Omega \left( 1- \tau^2 \frac{\kappa^2}{\omega^2} \right),
    \label{Conformal-factor-Theta}
\end{equation}
where $\omega$ is given by
\begin{equation}
    \omega = \frac{2 \Omega}{ \sqrt{- D_{\bmA \bmB} \Omega D^{\bmA \bmB} \Omega}}.
    \label{Definition-omega-function}
\end{equation}
Here, $D_{AB}$ denotes the intrinsic covariant derivative on the initial hypersurface $\mathcal{S}_a$. Moreover, one can show that $\Theta$ satisfies
\begin{subequations}
    \begin{eqnarray*}
        && \Theta > 0 \qquad \; \text{on } \mathcal{M}_{a,\kappa}, \qquad \{ \Theta =0 \} = \mathscr{I}^+_a \cup \mathcal{I}^- \cup \mathcal{I} \cup \mathcal{I}^+ \cup \mathscr{I}^+_a, \\
        && \bme_{\bmA \bmA'}(\Theta) \neq 0, \qquad \epsilon^{\bmA \bmB} \bar{\epsilon}^{\bmA' \bmB'} \bme_{\bmA \bmA'}(\Theta) \bme_{\bmB \bmB'}(\Theta) = 0 \qquad \; \text{on } \mathscr{I}^{\pm}_a.
    \end{eqnarray*}
\end{subequations}

For the rest of this article, we will refer to the frame $\{ \bme_{\bmA \bmA'} \}$ (or equivalently $\{ \bme_{\bma}  \}$) satisfying the conditions mentioned, the coordinates $(\tau, \rho, \bmt)$ and the conformal gauge defined above as the F-gauge. In the following section, Friedrich's approach to the conformal field equations will be introduced, where the aim is to encode the above gauge conditions, and their transport laws in the properties of the unknowns appearing in the conformal field equations.  

\subsection{Hyperbolic reduction using the conformal Gaussian gauge}
%comment: Stopped here 13/09/2023
\label{Section:The conformal field equations in the Conformal Gaussian gauge}
This section aims to introduce a hyperbolic reduction procedure of the ECFEs using the so-called conformal Gaussian gauge based on conformal geodesics, following the discussion of Chapter 13 in \cite{kroon16}.

In the following, let $\Tilde{\mathcal{S}} \equiv \{ \tau = 0 \}$ denote a spacelike hypersurface on the physical spacetime $(\Tilde{\mathcal{M}},\Tilde{\bmg})$ and let 
\begin{equation}
    \bme_{\bma_{\star}} = \bme_{\bma}|_{\Tilde{\mathcal{S}}}, \quad \bmbeta_{\star} = \bmbeta|_{\Tilde{\mathcal{S}}},
    \label{conformal-gauss-gauge-initial-data}
\end{equation}
denote a smooth initial data specified at each point $x_{\star} \in \Tilde{\mathcal{S}}$. Furthermore, assume that $(x(\tau), \bmbeta(\tau))|_{x_{\star}}$ is the unique solution to eqs. \eqref{conformal-geodesic-equations} passing through $x_{\star} \in \tilde{\mathcal{S}}$ such that
\begin{equation*}
    x(0) = x_{\star}, \qquad \dot{\bmx}(0) = \bme_{\bmzero\star}, \qquad \bmbeta(0) = \bmbeta_{ \star}.
\end{equation*} 
By varying $x_{\star}$, it can be shown that one obtains a smooth caustic free congruence of conformal geodesics in a \emph{small} neighbourhood $\tilde{\mathcal{U}}$ of $\Tilde{\mathcal{S}}$. 
\begin{remark}
    {\em In subsequent calculations, it will be assumed that $(x(\tau), \bmbeta (\tau))$ is a congruence of conformal geodesics satisfying eqs. \eqref{conformal-geodesic-equations} with initial data eq. \eqref{conformal-gauss-gauge-initial-data} specified on $\tilde{\mathcal{S}}$.} 
\end{remark}
The solution to the conformal geodesic equations allows us to fix the gauge freedom in the ECFEs associated with the choice of $\hat{\bmnabla}$. In particular, one can introduce the Weyl connection $\hat{\bmnabla}$ as 
$$
\hat{\bmnabla} - \Tilde{\bmnabla} = \bmS(\bmbeta).
$$
As mentioned in Section \ref{Section:Conformal geodesics}, the Weyl connection introduces a Weyl propagated frame field $\{ \bme_{\bma} \}$ satisfying eq. \eqref{Conformal-geodesics-Weyl-propagated-frame}. Moreover, one can obtain a canonical conformal factor $\Xi$ by imposing conditions eq. \eqref{conformal-geodesics-g-condition}. Applying $\Tilde{\nabla}_{\dot{\bmx}}$ to eq. \eqref{conformal-geodesics-g-condition} and using the conformal geodesic equations, one can show that $\Xi$ satisfies the evolution equation 
\begin{equation}
    \hat{\nabla}_{\dot{\bmx}} \Xi = \Xi \langle \bmbeta , \dot{\bmx} \rangle.
\end{equation}
Given $\{ \bme_{\bma} \}$ and $\Xi$ that satisfy the above mentioned equations and if the frame $\{ \bme_{\bma} \}$ is adapted to the conformal geodesics so that $\dot{\bmx} = \bme_{\bmzero}$, then the conformal metric $\bmg \equiv \Xi^{2} \tilde{\bmg}$ satisfies
$$
\bmg(\bme_{\bma},\bme_{\bmb}) = \eta_{\bma \bmb}.
$$
Finally, given local coordinates $( x^\alpha )$ on $\Tilde{\mathcal{S}}$ and setting $x^0 = \tau$, the local coordinates on $\Tilde{\mathcal{S}}$ can be dragged along the conformal geodesics to obtain a smooth coordinate system $(x^{\mu}) \equiv ( x^{0}, x^{\alpha} )$ on $\Tilde{\mathcal{U}}$. In this gauge, one can show that
\begin{equation}
    \dot{\bmx} = \bme_{\bmzero}= \partial_{\tau}, \quad \hat{\Gamma}_{\bmzero}{}^{\bma}{}_{\bmb} = 0, \quad \hat{L}_{\bmzero \bma} = 0 \qquad \text{ on } \Tilde{\mathcal{U}},
    \label{CGG-Gauge-Condiitons}
\end{equation}
and that the conformal factor $\Xi$ and 1-form $\bmd \equiv \Xi \bmbeta$ can be expressed in terms of the initial data as:
\begin{subequations}
    \begin{eqnarray}
        && \Xi(\tau) = \Xi|_{\Tilde{\mathcal{S}}} \left( 1 + \langle \bmbeta_{\star},\bme_{\bmzero \star} \rangle \tau + \frac{1}{4} \Tilde{\bmg}^{\sharp}( \bmbeta_{\star},\bmbeta_{\star} ) \tau^2 \right), \\
        && d_{\bmzero} = \dot{\Xi} = \partial_{\tau} \Xi, \quad d_{\bma} = \Xi|_{\Tilde{\mathcal{S}}} \langle \bmbeta_{\star}, \bme_{\bmzero \star}  \rangle \qquad \text{for } \quad \bma = \bmone, \bmtwo, \bmthree.
    \end{eqnarray}
    \label{Gaussian-Gauge-d-conformalfactor}
\end{subequations}
This choice of coordinates, frame fields and conformal factor will be known as the conformal Gaussian gauge system. 

To make use of this gauge in the ECFEs, let $\varphi$ denote an embedding map $\varphi: \Tilde{\mathcal{S}} \rightarrow \Tilde{\mathcal{M}}$ so that $\Tilde{\mathcal{S}}$ is a submanifold of $\Tilde{\mathcal{M}}$ and let $\phi$ denote the conformal transformation map $\phi: \Tilde{\mathcal{M}} \rightarrow \mathcal{M}$ so that the metric $\bmg \equiv \Xi^{2} \tilde{\bmg}$ on $\mathcal{M}$ satisfies eq. \eqref{conformal-geodesics-g-condition}. Then, the map $\phi \circ \varphi: \Tilde{\mathcal{S}} \rightarrow \mathcal{M}$ is also an embedding so that $\Tilde{\mathcal{S}}$ can be considered as a submanifold of $\mathcal{M}$. If $\Tilde{\bmh}$ is the metric induced by $\Tilde{\bmg}$ on $\Tilde{\mathcal{S}}$ and $\bmh$ is the metric induced by $\bmg$ on $\Tilde{\mathcal{S}}$, one has
\begin{equation}
    \bmh = \Omega^2 \Tilde{\bmh},
    \label{Inrinsic-metric-conformal-transformation}
\end{equation}
where $\Omega = \Xi|_{\Tilde{\mathcal{S}}}$. Given the smooth coordinate system $( x^{\mu} )$ on $\Tilde{\mathcal{U}}$ and if $\bme^{\mu}{}_{\bma}$ denote the coefficients of the frame field $\bme_{\bma}$ with respect to the coordinate system $( x^{\mu} )$ i.e. $\bme^{\mu}{}_{\bma} = \langle \bmd{x}^{\mu}, \bme_{\bma} \rangle$, then using the gauge conditions eq. \eqref{CGG-Gauge-Condiitons}, the evolution equations for $\bme^{\mu}{}_{\bma}, \hat{\Gamma}_{\bma}{}^{\bmb}{}_{\bmc}$ and $\hat{L}_{\bma \bmb}$ can be written as 
 \begin{subequations}
    \begin{eqnarray}
        && \partial_{\tau} \bme^{\mu}{}_{\bmb} = - \hat{\Gamma}_{\bmb}{}^{\bmc}{}_{\bmzero} \bme^{\mu}{}_{\bmc}, \\
        && \partial_{\tau} \hat{\Gamma}_{\bmb}{}^{\bmc}{}_{\bmd} = - \hat{\Gamma}_{\bmf}{}^{\bmc}{}_{\bmd} \hat{\Gamma}_{\bmb}{}^{\bmf}{}_{\bmzero} + \delta_{\bmzero}{}^{\bmc} \hat{L}_{\bmb \bmd} + \delta_{\bmd}{}^{\bmc} \hat{L}_{\bmb \bmzero} + \eta_{\bmzero \bmd} \eta^{\bmf \bmc} \hat{L}_{\bmb \bmf} + \Xi d^{\bmc}{}_{\bmd \bmzero \bmb}, \\
        && \partial_{\tau} \hat{L}_{\bmb \bmc} = - \hat{\Gamma}_{\bmb}{}^{\bmf}{}_{\bmzero} \hat{L}_{\bmf \bmc} + d_{\bmf} d^{\bmf}{}_{\bmc \bmzero \bmb}.
    \end{eqnarray}
 \end{subequations}
Note that these equations contain only derivatives with respect to $\tau$, so they can be considered as a transport system along the conformal geodesics $(x(\tau), \bmbeta (\tau))$. If $n_{a}$ is the $\bmg$-unit normal to $\tilde{\mathcal{S}}$, the rescaled Weyl tensor $d^{\bma}{}_{\bmb \bmc \bmd}$ can be decomposed in terms of the electric part $E_{\bma \bmb}$ and magnetic part $B_{\bma \bmb}$ defined as
\begin{equation*}
    E_{\bma \bmb} \equiv d_{\bma \bmc \bmb \bmd} n^{\bmc} n^{\bmd}, \qquad B_{\bma \bmb} = (*d)_{\bma \bmc \bmb \bmd} n^{\bmc} n^{\bmd},
\end{equation*}
where $(*d)_{\bma \bmc \bmb \bmd}$ are the components of the left Hodge dual of $d_{acbd}$. Using this decomposition, one can show that $E_{\bma \bmb}$ and $B_{\bma \bmb}$ satisfy
\begin{subequations}
    \begin{eqnarray*}
        && \partial_{\tau} (E_{\bmb \bmd}) + D_{\bma} B_{\bmc (\bmb} \epsilon_{\bmd)}{}^{\bma \bmc} + 2 a_{\bma} \epsilon^{\bma \bmc}{}_{(\bmb} B_{\bmd) \bmc} - 3 \chi_{(\bmb}{}^{\bmc} E_{\bmd) \bmc} - \epsilon_{\bmb}{}^{\bma \bmc} \epsilon_{\bmd}{}^{\bme \bmf} E_{\bma \bmc} \chi_{\bme \bmf} + \chi E_{\bmb \bmd} = 0, \\
        && \partial_{\tau}(B_{\bmb \bmd}) - D_{\bma}E_{\bmc (\bmb} \epsilon_{\bmd)}{}^{\bma \bmc} - 2 a_{\bma} \epsilon^{\bma \bmc}{}_{(\bmb} E_{\bmd) \bmc} - 3 \chi^{\bma}{}_{(\bmb} B_{\bmd)\bma} - \epsilon_{\bmb}{}^{\bma \bmc} \epsilon_{\bmd}{}^{\bme \bmf} B_{\bma \bmc} \chi_{\bme \bmf} + \chi B_{\bmb \bmd} = 0,
    \end{eqnarray*}
\end{subequations}
where 
$$
h_{\bma \bmb} \equiv g_{\bma \bmb} - \tau_{\bma} \tau_{\bmb}, \quad \chi_{\bma \bmb} \equiv h_{\bma}{}^{\bmc} \nabla_{\bmc} \tau_{\bmb}, \quad \chi \equiv h^{\bma \bmb} \chi_{\bma \bmb}, \quad a_{\bma} \equiv \tau^{\bmb} \nabla_{\bmb}\tau_{\bma},
$$
and $\tau_{\bma} = \sqrt{2} n_{\bma}$.

\subsubsection{Spinor formulation of the extended conformal field equations}
The ECFEs obtained from the hyperbolic reduction procedure discussed in the previous section can be expressed in terms of spinors. In particular, if $\tau^{AA'}$ denotes the spinor counterpart of the vector $\tau^a$ and $\{ \bmepsilon_{\bmA}{}^A \}$ denotes the spin frame introduced in Section \ref{Section:Friedrich formulation of spatial infinity} so that $\tau^{AA'}$ is given by eq. \eqref{Tau-adapted-frame}, then the evolution equations for $\bme_{\bmA \bmA'}, \hat{\Gamma}_{\bmA \bmA'}{}^{\bmB}{}_{\bmC}$, $\hat{L}_{\bmA \bmA' \bmB \bmB'}$ and $d_{\bmA \bmA' \bmB \bmB' \bmC \bmC' \bmD \bmD'}$ can be written as
\begin{subequations}
    \begin{eqnarray}
        && \tau^{\bmA \bmA'} \hat{\Sigma}_{\bmA \bmA' \bmB \bmB'} =0, \qquad \tau^{\bmC \bmC'} \hat{\Xi}_{\bmA \bmB \bmC \bmC' \bmD \bmD'} =0, \\
        && \tau^{\bmA \bmA'} \hat{\Delta}_{\bmA \bmA' \bmB \bmB' \bmC \bmC'} =0, \qquad \tau_{(\bmA}{}^{\bmA'} \hat{\Lambda}_{|\bmA'|\bmB \bmC \bmD)}=0,
    \end{eqnarray}
    \label{Reduced-Extended-Conf-Field-Eq-Spinor}
\end{subequations}
where we choose to omit the explicit form of these equations --- see \cite{kroon16} for a detailed expression of these equations. 

In the following, let $\hat{\Gamma}_{\bmA \bmB \bmC \bmD}, f_{\bmA \bmB}$ and $\Theta_{\bmA \bmB \bmC \bmD}$ denote the components of the space-spinor counterparts of $\hat{\bmnabla}$-spin connection coefficients, the 1-form $\bmf$ and the $\hat{\bmnabla}$-Schouten tensor, as introduced in Section \ref{Section:Friedrich formulation of spatial infinity}, i.e., 
\begin{equation*}
    \hat{\Gamma}_{\bmA \bmB \bmC \bmD} \equiv \tau_{(\bmB}{}^{\bmA'} \hat{\Gamma}_{\bmA) \bmA' \bmC \bmD}, \qquad \; f_{\bmA \bmB} \equiv \tau_{(\bmB}{}^{\bmA'} f_{\bmA) \bmA'}, \qquad \; \Theta_{\bmA \bmB \bmC \bmD} \equiv \tau_{\bmB}{}^{\bmA'} \tau_{\bmD}{}^{\bmC'} \hat{L}_{\bmA \bmA' \bmC \bmC'}.
\end{equation*}
Note that $\hat{\Gamma}_{\bmA \bmB \bmC \bmD}$ can be further decomposed in terms of $\xi_{\bmA \bmB \bmC \bmD}$ and $\chi_{\bmA \bmB \bmC \bmD}$ using eq. \eqref{Decomposition-Connection-Coefficients} while the spinorial counterpart of the rescaled Weyl tensor $d_{\bmA \bmA' \bmB \bmB' \bmC \bmC' \bmD \bmD'}$ can be decomposed in terms of the rescaled Weyl spinor $\phi_{\bmA \bmB \bmC \bmD}$ as follows:
\begin{equation}
    d_{\bmA \bmA' \bmB \bmB' \bmC \bmC' \bmD \bmD'} = - \phi_{\bmA \bmB \bmC \bmD} \epsilon_{\bmA' \bmB'} \epsilon_{\bmC' \bmD'} - \bar{\phi}_{\bmA' \bmB' \bmC' \bmD'} \epsilon_{\bmA \bmB} \epsilon_{\bmC \bmD}.
    \label{Rescaled-weyl-decomposition}
\end{equation}
Now, define the electric and magnetic parts of the rescaled Weyl spinor $\phi_{\bmA \bmB \bmC \bmD}$ as
\begin{equation*}
    \eta_{\bmA \bmB \bmC \bmD} \equiv \frac{1}{2} \left( \phi_{\bmA \bmB \bmC \bmD} + \phi^{+}_{\bmA \bmB \bmC \bmD}, \right), \qquad \mu_{\bmA \bmB \bmC \bmD} \equiv - \frac{i}{2} \left( \phi_{\bmA \bmB \bmC \bmD} - \phi^{+}_{\bmA \bmB \bmC \bmD} \right),
\end{equation*}
where the ${}^+$ sign denotes the Hermitian conjugation, so that
$$
\phi^{+}_{\bmA \bmB \bmC \bmD} \equiv \tau_{\bmA}{}^{\bmA'} \tau_{\bmB}{}^{\bmB'} \tau_{\bmC}{}^{\bmC'} \tau_{\bmD}{}^{\bmD'} \bar{\phi}_{\bmA' \bmB' \bmC' \bmD'}. 
$$
In terms of the above-mentioned fields, the evolution equations for 
$$(\bme^{0}{}_{\bmA \bmB}, \bme^{\alpha}{}_{\bmA \bmB}, \xi_{\bmA \bmB \bmC \bmD}, f_{\bmA \bmB}, \chi_{(\bmA \bmB) \bmC \bmD}, \Theta_{\bmC \bmD (\bmA \bmB)},\Theta_{\bmA \bmB \bmQ}{}^{\bmQ})$$ 
can be written as 
\begin{subequations}
\begin{eqnarray}
    && \partial_{\tau} \bme^{0}{}_{\bmA \bmB} = - \chi_{(\bmA \bmB)}{}^{\bmP \bmQ} \bme^{0}{}_{\bmP \bmQ} - f_{\bmA \bmB}, \\
    && \partial_{\tau} \bme^{\alpha}{}_{\bmA \bmB} = - \chi_{(\bmA \bmB)}{}^{\bmP \bmQ} \bme^{\alpha}{}_{\bmP \bmQ}, \\
    && \partial_{\tau} \xi_{\bmA \bmB \bmC \bmD} = - \chi_{(\bmA \bmB)}{}^{\bmP \bmQ} \xi_{\bmP \bmQ \bmC \bmD} + \frac{1}{\sqrt{2}} \left( \epsilon_{\bmA \bmC} \chi_{(\bmB \bmD) \bmP \bmQ} + \epsilon_{\bmB \bmD} \chi_{(\bmA \bmC) \bmP \bmQ}  \right) f^{\bmP \bmQ}  \\
    && \phantom{\partial_{\tau} \xi_{\bmA \bmB \bmC \bmD} =} - \sqrt{2} \chi_{(\bmA \bmB)(\bmC}{}^{\bmE} f_{\bmD) \bmE} - \frac{1}{2} \left( \epsilon_{\bmA \bmC} \Theta_{\bmB \bmD \bmQ}{}^{\bmQ} + \epsilon_{\bmB \bmD} \Theta_{\bmA \bmC \bmQ}{}^{\bmQ} \right) - i \Xi \mu_{\bmA \bmB \bmC \bmD}, \nonumber \\
    && \partial_{\tau} f_{\bmA \bmB} = - \chi_{(\bmA \bmB)}{}^{\bmP \bmQ} f_{\bmP \bmQ} + \frac{1}{\sqrt{2}} \Theta_{\bmA \bmB \bmQ}{}^{\bmQ}, \\
    && \partial_{\tau} \chi_{(\bmA \bmB) \bmC \bmD} = - \chi_{(\bmA \bmB)}{}^{\bmP \bmQ} \chi_{\bmP \bmQ \bmC \bmD} - \Theta_{\bmA \bmB (\bmC \bmD)} + \Xi \eta_{\bmA \bmB \bmC \bmD}, \\
    && \partial_{\tau} \Theta_{\bmC \bmD (\bmA \bmB)} = - \chi_{(\bmA \bmB)}{}^{\bmP \bmQ} \Theta_{\bmP \bmQ (\bmA \bmB)} - \partial_{\tau}\Xi \eta_{\bmA \bmB \bmC \bmD} + i \sqrt{2} d^{\bmP}{}_{(\bmA} \mu_{\bmB) \bmC \bmD \bmP}, \\ 
    && \partial_{\tau} \Theta_{\bmA \bmB \bmQ}{}^{\bmQ} = - \chi_{(\bmA \bmB)}{}^{\bmE \bmF} \Theta_{\bmE \bmF \bmQ}{}^{\bmQ} + \sqrt{2} d^{\bmP \bmQ} \eta_{\bmA \bmB \bmP \bmQ}.
\end{eqnarray}
\label{Evolution-system-in-space-spinors}
\end{subequations}
\noindent
In the above, $d_{\bmA \bmB} \equiv \tau_{(\bmB}{}^{\bmA'} d_{\bmA) \bmA'}$ is the space-spinor counterpart of the 1-form $\bmd$ introduced in the previous section. Note that the conformal factor $\Xi$, its derivatives $\partial_{\tau} \Xi$ and $d_{\bmA \bmB}$ can be expressed in terms of the initial data as discussed earlier ---see eqs. \eqref{Gaussian-Gauge-d-conformalfactor}. 

Finally, let $\mathcal{D}_{\bmA \bmB}$ and $\mathcal{P}$ denote the Sen connection and the Fermi derivative associated with the Levi-Civita connection $\bmnabla$, so that
\begin{equation*}
    \mathcal{D}_{\bmA \bmB} \equiv \tau_{(\bmA}{}^{\bmA'} \nabla_{\bmB) \bmA'}, \qquad \; \mathcal{P} \equiv \tau^{\bmA \bmA'} \nabla_{\bmA \bmA'},
\end{equation*}
then the so-called boundary-adapted evolution system for $\phi_{\bmA \bmB \bmC \bmD}$ can be written as
\begin{subequations}
    \begin{equation}
    - 2 P_{\bmzero \bmzero \bmzero \bmzero} =0, \qquad \; -2 P_{\bmzero \bmzero \bmzero \bmone} - \frac{1}{2} C_{\bmzero \bmzero} =0, \qquad \; -2 P_{\bmzero \bmzero \bmone \bmone} =0,
    \end{equation}
    \begin{equation}
        -2 P_{\bmzero \bmone \bmone \bmone} + \frac{1}{2} C_{\bmone \bmone} =0, \qquad \; -2 P_{\bmone \bmone \bmone \bmone} =0.
    \end{equation}
    \label{The-Bianchi-evolution-system-in-space-spinors}
\end{subequations}
where 
$$
P_{\bmA \bmB \bmC \bmD} = -\frac{1}{2} \left( \mathcal{P} \phi_{\bmA \bmB \bmC \bmD} - \mathcal{D}_{(\bmD}{}^{\bmF} \phi_{\bmA \bmB \bmC) \bmF} \right), \qquad \; C_{\bmA \bmB} = \mathcal{D}^{\bmE \bmF} \phi_{\bmA \bmB \bmE \bmF}.
$$
The constraint equations for $\phi_{\bmA \bmB \bmC \bmD}$ can be written as
\begin{equation}
    C_{\bmA \bmB} =0.
    \label{The-Bianchi-constraint-system-in-space-spinors}
\end{equation}
Given the evolution equations for the background fields $\bme^{0}{}_{\bmA \bmB}$, $\bme^{\alpha}{}_{\bmA \bmB}$, $\xi_{\bmA \bmB \bmC \bmD}$, $f_{\bmA \bmB}$, $\chi_{(\bmA \bmB) \bmC \bmD}$, $\Theta_{\bmC \bmD (\bmA \bmB)}$, $\Theta_{\bmA \bmB \bmQ}{}^{\bmQ}$ and the boundary-adapted evolution and constraint equations for the rescaled Weyl tensor $\phi_{\bmA \bmB \bmC \bmD}$, the next step is to obtain a scalar version of these equations that can be solved for the components of the background fields and the rescaled Weyl tensor. 

\subsubsection{Scalarising the extended conformal field equations}
\label{Section: Scalarising the space-spinor fields}
To obtain a scalar version of the extended conformal field equations, we consider the irreducible decomposition of spinors with two or four unprimed indices. 

In the following, let $\{ \bmepsilon_{\bmA}{}^{A} \}$ denote some arbitrary spin frame and introduce the primary spinors $x_{AB}, y_{AB}, z_{AB}, \epsilon^{s}_{ABCD}$ and $h_{ABCD}$ as 
\begin{subequations}
    \begin{eqnarray}
        && x_{AB} = \sqrt{2} \bmepsilon^{\bmzero}{}_{(A} \bmepsilon^{\bmone}{}_{B)}, \qquad \; y_{AB} = - \frac{1}{\sqrt{2}} \bmepsilon^{\bmone}{}_{(A} \bmepsilon^{\bmone}{}_{B)},\qquad \; z_{AB} =  \frac{1}{\sqrt{2}} \bmepsilon^{\bmzero}{}_{(A} \bmepsilon^{\bmzero}{}_{B)}, \hspace{1cm}
        \label{Primary-spinors-set1} \\
        && \epsilon^{s}_{ABCD} = \bmepsilon^{(\bmA}{}_{(A} \bmepsilon^{\bmB}{}_{B} \bmepsilon^{\bmC}{}_{C} \bmepsilon^{\bmD)_{s}}{}_{D)}, \qquad \; h_{ABCD} = - \epsilon_{A(C} \epsilon_{D)B},
        \label{Primary-spinors-set2}
    \end{eqnarray}
    \label{Primary-spinors}
\end{subequations}
where the indices with a bracket are symmetrised, and the lower index $s \leq 4$ is used to indicate that $s$ of the indices are set to equal $\bmone$ while the remaining are set to equal $\bmzero$. The $\epsilon$-spinor can be written in terms of  $\{ \bmepsilon_{\bmA}{}^{A} \}$ as
\begin{equation*}
    \epsilon_{AB} = \bmepsilon^{\bmzero}{}_{A} \bmepsilon^{\bmone}{}_{B} - \bmepsilon^{\bmzero}{}_{B} \bmepsilon^{\bmone}{}_{A}.
\end{equation*}
The spinors $\epsilon^{s}_{ABCD}$ can be written in terms of $x_{AB}, y_{AB}$ and $z_{AB}$ as
\begin{subequations}
    \begin{eqnarray*}
        && \epsilon^{0}_{ABCD} = \frac{2}{3} z_{AD} z_{BC} + \frac{2}{3} z_{AC} z_{BD} + \frac{2}{3} z_{AB} z_{CD}, \\
        && \epsilon^{1}_{ABCD} = \frac{1}{6} x_{CD} z_{AB} + \frac{1}{6} x_{BD} z_{AC} + \frac{1}{6} x_{BC} z_{AD} + \frac{1}{6} x_{AD} z_{BC} + \frac{1}{6} x_{AC} z_{BD} + \frac{1}{6} x_{AB} z_{CD}, \\
        && \epsilon^{2}_{ABCD} = \frac{1}{6} x_{AD} x_{BC} + \frac{1}{6} x_{AC} x_{BD} + \frac{1}{6} x_{AB} x_{CD}, \\
        && \epsilon^{3}_{ABCD} = - \frac{1}{6} x_{CD} y_{AB} - \frac{1}{6} x_{BD} y_{AC} - \frac{1}{6} x_{BC} y_{AD} - \frac{1}{6} x_{AD} y_{BC} - \frac{1}{6} x_{AC} y_{BD} - \frac{1}{6} x_{AB} y_{CD}, \\
        && \epsilon^{4}_{ABCD} = \frac{2}{3} y_{AD} y_{BC} + \frac{2}{3} y_{AC} y_{BD} + \frac{2}{3} y_{AB} y_{CD}.
    \end{eqnarray*}
\end{subequations}
The primary spinors satisfy a number of useful identities ---see appendix in \cite{FriedrichKannar00} for a full list. For example, the spinors $x_{AB}, y_{AB}$ and $z_{AB}$ satisfy
\begin{subequations}
    \begin{eqnarray*}
        && x_{AB} x^{AB} = -1, \qquad x_{AB} y^{AB} =0, \qquad x_{AB} z^{AB} =0, \\
        && y_{AB} y^{AB} =0, \qquad y_{AB} z^{AB} = -\frac{1}{2} \qquad z_{AB} z^{AB} =0.
    \end{eqnarray*}
\end{subequations}
An arbitrary spinor with 2 unprimed indices $l_{AB}$ can be decomposed as 
$$
l_{AB} = l_{x} x_{AB} + l_{y} y_{AB} + l_{z} z_{AB},
$$
where 
$$
l_{x} = - l_{AB} x^{AB}, \qquad \; l_{y}= -2 l_{AB} z^{AB}, \qquad \; l_{z} = -2 l_{AB} y^{AB}.
$$
Similarly, an arbitrary spinor with 4 unprimed indices $S_{ABCD}$ can be written as 
\begin{subequations}
    \begin{eqnarray*}
        && S_{ABCD} = S_{0} \epsilon^{0}_{ABCD} + S_{1} \epsilon^{1}_{ABCD} + S_{2} \epsilon^{2}_{ABCD} + S_{3} \epsilon^{3}_{ABCD} + S_{4} \epsilon^{4}_{ABCD} + S_{h} h_{ABCD} \nonumber \\
        && \phantom{S_{ABCD} } + S_{x} (x_{BD} \epsilon_{AC} + x_{AC} \epsilon_{BD}) + S_{y} (y_{BD} \epsilon_{AC} + y_{AC} \epsilon_{BD}) + S_{z} (z_{BD} \epsilon_{AC} + z_{AC} \epsilon_{BD}).
    \end{eqnarray*}
\end{subequations}
For totally symmetric spinors, $T_{ABCD} = T_{(ABCD)}$, the irreducible decomposition can be written as
$$
T_{ABCD} = T_{0} \epsilon^{0}_{ABCD} + T_{1} \epsilon^{1}_{ABCD} + T_{2} \epsilon^{2}_{ABCD} + T_{3} \epsilon^{3}_{ABCD} + T_{4} \epsilon^{4}_{ABCD}.
$$
Given the above-mentioned decomposition, we can list the components of the spinors appearing in the evolution eqs. \eqref{Evolution-system-in-space-spinors} for the background fields as
\begin{subequations}
    \begin{align*}
        &\bme^{0}{}_{\bmA \bmB} \rightarrow \left( \bme^{0}_{x} , \bme^{0}_{y}, \bme^{0}_{z} \right), \\
        &\bme^{\alpha}{}_{\bmA \bmB} \rightarrow \left( \bme^{1}_{x} , \bme^{1}_{y}, \bme^{1}_{z}, \bme^{2}_{x} , \bme^{2}_{y}, \bme^{2}_{z}, \bme^{3}_{x} , \bme^{3}_{y}, \bme^{3}_{z} \right), \\
        &\xi_{\bmA \bmB \bmC \bmD} \rightarrow \left( \xi_{0}, \xi_{1}, \xi_{2}, \xi_{3}, \xi_{4}, \xi_{h}, \xi_{x}, \xi_{y}, \xi_{z} \right), \\
        &f_{\bmA \bmB} \rightarrow \left( f_{x}, f_{y}, f_{z} \right), \\
        &\chi_{(\bmA \bmB) \bmC \bmD} \rightarrow \left( \chi_{0}, \chi_{1}, \chi_{2}, \chi_{3}, \chi_{4}, \chi_{h}, \chi_{x}, \chi_{y}, \chi_{z} \right), \\
        &\Theta_{\bmA \bmB (\bmC \bmD)} \rightarrow \left( \Theta_{0}, \Theta_{1}, \Theta_{2}, \Theta_{3}, \Theta_{4}, \Theta_{h}, \Theta_{x}, \Theta_{y}, \Theta_{z} \right), \\
        &\Theta_{\bmA \bmB} \equiv \Theta_{\bmA \bmB \bmQ}{}^{\bmQ} \rightarrow \left( \theta_{x}, \theta_{y}, \theta_{z} \right).
    \end{align*}
\end{subequations}
It is possible to obtain a scalarised version of the evolution eqs. \eqref{Evolution-system-in-space-spinors} by various contractions with the primary spinors. In total, one obtains 45 equations for the 45 components listed above. In our analysis, the xAct package \cite{xAct} for Wolfram language was used to obtain an explicit form of the scalarised equations. These will not be listed here for obvious reasons.

Given that $\phi_{\bmA \bmB \bmC \bmD}$ is totally symmetric, the components involved in the boundary-adapted evolution and constraint equations can be listed as follows:
$$
\phi_{\bmA \bmB \bmC \bmD} \rightarrow \left( \phi_{0}, \phi_{1}, \phi_{2}, \phi_{3}, \phi_{4} \right).
$$
In terms of the above, the evolution eqs. \eqref{The-Bianchi-evolution-system-in-space-spinors} and the constraint eqs. \eqref{The-Bianchi-constraint-system-in-space-spinors} for $\phi_{\bmA \bmB \bmC \bmD}$ consists of 5 and 3 equations, respectively, to be solved for the components $\phi_{0}, \phi_{1}, \phi_{2}, \phi_{3}$ and $\phi_{4}$ ---see Appendix \ref{Appendix: The boundary adapted evolution and constraint equations}. 

\section{The Newman-Penrose gauge}
\label{Section: The NP-gauge}
As mentioned in the introduction, the BMS-asymptotic charges used in this work are expressed in terms of the NP-gauge. Before introducing the expressions of the charges, we will discuss the NP-gauge conditions and the general relation between the NP-gauge frame with the F-gauge frame. As will become evident, the main distinction between the NP-gauge and the F-gauge is that the former is adapted to null infinity $\mathscr{I}^{\pm}$ while the latter is adapted to Cauchy hypersurfaces. 

In this section, the focus will be on the gauge conditions satisfied by the NP-gauge at future null infinity $\mathscr{I}^{+}$. Similar conditions can be formulated at past null infinity $\mathscr{I}^{-}$. Following the discussion in \cite{FriedrichKannar00}, introduce the conformal metric $\bmg^{\bullet}$ related to $\bmg$ by
\begin{equation}
    \bmg^{\bullet} = \theta^2 \bmg.
    \label{Conformal-Transformation-NP-Gauge}
\end{equation}
On a neighbourhood $\mathcal{U} \subset \mathcal{M}$ of $\mathscr{I}^+$, introduce the smooth adapted frame $\{ \bme^{\circ}_{\bmA \bmA'} \} $ satisfying the following conditions: 
\begin{enumerate}
    \item[i.] The frame field $\bme^{\circ}_{\bmone \bmone'}$ is tangent to and parallelly propagated along $\mathscr{I}^+$. Hence,
    \begin{equation*}
        \nabla^{\circ}_{\bmone \bmone'} \bme^{\circ}_{\bmone \bmone'} \simeq 0,
    \end{equation*}
    where $\simeq$ is used to denote equality on $\mathscr{I}^+$.
    \item[ii.] On $\mathcal{U}$, there exists a smooth function $u^{\circ}$ that induces an affine parameter on the null generators of $\mathscr{I}^+$ such that
    \begin{equation*}
        \bme^{\circ}_{\bmone \bmone'}(u^{\circ}) \simeq 1.
    \end{equation*}
    \item[iii.] The frame $\bme^{\circ}_{\bmzero \bmzero'}$ is tangent to the hypersurfaces transverse to $\mathscr{I}^+$ (on which $u^{\circ} = \text{constant}$), i.e.,
    \begin{equation*}
        \bme^{\circ}_{\bmzero \bmzero'} = \bmg(\bmd{u^{\circ}},\cdot).
    \end{equation*}
    \item[iv.] The frame fields $\bme^{\circ}_{\bmzero \bmone'}$ and $\bme^{\circ}_{\bmone \bmzero'}$ are tangent to the slices $\{ u^{\circ} = \text{constant} \} \cap \mathscr{I}^+$. These frame fields as well as $\bme^{\circ}_{\bmzero \bmzero'}$ and $\bme^{\circ}_{\bmone \bmone'}$ are parallelly propagated in the direction of $\bme^{\circ}_{\bmzero \bmzero'}$. 
\end{enumerate}
Let $(x^{\mu})$ denote a local coordinate system on $\mathcal{U}$, then the above gauge conditions can be expressed in terms of the spin connection coefficients $\Gamma^{\circ}_{\bmA \bmA'}{}^{\bmB}{}_{\bmC}$, which can be written in terms of the components of $\bme^{\circ}_{\bmA \bmA'}$ with respect to $(x^{\mu})$ as
\begin{equation*}
    \Gamma^{\circ}_{\bmA \bmA' \bmB \bmC} = \frac{1}{2} \left( \bme^{\circ \mu}{}_{\bmA \bmA'} \bme^{\circ \nu}{}_{\bmB \bmone'} \nabla_{\mu} \bme^{\circ}_{\nu \bmC \bmzero'} + \bme^{\circ \mu}{}_{\bmA \bmA'} \bme^{\circ \nu}{}_{\bmC \bmone'} \nabla_{\mu} \bme^{\circ}_{\nu \bmB \bmzero'} \right).
\end{equation*}
Then, the gauge conditions are given by
\begin{subequations}
    \begin{eqnarray*}
        && \Gamma^{\circ}_{\bmone \bmzero' \bmone \bmone} \simeq 0, \qquad \Gamma^{\circ}_{\bmone \bmone' \bmone \bmone} \simeq 0, \\
        && \Gamma^{\circ}_{\bmone \bmzero' \bmzero \bmzero} = \bar{\Gamma}^{\circ}_{\bmzero \bmone' \bmzero' \bmzero'}, \qquad \Gamma^{\circ}_{\bmone \bmone' \bmzero \bmzero} = \bar{\Gamma}^{\circ}_{\bmzero \bmone' \bmzero' \bmone'} + \Gamma^{\circ}_{\bmzero \bmone' \bmzero \bmone}, \qquad \Gamma^{\circ}_{\bmzero \bmzero' \bmB \bmC} = 0 \qquad  \text{  on } \mathcal{U}.
    \end{eqnarray*}
\end{subequations}
To introduce the NP-gauge frame $\{ \bme^{\bullet}_{\bmA \bmA'} \}$ and the NP-gauge conditions, consider the conformal rescaling $\bmg \to \bmg^{\bullet}$ where $\bmg^{\bullet}$ is defined by eq. \eqref{Conformal-Transformation-NP-Gauge}. Given $\theta > 0$ and an arbitrary function $p > 0$ which is constant on the generators of $\mathscr{I}^+$, the NP-gauge frame $\{ \bme^{\bullet}_{\bmA \bmA'} \}$ can be introduced as follows: let $\bme^{\bullet}_{\bmone \bmone'} \simeq \theta^{-2} p \bme^{\circ}_{\bmone \bmone'}$ and introduce the affine parameter $u^{\bullet}$ along the generators of $\mathscr{I}^+$ as follows 
\begin{equation*}
    u^{\bullet} (u^{\circ}) = \int^{u^{\circ}}_{u^{\circ}_{\star}} \theta^2(u') p^{-1}(u') du' + u^{\bullet}_{\star} \qquad \text{on } \mathscr{I}^+.
\end{equation*}
Hence, the frame $\bme^{\bullet}_{\bmone \bmone'} $ is parallelly propagated along $\mathscr{I}^+$ and it satisfies $\bme^{\bullet}_{\bmone \bmone'} (u^{\bullet}) \simeq 0$. If  $\mathcal{C}$ denotes a cut on $\mathscr{I}^+$, which is diffeomorphic to $\mathbb{S}^2$, then the coordinates $( \vartheta, \varphi )$ on $\mathcal{C}$ and the conformal factor $\theta$ can be fixed such that the metric on $\mathcal{C}$ is the standard metric on $\mathbb{S}^2$.

In the following, assume that $u^{\circ}= u^{\circ}_{\star}$ and $u^{\bullet}= u^{\bullet}_{\star}$ on $\mathcal{C}$ and set 
\begin{equation*}
    \bme^{\bullet}_{\bmzero \bmzero'} = p^{-1} \bme^{\circ}_{\bmzero \bmzero'}, \qquad \bme^{\bullet}_{\bmone \bmone'} = \theta^{-2} p \bme^{\circ}_{\bmone \bmone'}, \qquad \bme^{\bullet}_{\bmzero \bmone'} = \theta^{-1} \bme^{\circ}_{\bmzero \bmone'} \qquad \text{on } \mathcal{C}.
\end{equation*}
Given that $\bmg = \Theta^{2} \tilde{\bmg}$, the transformation $\bmg \mapsto \bmg^{\bullet}$ implies $\Theta \mapsto \Theta^{\bullet} = \theta \Theta$ and the transformation laws for the spin connection coefficients can be written on $\mathcal{C}$ as 
\begin{eqnarray*}
    && \Gamma^{\bullet}_{\bmone \bmzero' \bmzero \bmzero}= p^{-1} \left(\Gamma^{\circ}_{\bmone \bmzero' \bmzero \bmzero} - \bme^{\circ}_{\bmzero \bmzero'} (\log(\theta)) \right), \\
    && \Gamma^{\bullet}_{\bmzero \bmone' \bmone \bmone} = p \theta^{-2} \left( \Gamma^{\circ}_{\bmzero \bmone' \bmone \bmone} + \bme^{\circ}_{\bmone \bmone'} (\log(\theta)) \right).
\end{eqnarray*}
Thus, with a suitable choice of $d \theta$ and $p$, one can achieve the following
\begin{equation}
    \Gamma^{\bullet}_{\bmone \bmzero' \bmzero \bmzero} =0, \qquad \Gamma^{\bullet}_{\bmzero \bmone' \bmone \bmone} =0, \qquad \bme^{\bullet}_{\bmzero \bmzero'} (\Theta^{\bullet}) = \text{const.} \neq 0 \qquad  \qquad \text{ on } \mathcal{C}.
\end{equation}
The conformal rescaling given by eq. \eqref{Conformal-Transformation-NP-Gauge} implies a transformation of the trace-free part of the $\bmg$-Ricci tensor, denoted by $s_{ab}$. In particular, one has that
\begin{equation*}
    s^{\bullet}_{\mu \nu} = s_{\mu \nu} - \frac{2}{\theta} \left( \nabla_{\mu} \nabla_{\nu} \theta - \frac{2}{\theta} \nabla_{\mu} \theta \nabla_{\nu} \theta  - \frac{1}{4} g_{\mu \nu} (\nabla_{\lambda}\nabla^{\lambda}\theta - \frac{2}{\theta} \nabla_{\lambda} \theta \nabla^{\lambda}\theta)\right),
\end{equation*} 
where $s_{\mu \nu}$ are the components of $\bms$ with respect to $( x^{\mu} )$. Now, define $\Phi^{\circ}_{22}$ and $\Phi^{\bullet}_{22}$ as
\begin{equation}
    \Phi^{\circ}_{22} \equiv \frac{1}{2} s_{\mu \nu} \bme^{\circ \mu}{}_{\bmone \bmone'} \bme^{\circ \nu}{}_{\bmone \bmone'}, \qquad \Phi^{\bullet}_{22} \equiv \frac{1}{2} s^{\bullet}_{\mu \nu} \bme^{\bullet \mu}{}_{\bmone \bmone'} \bme^{\bullet \nu}{}_{\bmone \bmone'}.
    \label{Definition-Phi-22}
\end{equation}
The condition $\Phi^{\bullet}_{22} = 0$ on $\mathscr{I}^+$ implies a linear ordinary differential equation (ODE) for $\theta^{-1}$ on the generators of $\mathscr{I}^{+}$
\begin{equation}
    \bme^{\circ}_{\bmone \bmone'}(\bme^{\circ}_{\bmone \bmone'}(\theta^{-1})) + \theta^{-1} \Phi^{\circ}_{22} = 0. 
    \label{theta-equation}
\end{equation}
The initial data for $\theta$, $\bme^{\circ}_{\bmone \bmone'}(\theta)$ on $\mathcal{C}$ can be used to solve for $\theta$ and obtain
\begin{equation*}
    \Phi^{\bullet}_{22}\simeq 0, \qquad \Gamma^{\bullet}_{\bmzero \bmone' \bmone \bmone} \simeq 0.
\end{equation*}
By fixing $\theta$ and $\bme^{\bullet}_{\bmone \bmone'}$ on $\mathscr{I}^+$, the frame fields $\bme^{\bullet}_{\bmzero \bmone'}$ and $\bme^{\bullet}_{\bmone \bmzero'}$ can be determined up to a rotation. Then, the phase parameter $c$ can be determined by solving 
\begin{equation}
    \bme^{\bullet}_{\bmone \bmone'} (c) = - i \bme^{\bullet \mu}{}_{\bmone \bmzero'} \bme^{\bullet \nu}{}_{\bmone \bmone'} \nabla^{\bullet}_{\mu} \bme^{\bullet}_{\nu \bmzero \bmone'},
    \label{phase-parameter-equation}
\end{equation}
along the generators of $\mathscr{I}^+$ with initial data $c=0$ on $\mathcal{C}$. Making the replacement $\bme^{\bullet}_{\bmzero \bmone'} \to e^{i c} \bme^{\bullet}_{\bmzero \bmone'}$ allows us to obtain
\begin{equation*}
    \Gamma^{\bullet}_{\bmone \bmone' \bmzero \bmone} \simeq 0.
\end{equation*}
Given that $s^{\bullet}_{\mu \nu}$ is defined by
\begin{equation}
    \Theta^{\bullet} s^{\bullet}_{\mu \nu} = \frac{1}{2} g^{\bullet}_{\mu \nu} \nabla^{\bullet}_{\lambda} \nabla^{\bullet \lambda} \Theta^{\bullet} - 2 \nabla^{\bullet}_{\mu} \nabla^{\bullet}_{\nu} \Theta^{\bullet},
    \label{Trace-free-Ricci-tensor-bullet-g}
\end{equation}
contracting the above with $\bme^{\bullet \mu}{}_{\bmzero \bmone'} \bme^{\bullet \nu}{}_{\bmone \bmzero'}$ implies that
\begin{equation*}
    \nabla^{\bullet}_{\lambda} \nabla^{\bullet \lambda} \Theta^{\bullet} \simeq 0.
\end{equation*}
While a contraction with $\bme^{\bullet \mu}{}_{\bmzero \bmzero'} \bme^{\bullet \nu}{}_{\bmone \bmone'}$ and $\bme^{\bullet \mu}{}_{\bmzero \bmzero'} \bme^{\bullet \nu}{}_{\bmzero \bmone'}$ gives
\begin{equation*}
    \bme^{\bullet}_{\bmone \bmone'} (\bme^{\bullet}_{\bmzero \bmzero'} (\Theta^{\bullet})) \simeq 0, \qquad \bme^{\bullet}_{\bmzero \bmzero'} (\Theta^{\bullet}) \simeq \text{const.} 
\end{equation*}
and 
\begin{equation*}
    \bme^{\bullet}_{\bmzero \bmone'} (\bme^{\bullet}_{\bmzero \bmzero'} (\Theta^{\bullet})) = \Gamma^{\bullet}_{\bmone \bmone' \bmzero \bmzero} \bme^{\bullet}_{\bmzero \bmzero'}(\Theta^{\bullet}),
\end{equation*}
respectively. Thus, one concludes
$$
\Gamma^{\bullet}_{\bmone \bmone' \bmzero \bmzero} \simeq 0.
$$
The conformal rescaling given by eq. \eqref{Conformal-Transformation-NP-Gauge} implies a transformation law for the Ricci scalars associated with $\bmnabla^{\bullet}$ and $\bmnabla$
\begin{equation*}
    R^{\bullet} = \frac{1}{\theta^2} R + \frac{12}{\theta^2} \nabla^{\bullet}_{\alpha} \theta \nabla^{\bullet \alpha} \theta - \frac{6}{\theta} \nabla^{\bullet}_{\alpha} \nabla^{\bullet \alpha} \theta.
\end{equation*}
By requiring that $R^{\bullet} \simeq 0$, one obtains the following linear ODE for $\bme^{\bullet}_{\bmzero \bmzero'} (\theta)$ on the generators of $\mathscr{I}^{+}$
\begin{equation*}
    \bme^{\bullet}_{\bmone \bmone'} (\bme^{\bullet}_{\bmzero \bmzero'}(\theta)) - \frac{2}{\theta} \bme^{\bullet}_{\bmone \bmone'}(\theta) \bme^{\bullet}_{\bmzero \bmzero'}(\theta) = \mathcal{F}^{\bullet},
\end{equation*}
where $\mathcal{F}^{\bullet}$ can be determined using the quantities already obtained on $\mathscr{I}^+$. Given the initial data for $\bme^{\bullet}_{\bmzero \bmzero'} (\theta)= p^{-1} \theta \Gamma^{\circ}_{\bmone \bmzero' \bmzero \bmzero}$ on $\mathcal{C}$, the ODE can be integrated to obtain 
\begin{equation*}
    R^{\bullet} \simeq 0, \qquad \Gamma^{\bullet}_{\bmone \bmzero' \bmzero \bmzero} \simeq 0.
\end{equation*}
To introduce coordinates on $\mathcal{U}$, let $r^{\bullet}$ denote the affine parameter along the null generators of the hypersurfaces $\{ u^{\bullet} = \text{const.} \}$ such that $\bme^{\bullet}_{\bmzero \bmzero'}(r^{\bullet}) = 0$ and $r^{\bullet} \simeq 0$. Then, the coordinates $( \vartheta, \varphi )$ on $\mathcal{C}$ can be extended to $\mathscr{I}^+$ and the hypersurfaces $\{ u^{\bullet} = \text{const.} \}$ by requiring them to be constant along the null generators of $\mathscr{I}^+$ and $\{ u^{\bullet} = \text{const.} \}$ hypersurfaces, respectively. This allows us to obtain Bondi coordinates $(u^{\bullet}, r^{\bullet}, \vartheta, \varphi)$ in a neighbourhood $\mathcal{U}$ of $\mathscr{I}^+$. 

In the following, the conditions on the conformal rescaling, the frame field and the coordinates will be referred to as the NP-gauge. We will assume that the NP-gauge frame $\{ \bme^{\bullet}_{\bmA \bmA'} \}$ is a frame field that satisfies the conditions mentioned above. The term NP-spin-frame will be used to refer to the normalised spin frame $\{ \bmepsilon^{\bullet}_{\bmA}{}^{A} \}$ implying a NP-gauge frame. Other quantities in the NP-gauge will also be denoted by ${}^\bullet$. 

\begin{remark}
    {\em In subsequent calculations, the spin connection coefficients $\Gamma^{\bullet}_{\bmA \bmA'}{}^{\bmC}{}_{\bmD}$ will be referred to as the NP connection coefficients. It will be useful to introduce the following shorthand notation:
    \begin{subequations}
    \begin{eqnarray}
        && \sigma^{\bullet} \equiv -\Gamma^{\bullet}_{\bmzero \bmone'}{}^{\bmone}{}_{\bmzero}, \qquad \mu^{\bullet} \equiv -\Gamma^{\bullet}_{\bmzero \bmone'}{}^{\bmzero}{}_{\bmone}, \qquad \gamma^{\bullet} \equiv \Gamma^{\bullet}_{\bmone \bmone'}{}^{\bmzero}{}_{\bmzero}, \\
        && \lambda^{\bullet} \equiv \Gamma^{\bullet}_{\bmone \bmzero'}{}^{\bmzero}{}_{\bmone}, \qquad \rho^{\bullet} \equiv -\Gamma^{\bullet}_{\bmone \bmzero'}{}^{\bmone}{}_{\bmzero}, \qquad \epsilon^{\bullet} \equiv \Gamma^{\bullet}_{\bmzero \bmzero'}{}^{\bmzero}{}_{\bmzero}.
    \end{eqnarray}
    \label{NP-connection-ceofficients}
\end{subequations}}
\end{remark}
Given the conformal relation between $\bmg^{\bullet}$ and $\bmg$ in eq. \eqref{Conformal-Transformation-NP-Gauge}, one has 
\begin{equation}
    g^{\bullet}_{\bma \bmb} = \theta^2 g_{\bma \bmb},
    \label{F-NP-gauge-metric-relation}
\end{equation}
where $g^{\bullet}_{\bma \bmb} \equiv \bmg^{\bullet} (\bme^{\bullet}_{\bma},\bme^{\bullet}_{\bmb})$ and $g_{\bma \bmb} \equiv \bmg(\bme_{\bma},\bme_{\bmb})$. Then, the relation between the NP-gauge frame $\{ \bme^{\bullet}_{\bma} \}$ and the F-gauge frame $\{ \bme_{\bma} \}$, parameterised in terms of $\theta$ and $\Lambda^{\bma}{}_{\bmb} \in O(1,3)$, can be written as
\begin{equation}
    \bme^{\bullet}_{\bma} = \theta^{-1} \Lambda^{\bmb}{}_{\bma} \bme_{\bmb}, \qquad \bme_{\bma} = \theta \Lambda_{\bma}{}^{\bmb} \bme^{\bullet}_{\bmb}.
    \label{F-NP-gauge-Tensor-Frame-relation}
\end{equation}
Moreover, the NP-gauge spin frame $\{ \bmepsilon^{\bullet}_{\bmA}{}^{A} \}$ and the F-gauge spin frame $\{ \bmepsilon^{\bullet}_{\bmA}{}^{A} \}$ are related by
\begin{equation}
    \bmepsilon^{\bullet}_{\bmA}{}^A = \theta^{- \frac{1}{2}} \Lambda^{\bmB}{}_{\bmA} \bmepsilon_{\bmB}{}^A, \qquad \bmepsilon^{\bullet \bmA}{}_{A} = \theta^{\frac{1}{2}} \Lambda_{\bmB}{}^{\bmA} \bmepsilon^{\bmB}{}_{A}.
    \label{NP-F-gauge-spin-frame}
\end{equation}
where $\Lambda^{\bmB}{}_{\bmA} \in SL(2,\mathbb{C})$.
\section{BMS-supertranslations charges}

\label{Section: BMS-supertranslations charges}
Now that we have defined the NP-gauge and its relation to the F-gauge, we can introduce BMS-asymptotic charges expressed in terms of the NP-gauge. Following the discussion in \cite{Prabhu19}, for every $f \in C^{\infty}(\mathbb{S}^2)$, the associated BMS-supertranslation charge $\mathcal{Q}$ can be written as an integral over some cut  $\mathcal{C}$ of $\mathscr{I}^{\pm}$
\begin{equation}
    \mathcal{Q}(f;\mathcal{C}) \equiv  \oint_{\mathcal{C}}
\bm{\varepsilon}_2 f (\mathcal{P}^{\bullet} - i (*\mathcal{P}^{\bullet}) +\tfrac{1}{2}\sigma^{\bullet ab}N^{\bullet}_{ab}),
\label{Full-GR-Charges}
\end{equation}
where $\bm{\varepsilon}_2$ is the area element on $\mathcal{C}$, $\sigma^{\bullet ab}$ is the shear tensor, $N^{\bullet}_{ab}$ is the news tensor. $\mathcal{P}^{\bullet}$ and $(*\mathcal{P}^{\bullet})$ are defined in terms of the rescaled Weyl tensor $d^{\bullet}_{abcd}$ and its left Hodge dual $(*d^{\bullet})_{abcd}$ as follows:
\begin{subequations}
    \begin{eqnarray*}
        && \mathcal{P}^{\bullet} \equiv d^{\bullet}_{cdef} l^c n^d l^e n^f, \\
        && (*\mathcal{P}^{\bullet}) \equiv (*d^{\bullet})_{cdef}  l^c n^d l^e n^f,
    \end{eqnarray*}
\end{subequations}
where the vectors $ l^a$ and $n^a$ are identified as follows:
\begin{subequations}
    \begin{eqnarray}
        && l^{a} \equiv \bme^{\bullet}_{\bmzero \bmzero'}{}^{a}, \qquad n^{a} \equiv \bme^{\bullet}_{\bmone \bmone'}{}^{a} \qquad \text{on } \mathscr{I}^{+}, \\
        && l^{a} \equiv \bme^{\bullet}_{\bmone \bmone'}{}^{a}, \qquad n^{a} \equiv \bme^{\bullet}_{\bmzero \bmzero'}{}^{a} \qquad \text{on } \mathscr{I}^{-}.
    \end{eqnarray}
    \label{null-tetrad-l-and-n}
\end{subequations}
To translate eq. \eqref{Full-GR-Charges} to the F-gauge, our strategy is to obtain an expression for $\mathcal{P}^{\bullet}$, $(*\mathcal{P}^{\bullet})$ and $\sigma^{\bullet ab}N^{\bullet}_{ab})$ in terms of scalars for which it would be simpler to compute the transformation to the F-gauge. Given that the spinorial counterpart of $d^{\bullet}_{abcd}$ can be decomposed in terms of the rescaled Weyl spinor $\phi^{\bullet}_{ABCD}$ as 
\begin{equation*}
    d^{\bullet}_{A A' B B' C C' D D'} = - \phi^{\bullet}_{A B C D} \epsilon^{\bullet}_{A' B'} \epsilon^{\bullet}_{C' D'} - \bar{\phi}^{\bullet}_{A' B' C' D'} \epsilon^{\bullet}_{A B} \epsilon^{\bullet}_{C D}.
\end{equation*}
One has 
\begin{eqnarray*}
    && \mathcal{P}^{\bullet} = d^{\bullet}_{C C' D D' E E' F F'} o^{\bullet C} \bar{o}^{\bullet C'} \iota^{\bullet D} \bar{\iota}^{\bullet D'} o^{\bullet E} \bar{o}^{\bullet E'} \iota^{\bullet F} \bar{\iota}^{\bullet F'}, \\
    && \phantom{\mathcal{P}^{\bullet}} = -(\phi^{\bullet}_2 + \bar{\phi}^{\bullet}_2),
\end{eqnarray*}
where $\{ \bmo^{\bullet}, \bmiota^{\bullet} \}$ is the spin dyad adapted to the NP-gauge and 
\begin{equation*}
    \phi^{\bullet}_2 \equiv \phi^{\bullet}_{ABCD} o^{ \bullet A} \iota^{\bullet B} o^{\bullet C} \iota^{\bullet D}, \qquad \bar{\phi}^{\bullet}_2 \equiv \bar{\phi}^{\bullet}_{A'B'C'D'} \bar{o}^{\bullet A'} \bar{\iota}^{\bullet B'} \bar{o}^{\bullet C'} \bar{\iota}^{\bullet D'}.
\end{equation*}
For $(*\mathcal{P}^{\bullet})$, one has
\begin{equation*}
    (*\mathcal{P}^{\bullet}) = (*d^{\bullet})_{C C' D D' E E' F F'} o^{\bullet C} \bar{o}^{\bullet C'} \iota^{\bullet D} \bar{\iota}^{\bullet D'} o^{\bullet E} \bar{o}^{\bullet E'} \iota^{\bullet F} \bar{\iota}^{\bullet F'}.
\end{equation*}
But, given that
\begin{equation*}
    (*d^{\bullet})_{A A' B B' C C' D D'} = i \left( \phi^{\bullet}_{ABCD} \epsilon^{\bullet}_{A' B'} \epsilon^{\bullet}_{C' D'} - \bar{\phi}^{\bullet}_{A' B' C' D'} \epsilon^{\bullet}_{A B} \epsilon^{\bullet}_{C D} \right),
\end{equation*}
One can show that $(*\mathcal{P}^{\bullet})$ is given by
\begin{equation*}
    (*\mathcal{P}^{\bullet}) = i (\phi^{\bullet}_2 - \bar{\phi}^{\bullet}_2).
\end{equation*}
Hence, one has that 
\begin{equation}
    \mathcal{P}^{\bullet} - i (*\mathcal{P}^{\bullet}) = - 2 \bar{\phi}^{\bullet}_2.
    \label{P-*P-terms}
\end{equation}
For the background term involving $\sigma^{\bullet ab} N^{\bullet}_{ab}$, note that $\sigma^{\bullet ab}$ and $N^{\bullet}_{ab}$ are defined as 
\begin{subequations}
    \begin{eqnarray*}
        && N^{\bullet}_{ab} \equiv 2(\mathcal{L}_n -\Phi)\sigma^{\bullet}_{ab}, \\
        && \sigma^{\bullet}_{ab} \equiv \left( q^{\bullet}_{a}{}^{c} q^{\bullet}_{b}{}^{d} - \frac{1}{2} q^{\bullet}_{ab} q^{\bullet cd} \right) \nabla^{\bullet}_c l_d,
    \end{eqnarray*}
\end{subequations}
where $q^{\bullet}_{ab}$ is the induced metric on $\mathscr{I}= \mathscr{I}^{+} \cup \mathscr{I}^-$, and $\Phi$ is defined by $\Phi\equiv \frac{1}{4}\nabla^{\bullet}_a n^a$ on $\mathscr{I}$ and it satisfies the following:
\begin{equation*}
    \Phi|_{i}=2. 
\end{equation*}
Note that the metric $\bmg^{\bullet}$ and the covariant derivative $\bmnabla^{\bullet}$ can be decomposed in terms of the null tetrad $(l^{a},n^{a},m^{a}, \bar{m}^{a})$ as
\begin{subequations}
    \begin{eqnarray*}
        && g^{\bullet}_{ab}= n_a  l_b +  l_a n_b -\bar{m}_a m_b -m_a \bar{m}_b, \\
        && \nabla^{\bullet}_a = n_a D +  l_a \Delta - \bar{m}_a \delta - m_a \bar{\delta}, 
    \end{eqnarray*}
\end{subequations}
where $D\equiv   l^a \nabla^{\bullet}_a,\;  \Delta\equiv n^a \nabla^{\bullet}_a,\; \delta \equiv  m^a \nabla^{\bullet}_a$ and $\bar{\delta}\equiv  \bar{m}^a \nabla^{\bullet}_a$, $l^{a}$ and $n^{a}$ are defined by eq. \eqref{null-tetrad-l-and-n} and 
\begin{equation}
    m^{a} \equiv \bme^{\bullet}_{\bmzero \bmone'}{}^{a}, \qquad \bar{m}^{a} \equiv \bme^{\bullet}_{\bmone \bmzero'}{}^{a} \qquad \text{on } \mathscr{I}^{\pm}.
    \label{Definition-m-barm}
\end{equation}
From the above, one can show that 
\begin{subequations}
    \begin{eqnarray*}
        && \sigma^{\bullet}_{ab} = \sigma^{\bullet} \bar{m}_a \bar{m}_b +\bar{\sigma}^{\bullet} m_a m_b, \qquad \text{on } \mathscr{I}^{+},  \\
        && \sigma^{\bullet}_{ab} = -\lambda^{\bullet} m_a m_b -\bar{\lambda}^{\bullet} \bar{m}_a \bar{m}_b, \qquad \text{on } \mathscr{I}^{-},
    \end{eqnarray*}
\end{subequations}
and 
\begin{subequations}
    \begin{eqnarray*}
        && \Phi=\frac{1}{4}(\mu^{\bullet}+\bar{\mu}^{\bullet}-\gamma^{\bullet}-\bar{\gamma}^{\bullet}), \qquad \text{on } \mathscr{I}^{+}, \\
        && \Phi = \frac{1}{4} (\epsilon^{\bullet} + \bar{\epsilon}^{\bullet} - \rho^{\bullet} - \bar{\rho}^{\bullet}), \qquad \text{on } \mathscr{I}^{-}.
    \end{eqnarray*}
\end{subequations}
Given that
\begin{equation*}
    \mathcal{L}_n \sigma^{\bullet}_{ab} = \Delta \sigma^{\bullet}_{ab} +\sigma^{\bullet}_{cb}
    \nabla^{\bullet}_a n^c + \sigma^{\bullet}_{ac} \nabla^{\bullet}_b n^c,
\end{equation*}
a long computation yields
\begin{subequations}
    \begin{eqnarray}
        && \sigma^{\bullet ab} N^{\bullet}_{ab}  = 2 \Delta |\sigma^{\bullet}|^2 -|\sigma^{\bullet}|^2 \big(3\mu^{\bullet}+3\bar{\mu}^{\bullet}+\gamma^{\bullet}+\bar{\gamma}^{\bullet} \big), \qquad \text{on } \mathscr{I}^{+}, \label{The-background-term-scri+} \\
        && \sigma^{\bullet ab}N^{\bullet}_{ab} = 2 \Delta |\lambda^{\bullet}|^2 - |\lambda^{\bullet}|^2 \big(3\rho^{\bullet}+3\bar{\rho}^{\bullet}+\epsilon^{\bullet}+\bar{\epsilon}^{\bullet} \big), \qquad \text{on } \mathscr{I}^{-}. \label{The-background-term-scri-}
    \end{eqnarray}
    \label{The-background-term}
\end{subequations}
Using eqs. \eqref{P-*P-terms} and \eqref{The-background-term} and substituting in eq. \eqref{Full-GR-Charges}, one obtains
\begin{subequations}
    \begin{eqnarray*}
        && \mathcal{Q}(f,\mathcal{C}) = \oint_{\mathcal{C}} \bm{\varepsilon}_2 f \bigg( - 2 \bar{\phi}^{\bullet}_2 + \Delta|\sigma^{\bullet}|^2 - \frac{1}{2}|\sigma^{\bullet}|^2(3\mu^{\bullet}+3\bar{\mu}^{\bullet}+\gamma^{\bullet}+\bar{\gamma}^{\bullet}) \bigg), \qquad \text{on } \mathscr{I}^{+}, \\
        && \mathcal{Q}(f,\mathcal{C}) = \oint_{\mathcal{C}} \bm{\varepsilon}_2 f \bigg( - 2 \bar{\phi}^{\bullet}_2 + \Delta |\lambda^{\bullet}|^2 - \frac{1}{2}|\lambda^{\bullet}|^2 (3\rho^{\bullet}+3\bar{\rho}^{\bullet}+\epsilon^{\bullet}+\bar{\epsilon}^{\bullet} ) \bigg), \qquad \text{on } \mathscr{I}^{-},
    \end{eqnarray*}
\end{subequations}
for which it will be simpler to compute the transformation to the F-gauge.

\subsection{BMS-supertranslation charges in the F-gauge}
The next step in this analysis is to express $\bar{\phi}^{\bullet}_2,\; \sigma^{\bullet},\; \mu^{\bullet},\; \gamma^{\bullet},\; \lambda^{\bullet}, \; \rho^{\bullet}$ and $\epsilon^{\bullet}$ in terms of F-gauge quantities. Given that 
\begin{subequations}
    \begin{eqnarray*}
        && \sigma^{\bullet} \equiv \bmepsilon^{\bullet}_{\bmzero}{}^{A} \nabla^{\bullet}_{\bmzero \bmone'} \bmepsilon^{\bullet \bmone}{}_{A}, \qquad \mu^{\bullet} \equiv - \bmepsilon^{\bullet}_{\bmone}{}^{A} \nabla^{\bullet}_{\bmzero \bmone'} \bmepsilon^{\bullet \bmzero}{}_{A}, \qquad \gamma^{\bullet} \equiv - \bmepsilon^{\bullet}_{\bmzero}{}^{A} \nabla^{\bullet}_{\bmone \bmone'} \bmepsilon^{\bullet \bmzero}{}_{A}, \\
        && \lambda^{\bullet} \equiv -\bmepsilon^{\bullet}_{\bmone}{}^{A} \nabla^{\bullet}_{\bmone \bmzero'} \bmepsilon^{\bullet \bmzero}{}_{A}, \qquad \rho^{\bullet} \equiv  \bmepsilon^{\bullet}_{\bmzero}{}^{A} \nabla^{\bullet}_{\bmone \bmzero'} \bmepsilon^{\bullet \bmone}{}_{A}, \qquad \epsilon^{\bullet} \equiv  \bmepsilon^{\bullet}_{\bmzero}{}^{A} \nabla^{\bullet}_{\bmzero \bmzero'} \bmepsilon^{\bullet \bmzero}{}_{A},
    \end{eqnarray*}
\end{subequations}
and using eq. \eqref{NP-F-gauge-spin-frame}, a long computation yields
\begin{subequations}
    \begin{eqnarray}
        && \sigma^{\bullet} = \Lambda^{\bmA}{}_{\bmzero} \bme^{\bullet}_{\bmzero \bmone'} (\Lambda_{\bmA}{}^{\bmone}) - \theta^{-1} \Lambda_{\bmA}{}^{\bmone} \Lambda^{\bmC}{}_{\bmzero} \Lambda^{\bmB}{}_{\bmzero} \bar{\Lambda}^{\bmB'}{}_{\bmone'} \Gamma_{\bmB \bmB'}{}^{\bmA}{}_{\bmC}, \\
        && \mu^{\bullet} = \theta^{-1} \bme^{\bullet}_{\bmone \bmone'} (\theta) - \Lambda^{\bmA}{}_{\bmone} \bme^{\bullet}_{\bmzero \bmone'}(\Lambda_{\bmA}{}^{\bmzero})+ \theta^{-1} \Lambda_{\bmA}{}^{\bmzero} \Lambda^{\bmC}{}_{\bmone} \Lambda^{\bmB}{}_{\bmzero} \bar{\Lambda}^{\bmB'}{}_{\bmone'} \Gamma_{\bmB \bmB'}{}^{\bmA}{}_{\bmC}, \\
        && \gamma^{\bullet} = - \frac{1}{2} \theta^{-1} \bme^{\bullet}_{\bmone \bmone'}(\theta) - \Lambda^{\bmA}{}_{\bmzero} \bme^{\bullet}_{\bmone \bmone'}(\Lambda_{\bmA}{}^{\bmzero}) + \theta^{-1} \Lambda_{\bmA}{}^{\bmzero} \Lambda^{\bmC}{}_{\bmzero} \Lambda^{\bmB}{}_{\bmone} \bar{\Lambda}^{\bmB'}{}_{\bmone'} \Gamma_{\bmB \bmB'}{}^{\bmA}{}_{\bmC}, \\
        && \lambda^{\bullet} = - \Lambda^{\bmA}{}_{\bmone} \bme^{\bullet}_{\bmone \bmzero'} (\Lambda_{\bmA}{}^{\bmzero}) + \theta^{-1} \Lambda_{\bmA}{}^{\bmzero} \Lambda^{\bmC}{}_{\bmone} \Lambda^{\bmB}{}_{\bmone} \bar{\Lambda}^{\bmB'}{}_{\bmzero'} \Gamma_{\bmB \bmB'}{}^{\bmA}{}_{\bmC}, \\
        && \rho^{\bullet} = \theta^{-1} \bme^{\bullet}_{\bmzero \bmzero'}(\theta) + \Lambda^{\bmA}{}_{\bmzero} \bme^{\bullet}_{\bmone \bmzero'} (\Lambda_{\bmA}{}^{\bmone}) - \theta^{-1} \Lambda_{\bmA}{}^{\bmone} \Lambda^{\bmC}{}_{\bmzero} \Lambda^{\bmB}{}_{\bmone} \bar{\Lambda}^{\bmB'}{}_{\bmzero'} \Gamma_{\bmB \bmB'}{}^{\bmA}{}_{\bmC}, \\
        && \epsilon^{\bullet} = \frac{1}{2} \theta^{-1} \bme^{\bullet}_{\bmzero \bmzero'} (\theta) - \Lambda^{\bmA}{}_{\bmzero} \bme^{\bullet}_{\bmzero \bmzero'}(\Lambda_{\bmA}{}^{\bmzero}) + \theta^{-1} \Lambda_{\bmA}{}^{\bmzero} \Lambda^{\bmC}{}_{\bmzero} \Lambda^{\bmB}{}_{\bmzero} \bar{\Lambda}^{\bmB'}{}_{\bmzero'} \Gamma_{\bmB \bmB'}{}^{\bmA}{}_{\bmC},
    \end{eqnarray}
    \label{NP-connection-coefficient-transformation}
\end{subequations}
where the LHS in the above expressions is written in terms of F-gauge quantities, except for $\bme^{\bullet}_{\bmA \bmA'}$, which will be explicitly computed in later sections. The component $\bar{\phi}^{\bullet}_2$ can be written in terms of F-gauge quantities as
\begin{eqnarray}
    && \bar{\phi}^{\bullet}_2 = - \frac{3}{2}  \theta^{-3} \Lambda^{\bmA}{}_{\bmone} \Lambda^{\bmB}{}_{\bmzero} \Lambda^{\bmC}{}_{\bmone} \Lambda^{\bmD}{}_{\bmzero} \bar{\Lambda}^{\bmA'}{}_{\bmzero'} \bar{\Lambda}^{\bmB'}{}_{\bmone'} \bar{\Lambda}^{\bmC'}{}_{\bmzero'} \bar{\Lambda}^{\bmD'}{}_{\bmone'} ( \bar{d}_{\bmA \bmA' \bmB \bmB' \bmC \bmC' \bmD \bmD'} \label{phi-bullet-transformation} \\
    && \phantom{\bar{\phi}^{\bullet}_2 } -\bar{d}_{\bmB \bmA' \bmA \bmB' \bmC \bmC' \bmD \bmD'} + \bar{d}_{\bmB \bmA' \bmA \bmB' \bmD \bmC' \bmC \bmD'} - \bar{d}_{\bmA \bmA' \bmB \bmB' \bmD \bmC' \bmC \bmD'} ). \nonumber
\end{eqnarray}
Here, $\bar{d}_{\bmA \bmA' \bmB \bmB' \bmC \bmC' \bmD \bmD'}$ denotes the components of the complex conjugate of $d_{AA'BB'CC'DD'}$ in the F-gauge. The discussion above indicates that $\mathcal{Q}$ can be written as
\begin{equation}
    \mathcal{Q} (f, \mathcal{C}) \equiv \mathcal{Q} (\theta, \Lambda^{\bmA}{}_{\bmB}, \Lambda_{\bmB}{}^{\bmA}, \bar{\Lambda}^{\bmA'}{}_{\bmB'}, \bar{\Lambda}_{\bmB'}{}^{\bmA'}, \bme^{\bullet}_{\bmA \bmA'}, \Gamma_{\bmA \bmA'}{}^{\bmC}{}_{\bmD},\bar{d}_{\bmA \bmA' \bmB \bmB' \bmC \bmC' \bmD \bmD'}).
    \label{Asymptotic-BMS-charges-F-gauge}
\end{equation}
Accordingly, the evaluation of $\mathcal{Q}$ at $\mathcal{I}^{\pm}$ requires a solution for $\bme_{\bmA \bmA'}$, $\Gamma_{\bmA \bmA'}{}^{\bmC}{}_{\bmD}$ and $\bar{d}_{\bmA \bmA' \bmB \bmB' \bmC \bmC' \bmD \bmD'}$ at $\mathcal{I}^{\pm}$. As will become evident, a solution of the conformal field equations is also necessary for obtaining an asymptotic expression of $\theta, \Lambda^{\bmA}{}_{\bmB}$ and $\bme^{\bullet}_{\bmA \bmA'}$.

\section{The initial data and the constraint equations}
\label{Section: Initial data}
This section discusses the initial data used in our analysis, which is required to obtain a solution to the conformal field equations and, more precisely, to obtain non-trivial BMS asymptotic charges at $\mathscr{I}^{\pm}$. The conformal constraint equations, satisfied by our initial data, will be introduced in terms of the Levi-Civita connection associated with $\bmg$, contrary to the extended conformal field equations discussed in Section \ref{Section: The extended conformal field equations}.

\subsection{The conformal constraint equations}
To introduce the conformal constraint equations, let $(\tilde{\mathcal{M}},\tilde{\bmg})$ denote the physical spacetime satisfying eq. \eqref{vacuum-Einstein-field-equations}, and introduce the unphysical spacetime $(\mathcal{M}, \bmg)$ with $\bmg$ given by eq. \eqref{metric-conformal-transformation}. Moreover, let $\tilde{\mathcal{S}}$ denote a three-dimensional hypersurface on $\tilde{\mathcal{M}}$ with an induced metric denoted by $\tilde{\bmh}$. As mentioned earlier, the hypersurface $\tilde{\mathcal{S}}$ can be regarded as a hypersurface on $\mathcal{M}$ given the composition map $\phi \circ \varphi : \tilde{\mathcal{S}} \to \mathcal{M}$ where $\phi: \tilde{\mathcal{M}} \to \mathcal{M}$ and $\varphi: \tilde{\mathcal{S}} \to \tilde{\mathcal{M}}$. The metric $\bmg$ on $\mathcal{M}$ also induces an intrinsic metric $\bmh$ on $\tilde{\mathcal{S}}$ related to $\tilde{\bmh}$ by eq. \eqref{Inrinsic-metric-conformal-transformation}. If $\tilde{\bmn}$ and $\bmn$ are the $\tilde{\bmg}$ and $\bmg$ unit normals of $\tilde{\mathcal{S}}$, then 
\begin{equation*}
    \epsilon \equiv \tilde{\bmg}(\tilde{\bmn},\tilde{\bmn}) = \bmg(\bmn,\bmn) = 1,
\end{equation*}
since $\tilde{\mathcal{S}}$ is a spacelike hypersurface.  Moreover, if $\bmu, \bmv$ are some arbitrary vectors on $\tilde{\mathcal{S}}$, then the extrinsic curvatures $\tilde{\bmK}$ and $\bmK$ are defined by
\begin{equation*}
    \tilde{\bmK}(\bmu, \bmv) = \langle \tilde{\bmnabla}_{\bmu} \tilde{\bmn},\bmv \rangle, \qquad \bmK(\bmu, \bmv) = \langle \bmnabla_{\bmu} \bmn,\bmv \rangle.
\end{equation*}
It can be shown that the relation between $\tilde{\bmK}$ and $\bmK$ is given by
\begin{equation*}
    \bmK = \Omega(\tilde{\bmK} + \Sigma \tilde{\bmh}),
\end{equation*}
where
\begin{equation*}
    \Sigma = \bmg^{\sharp}(\bmd{\Xi},\bmn) = \Xi^{-1} \tilde{\bmg}^{\sharp}(\bmd{\Xi},\tilde{\bmn}).
\end{equation*}
Let $\{ \bme_{\bmi} \}$ denote an $\bmh$-orthonormal frame. Then, the vacuum conformal constraint equations on $\tilde{\mathcal{S}}$ are given by
\begin{subequations}
    \begin{eqnarray}
        && D_{\bmi} D_{\bmj} \Omega + \Sigma K_{\bmi \bmj} + \Omega  L_{\bmi  \bmj }  - s h_{\bmi  \bmj } =0, \label{ConformalConstraint1}\\
        && D_{\bmi} \Sigma - K_{\bmi}{}^{ \bmk } D_{ \bmk } \Omega + \Omega L_{\bmi  } =0, \label{ConformalConstraint2}\\
        && D_{\bmi} s + \Sigma L_{\bmi  } + L_{\bmi  \bmk } D^{ \bmk } \Omega =0, \label{ConformalConstraint3}\\
        && D_{\bmi } L_{ \bmj   \bmk } - D_{ \bmj } L_{\bmi   \bmk }  + \Sigma d_{\bmk   \bmi     \bmj } - D^{ \bml }\Omega d_{ \bml  \bmk  \bmi  \bmj } + K_{ \bmi  \bmk } L_{ \bmj   } - K_{ \bmj  \bmk } L_{ \bmi   } =0, \label{ConformalConstraint4}\\
        && D_{ \bmi } L_{ \bmj   } - D_{ \bmj } L_{ \bmi   } + K_{ \bmj }{}^{ \bmk } L_{ \bmi   \bmk } - K_{ \bmi }{}^{ \bmk } L_{ \bmj   \bmk } - D^{ \bml } \Omega d_{ \bml     \bmi  \bmj } =0, \label{ConformalConstraint5}\\
        && D^{ \bmk } d_{ \bmk     \bmi   \bmj } - K^{ \bmk }{}_{ \bmi } d_{ \bmj     \bmk   } + K^{ \bmk }{}_{ \bmj } d_{ \bmi     \bmk   }=0,\label{ConformalConstraint6}\\
        && D^{ \bmk } d_{ \bmk     \bmj   } -K^{ \bmi   \bmk }d_{ \bmi     \bmj   \bmk }=0, \label{ConformalConstraint7}\\
        && 6 \Omega s - 3 \Sigma^2 - 3 D_{ \bmi } \Omega D^{ \bmi } \Omega =0. \label{ConformalConstraint8}
    \end{eqnarray}
    \label{Frame-version-conformal-constraint-equations}
\end{subequations}
In the above, $\bmD$ denotes the Levi-Civita connection associated with $\bmh$ and $D_{\bmi} \equiv \bme_{\bmi}{}^{i} D_{i}$ while $h_{\bmi \bmj} \equiv \bmh (\bme_{\bmi}, \bme_{\bmj})$ denotes the components of $\bmh$ with respect to $\{ \bme_{\bmi} \}$. Similarly, $K_{\bmi \bmj}$ and $l_{\bmi \bmj}$ denote the components of the extrinsic curvature and intrinsic Schouten tensor with respect to $\{ \bme_{\bmi} \}$, respectively. Furthermore, $L_{\bmi}, d_{\bmi \bmj}$ and $d_{\bmi \bmj \bmk}$ are the spatial components of the contraction of $L_{\bma \bmb}, d_{\bma \bmb \bmc \bmd}$ with $n^{\bma}$. In particular, $L_{\bmi}, d_{\bmi \bmj}$ and $d_{\bmi \bmj \bmk}$ are defined as
\begin{equation*}
    L_{\bmi} \equiv L_{\bmi \bma} n^{\bma}, \qquad \; d_{\bmi \bmj} \equiv d_{\bmi \bma \bmj \bmb} n^{\bma} n^{\bmb}, \qquad \; d_{\bmi \bmj \bmk} \equiv d_{\bmi \bma \bmj \bmk} n^{\bma}.
\end{equation*}
\begin{remark}
     {\em The spatial components of $E_{ab}$, introduced in Section \ref{Section:The conformal field equations in the Conformal Gaussian gauge} are equivalent to $d_{\bmi \bmj}$ while the spatial components of $B_{ab}$ are related to $d_{\bmi \bmj \bmk}$ by
    \begin{equation*}
        d_{\bmi \bmj \bmk} = \epsilon^{\bml}{}_{\bmj \bmk} B_{\bmi \bml},
    \end{equation*}
    where $\epsilon^{\bml}{}_{\bmj \bmk}$ denotes the components of the three-dimensional volume form on $\tilde{\mathcal{S}}$.} 
\end{remark}

\subsection{The initial data}
To introduce the initial data for the ECFEs, the starting point is to consider an initial data set satisfying the Hamiltonian and momentum constraints implied by the vacuum Einstein field eq. \eqref{vacuum-Einstein-field-equations}. More precisely, we will be interested in a vacuum initial data set $(\tilde{\mathcal{S}},\tilde{\bmh},\tilde{\bmK})$ satisfying 
\begin{subequations}
    \begin{eqnarray}
        && \tilde{r} + \tilde{K}^2 - \tilde{K}_{jl} \tilde{K}^{jl} =0,
        \label{Hamiltonian-constraint} \\
        && \tilde{D}^{j} \tilde{K}_{kj} - \tilde{D}_{k}\tilde{K} =0.
        \label{Momentum-constraint}
    \end{eqnarray}
    \label{Vacuum-constraint-equations}
\end{subequations}
To obtain non-trivial BMS asymptotic charges at $\mathscr{I}^{\pm}$, let us consider the vacuum initial data prescribed in \cite{Huang10}:

\begin{proposition}
\label{Proposition:DataLan}
    For any $\xi, \zeta \in C^{2}(\mathbb{S}^2)$ and $q \geq 1$, there exists a vacuum initial data set $(\tilde{\bmh}, \tilde{\bm{\pi}})$ where the components of the intrinsic metric $\tilde{\bmh}$ and the momentum tensor $\tilde{\bm{\pi}}$ with respect to the standard Euclidean coordinate chart $( x^{\alpha} )$ have the asymptotics
    \begin{subequations}
        \begin{eqnarray*}
            && \tilde{h}_{\alpha \beta} = - \left(1+ \frac{A}{r} \right) \delta_{\alpha \beta} - \frac{\xi}{r} \left( \frac{x_{\alpha} x_{\beta}}{r^2} - \frac{1}{2} \delta_{\alpha \beta} \right) + O_2 (r^{-1-q}), \\
            && \tilde{\pi}_{\alpha \beta} = \frac{\zeta}{r^2} \frac{x_{\alpha} x_{\beta}}{r^2} + \frac{1}{r^3} \left(- B_{\alpha} x_{\beta} - B_{\beta} x_{\alpha} + (B^{\gamma} x_{\gamma}) \delta_{\alpha \beta} \right) + O_1 (r^{-2-q}),
        \end{eqnarray*}
    \end{subequations}
    where $A$, $\{ B_{\alpha}\}_{\alpha=1}^{3}$ are some constants, and $r \equiv \sqrt{(x^1)^2 + (x^2)^2 + (x^3)^2}$. The momentum tensor $\tilde{\bm{\pi}}$ is defined as
    \begin{equation}
        \tilde{\pi}_{ij} \equiv \tilde{K}_{ij} - \tilde{K} \tilde{h}_{ij}.
        \label{Momentum-tensor}
    \end{equation}
    \label{Proposition:physical-initial-data}
\end{proposition}
To simplify the analysis in this work, we set $q=1$ so that the components of $\tilde{\bmh}$ and $\tilde{\bm{\pi}}$ are written as  
\begin{subequations}
        \begin{eqnarray*}
            && \tilde{h}_{\alpha \beta} = -\delta_{\alpha \beta} - \frac{1}{r} \left[ \left( A - \frac{\xi}{2} \right) \delta_{\alpha \beta} + \xi \frac{x_{\alpha} x_{\beta}}{r^2} \right] + O_2 (r^{-2}), \\
            && \tilde{\pi}_{\alpha \beta} = \frac{1}{r^2} \left[ \frac{1}{r} \left( - B_{\alpha} x_{\beta} - B_{\beta} x_{\alpha} + (B^{\gamma} x_{\gamma}) \delta_{\alpha \beta}\right) + \zeta \frac{x_{\alpha} x_{\beta}}{r^2} \right] + O_1 (r^{-3}),
        \end{eqnarray*}
        \label{vacuum-initial-data}
    \end{subequations}
where the expressions above have been rearranged to group terms of similar orders in $r$. Given eq. \eqref{Momentum-tensor}, the components of $\tilde{\bmK}$ with respect to $( x^{\alpha} )$ can be written as
\begin{equation*}
    \tilde{K}_{\alpha \beta} = \frac{1}{r^2} \left[  - \frac{1}{2} \zeta \delta_{\alpha \beta} + \frac{1}{r} \left( - B_{\alpha} x_{\beta} - B_{\beta} x_{\alpha} + (B^{\gamma} x_{\gamma}) \delta_{\alpha \beta}\right) + \zeta \frac{x_{\alpha} x_{\beta}}{r^2} \right] + O_1 (r^{-3})
\end{equation*}
In order to discuss the region near spatial infinity, we introduce the inverse coordinates $( y^{\alpha} )$ related to $( x^{\alpha} )$ by
\begin{equation*}
    y^{\alpha} = - \frac{x^{\alpha}}{r^2}, \qquad y_{\alpha} = - \frac{x_{\alpha}}{r^2}.
\end{equation*}
Then, the inverse coordinate transformation is given by
\begin{equation*}
    x^{\alpha} = - \frac{y^{\alpha}}{\varrho^2}, \qquad x_{\alpha} = - \frac{y_{\alpha}}{\varrho^2}, \qquad \varrho = \frac{1}{r},
\end{equation*}
where $\varrho = \sqrt{(y^{1})^2 + (y^{2})^2+(y^{3})^2}$. In terms of $( y^{\alpha} )$, the components of $\tilde{\bmh}$ and $\tilde{\bmK}$ can be written as
\begin{subequations}
    \begin{eqnarray*}
        && \tilde{h}_{\alpha \beta} = - \frac{(1+ A \varrho)}{\varrho^4} \delta_{\alpha \beta} - \frac{\xi}{\varrho^{3}} \left( \frac{y_{\alpha} y_{\beta}}{\varrho^2} - \frac{\xi}{2} \delta_{\alpha \beta} \right)+ O_2(\varrho^{-2}), \\
        && \tilde{K}_{\alpha \beta} = - \frac{\zeta}{2 \varrho^2} \delta_{\alpha \beta} - \frac{1}{\varrho^3} \left( 2B_{(\alpha} y_{\beta)} + \frac{1}{2} (B^{\gamma}y_{\gamma}) \delta_{\alpha \beta} \right) +  \frac{y_{\alpha} y_{\beta}}{\varrho^4} \left( \zeta - 4 \frac{(B^{\gamma}y_{\gamma})}{\varrho}  \right) + O_1(\varrho^{-1}). 
    \end{eqnarray*}
\end{subequations}
In the following, let $\mathcal{S}'$ denote a three-dimensional compact manifold with a spatial infinity point $i$ and let $\phi$ denote the diffeomorphic map from $\mathcal{S}' \setminus \{ i \}$ to $\tilde{\mathcal{S}}$ with the conformal factor $\Omega'$ given by
\begin{equation*}
    \Omega' = \frac{\varrho^2}{\sqrt{1 + A \varrho}}.
\end{equation*}
Then, the components of $\bmh'\equiv \Omega'^{2} \tilde{\bmh}$ and $\bmK'\equiv \Omega' \tilde{\bmK}$ with respect to $(y^{\alpha})$ are related to $\tilde{h}_{\alpha \beta}$ and $\tilde{K}_{\alpha \beta}$ by
\begin{equation*}
    h'_{\alpha \beta} = \Omega'^2 \tilde{h}_{\alpha \beta}, \qquad K'_{\alpha \beta} = \Omega' \tilde{K}_{\alpha \beta}.
\end{equation*}
By expanding $\Omega'$ around $\varrho=0$, the components $ h'_{\alpha \beta}$ and $K'_{\alpha \beta}$ can be written as
\begin{subequations}
    \begin{align}
        h'_{\alpha \beta} &= -\delta_{\alpha \beta} - \xi \varrho \left( \frac{y_{\alpha} y_{\beta}}{\varrho^2} -\frac{1}{2} \delta_{\alpha \beta} \right) + O_2 (\varrho^2), \label{Initial-data-h'} \\
        \hspace{1cm} K'_{\alpha \beta} &= - \frac{\zeta}{2} \delta_{\alpha \beta} - \frac{1}{\varrho} \left( 2 B_{(\alpha} y_{\beta)} + \frac{1}{2} (B^{\gamma}y_{\gamma}) \delta_{\alpha \beta} \right) + \left( \zeta - 4 \frac{(B^{\gamma}y_{\gamma})}{\varrho}  \right) \frac{y_{\alpha} y_{\beta}}{\varrho^2} + O_1(\varrho).\label{Initial-data-K'}
    \end{align}
    \label{Initial-data-h'andK'}
\end{subequations}
The above initial data $(\bmh',\bmK')$ is said to be asymptotically Euclidean and regular in the spirit of Definition \ref{Definition:AsympEuclideanAndRegular}. To analyse the conformal constraint equations, it will be convenient to express $\bmh'$ and $ \bmK'$ in terms of the so-called normal coordinates and to introduce the conformal normal initial data. 

\subsubsection{Normal coordinates and conformal normal initial data}
\label{Section:Normal coordinates and conformal normal initial data}
Consider the $\bmh'$-geodesics emanating from $i \in \mathcal{S}'$ to nearby points in a neighbourhood $\mathcal{U}' \in \mathcal{S}'$ and introduce the subset $\mathcal{T}$ of the tangent space at $i$
\begin{equation*}
    \mathcal{T} \equiv \{ \bmv \in T|_{i}(\mathcal{S}') | \hspace{2mm} \gamma_{\bmv} \text{ is defined on an interval containing } [0,1] \},
\end{equation*}
where $\gamma_{\bmv} \equiv \gamma_{\bmv}(t)$ is the geodesic starting at $i$ (i.e. $\gamma_{\bmv}(0) = i$) with an initial tangent vector $\bmv$. Let $\text{exp}_{i}$ denote the exponential map at i, $\text{exp}_{i}: \mathcal{T} \to \mathcal{S}'$, defined such that $\text{exp}_{i}(\bmv) = \gamma_{\bmv}(1)$. Then, the neighbourhood $\mathcal{U}'$ is said to be a normal neighbourhood of $i$ if $\mathcal{U}' =  \text{exp}_{i}(\mathcal{Q})$, where $\mathcal{Q} \subset T|_{i}(\mathcal{S}')$ is a neighbourhood of the zero vector $\bmzero$, and if for all $t \in [0,1]$, 
\begin{equation*}
    \bmv \in \mathcal{Q} \rightarrow t \bmv \in \mathcal{Q}.
\end{equation*}
For any point $p = \text{exp}_{i}(\bmv) \in \mathcal{U}'$, the normal coordinates $(z^{\alpha})$ is given by $z^{\alpha} = \bme^{\alpha}{}_{\bmi} v^{\bmi}$, where $v^{\bmi}$ are the components of $\bmv$ with respect to an orthonormal basis $\{ \bme_{\bmi} \}$ and $\bme^{\alpha}{}_{\bmi}$ are defined by
\begin{equation*}
    \bme^{\alpha}{}_{\bmi} \equiv \langle \bmd{y}^{\alpha}, \bme_{\bmi} \rangle.
\end{equation*}
Let $v^{\alpha} \equiv \langle \bmd{z}^{\alpha}, \bmv \rangle$ denote the components of $\bmv$ with respect to $(z^{\alpha})$, then the normal coordinates for any point $p$ on $\gamma_{\bmv}(t)$ is given by $x^{\alpha}(t) = t z^{\alpha}$. If $\gamma'_{\alpha}{}^{\beta}{}_{\gamma}$ denote the components of the $\bmh'$-Levi-Civita connection coefficients with respect to $(z^{\alpha})$, the geodesic equation can be written as
\begin{equation}
    \frac{d^2 x^{\beta}}{d t^2} + \gamma'_{\alpha}{}^{\beta}{}_{\gamma} \frac{d x^{\alpha}}{d t} \frac{d x^{\gamma}}{d t}=0,
\end{equation}
then one can show that $\gamma'_{\alpha}{}^{\beta}{}_{\gamma}$ vanish at $i$, i.e.,
\begin{equation*}
    \gamma'_{\alpha}{}^{\beta}{}_{\gamma} (i) =0.
\end{equation*}
This implies that the components of the metric $\bmh'$ in normal coordinates satisfy
\begin{equation}
    h'_{\alpha \beta,\gamma} =0\qquad \text{ at }i,
\end{equation}
where $h'_{\alpha \beta,\gamma}$ is the derivative of $h'_{\alpha \beta}$ with respect to $z^{\gamma}$. Taylor expanding the metric $\bmh'$ around $i$ gives
\begin{equation}
    h'_{\alpha \beta} (z) = h'^{(0)}_{\alpha \beta} + \frac{1}{2} h'_{\alpha \beta, \gamma \delta} z^{\gamma} z^{\delta} + O(|z|^{3}),
    \label{Metric-normal-coordinates}
\end{equation} 
where $|z|^{2} \equiv \delta_{\alpha \beta} z^{\alpha} z^{\beta}$ and $h'^{(0)}_{\alpha \beta} = - \delta_{\alpha \beta}$ is the metric at $i$ (i.e. at $|z|=0$). The non-vanishing $O(\varrho)$ terms in the initial data for $\bmh'$ given by eq. \eqref{Initial-data-h'andK'} implies that $(y^{\alpha})$ are not normal coordinates. However, the discussion in \cite{LBrewin} shows that the transformation between a generic set of coordinates $(y^{\alpha})$ and normal coordinates $(z^{\alpha})$ is given by
\begin{equation}
    y^{\alpha} = z^{\alpha} - \frac{1}{2} \gamma'^{(0)}_{\beta}{}^{\alpha}{}_{\gamma} z^{\beta} z^{\gamma} + O(|z|^3),
    \label{Original-to-normal-coordinates}
\end{equation}
where $\gamma'^{(0)}_{\beta}{}^{\alpha}{}_{\gamma}$ are the components of the $\bmh'$-Levi-Civita connection coefficients with respect to $(y^{\alpha})$ evaluated at $\varrho=0$. In a slight abuse of notation, let $h'_{\beta \delta,\gamma}$ denote the derivative of $h'_{\beta \delta}$ with respect to $y^{\gamma}$, then $\gamma'_{\beta}{}^{\alpha}{}_{\gamma}$ is given by
\begin{equation*}
    \gamma'_{\beta}{}^{\alpha}{}_{\gamma} = \frac{1}{2} h'^{\alpha \delta} \left( h'_{\beta \delta,\gamma} + h'_{\delta \gamma, \beta} - h'_{\beta \gamma, \delta} \right).
\end{equation*}
From eq. \eqref{Initial-data-h'andK'}, one has
\begin{subequations}
    \begin{eqnarray*}
        && \gamma'^{(0)}_{\beta}{}^{\alpha}{}_{\gamma} = \xi_{;( \gamma} \left( \vartheta_{\beta)} \vartheta^{\alpha} - \frac{1}{2} \delta_{\beta)}{}^{\alpha} \right) - \frac{1}{2} \delta^{\alpha \delta} \xi_{;\delta} \left( \vartheta_{\beta} \vartheta_{\gamma} - \frac{1}{2} \delta_{\beta \gamma} \right) - \frac{1}{2} \vartheta_{\beta} \vartheta^{\alpha} \vartheta_{\gamma} \left( \vartheta^{\delta} \xi_{; \delta} + \xi \right) \\
        && \phantom{\gamma'^{(0)}_{\beta}{}^{\alpha}{}_{\gamma}} + \frac{1}{2} \vartheta^{\delta} \xi_{;\delta} \left( \vartheta_{(\gamma} \delta_{\beta)}{}^{\alpha} - \frac{1}{2} \vartheta^{\alpha} \delta_{\beta \gamma} \right)- \frac{\xi}{2} \left( \vartheta_{(\gamma} \delta_{\beta)}{}^{\alpha} + \frac{3}{2} \vartheta^{\alpha} \delta_{\beta \gamma} \right), 
    \end{eqnarray*}
\end{subequations}
where $\vartheta^{\alpha} \equiv  y^{\alpha}/ \varrho$. In the above, $\xi_{;\gamma}$ is the derivative of $\xi$ with respect to the angular coordinates $\vartheta^{\gamma}$ and ${}_{(\gamma \beta)}$ indicates a symmetrisation over $\gamma$ and $\beta$. The transformation given by eq. \eqref{Original-to-normal-coordinates} implies that the components of $\bmh'$ with respect to $(z^{\alpha})$ admit the following expansion near $|z|=0$:
\begin{equation*}
     h'_{\alpha \beta} = - \delta_{\alpha \beta} + O(|z|^2).
\end{equation*}
Moreover, the components of the extrinsic curvature $K'_{\alpha \beta}$  with respect to $(z^{\alpha})$ can be written as
\begin{equation*}
    K'_{\alpha \beta} = - \frac{\zeta}{2} \delta_{\alpha \beta} - \frac{1}{2} \left( 2 B_{(\alpha} \vartheta_{\beta)} + \frac{1}{2} (B^{\gamma} \vartheta_{\gamma}) \delta_{\alpha \beta} \right) + \zeta \vartheta_{\alpha} \vartheta_{\beta} - 4 (B^{\gamma} \vartheta_{\gamma}) \vartheta_{\alpha} \vartheta_{\beta} + O(|z|).
\end{equation*}
Taylor expanding the conformal factor $\Omega'$ around $|z|=0$ gives
\begin{equation*}
    \Omega'= |z|^2 - \frac{A}{2} |z|^3 - \gamma'^{(0)}_{\beta}{}^{\alpha}{}_{\gamma} z^{\beta} z_{\alpha} z^{\gamma} + O(|z|^4).
\end{equation*}
The next step is to exploit the conformal freedom in Definition \ref{Definition:AsympEuclideanAndRegular} with the aim of simplifying upcoming calculations. In particular, introduce the conformal normal initial data $(\bar{\bmh}, \bar{\bmK})$, related to $(\bmh',\bmK')$ by
\begin{equation}
    \bar{\bmh} = \varpi^{2} \bmh', \qquad \bar{\bmK} = \varpi \bmK'.
    \label{Conformal-normal-initial-data-rescaling}
\end{equation}
In the following, let $l'_{\alpha \beta}(i)$ denote the components of the $\bmh'$-Schouten tensor with respect to $(z^{\alpha})$ evaluated at $i$. If the conformal factor $\varpi$ is given by
\begin{equation}
    \varpi \equiv e^f, \qquad \text{ with } f= \frac{1}{2} l'_{\alpha \beta}(i) z^{\alpha} z^{\beta},
    \label{Conforaml-normal-conformal-factor}
\end{equation}
then one can show that the Riemann curvature tensor associated with $\bar{\bmh}$ is vanishing at $i$. In particular, if $\bmD'$ denote the $\bmh'$-covariant derivative, the conformal factor $\varpi$ can be shown to satisfy
\begin{equation*}
    \varpi(i) =1, \qquad D'_{\alpha} \varpi(i) = 0, \qquad D'_{\alpha} D'_{\beta} \varpi(i) = l'_{\alpha \beta}(i).
\end{equation*}
Accordingly, the conformal rescaling given by eq. \eqref{Conformal-normal-initial-data-rescaling} implies that 
\begin{equation*}
    \bar{l}_{\alpha \beta}(i)=0.
\end{equation*}
However, the three-dimensional Riemann curvature tensor $\bar{r}_{\alpha \beta \gamma \delta}$ is fully determined by $\bar{l}_{\alpha \beta}$. Hence, one has
\begin{equation*}
    \bar{r}_{\alpha \beta \gamma \delta}(i)=0.
\end{equation*}
Note that the conformal rescaling in eq. \eqref{Conformal-normal-initial-data-rescaling} indicates that $(\bar{\bmh},\bar{\bmK})$ are related to $(\tilde{\bmh},\tilde{\bmK})$ by
\begin{equation*}
    \bar{\bmh} = \Omega^{2} \tilde{\bmh}, \qquad \bar{\bmK} = \Omega \tilde{\bmK},
\end{equation*}
where 
\begin{equation*}
    \Omega \equiv \varpi \Omega'.
\end{equation*}
In a slight abuse of notation, let 
\begin{equation*}
    \vartheta^{\alpha} \equiv \frac{z^{\alpha}}{\rho},
\end{equation*}
where $\rho \equiv |z|$. By Taylor expanding $\Omega$ around $\rho =0$, one gets
\begin{subequations}
    \begin{eqnarray}
        && \Omega = \rho^{2} - \left( \frac{A}{2} + \gamma'^{(0)}_{\beta}{}^{\alpha}{}_{\gamma} \vartheta^{\beta} \vartheta_{\alpha} \vartheta^{\gamma} \right) \rho^3 + O(\rho^4), \\
        && \phantom{\Omega} = \rho^{2} + \frac{1}{6} \Pi_{3}[\Omega] \rho^3 + O(\rho^4). \nonumber
    \end{eqnarray}
    \label{Omega-expansion}
\end{subequations}
In the above, $\Pi_{3}[\Omega]$ is defined by
\begin{equation}
    \Pi_{3}[\Omega] \equiv - 6 \left( \frac{A}{2} + \gamma'^{(0)}_{\beta}{}^{\alpha}{}_{\gamma} \vartheta^{\beta} \vartheta_{\alpha} \vartheta^{\gamma} \right).
    \label{Pi-3-Omega}
\end{equation}
A direct computation readily shows that 
\[
\gamma'^{(0)}_{\beta}{}^{\alpha}{}_{\gamma} \vartheta^{\beta} \vartheta_{\alpha} \vartheta^{\gamma} = -\frac{7}{4}\xi,
\]
where it is recalled that $\xi\in C^2(\mathbb{S}^2)$ is a freely specifiable function on the 2-sphere. Thus, one has that
\begin{equation}
\Pi_{3}[\Omega] = \frac{21}{2}\xi -3A. 
\label{Pi-3-Omega_ALT}
\end{equation}
Accordingly, one sees that the coefficient  $\Pi_{3}[\Omega]$ is completely determined by the freely specifiable data $A$ and $\xi$.
\begin{remark}
    {\em The $\Pi_{n}$ notation will be used frequently in later calculations. For any smooth function $\chi$, we use $\Pi_{n}[\chi]$ to denote the coefficient of $(1/n!) \rho^{n}$ in its Taylor series around $\rho=0$.}
\end{remark}
From the previous discussion, the components $\bar{\bmh}$ and $\bar{\bmK}$ with respect to $(z^{\alpha})$ are given by
\begin{subequations}
    \begin{eqnarray*}
        && \bar{h}_{\alpha \beta} = - \delta_{\alpha \beta} + O(\rho^3), \label{Conformal-normal-metric} \\
        && \bar{K}_{\alpha \beta} = - \frac{\zeta}{2} \delta_{\alpha \beta} - \frac{1}{2} \left( 2 B_{(\alpha} \vartheta_{\beta)} + \frac{1}{2} (B^{\gamma} \vartheta_{\gamma}) \delta_{\alpha \beta} \right) + \zeta \vartheta_{\alpha} \vartheta_{\beta} - 4 (B^{\gamma} \vartheta_{\gamma}) \vartheta_{\alpha} \vartheta_{\beta} + O(\rho).
    \end{eqnarray*}
\end{subequations}
In other words, $(\bar{\bmh}, \bar{\bmK})$ can be written as
\begin{subequations}
    \begin{eqnarray*}
        && \bar{\bmh} = \bar{h}_{\alpha \beta} \bmd{z}^{\alpha} \otimes \bmd{z}^{\beta}, \\
        && \bar{\bmK} = \bar{K}_{\alpha \beta} \bmd{z^{\alpha}} \otimes \bmd{z^{\beta}}.
    \end{eqnarray*}
\end{subequations}
Observe that the leading order term in $\bar{K}_{\alpha \beta}$ is equivalent to the leading order in $K'_{\alpha \beta}$ since the contribution from $\varpi$ is at higher orders. In upcoming calculations, the initial data $(\bar{\bmh}, \bar{\bmK})$ will be referred to as the conformal normal initial data. 

\subsubsection{The regular initial data at spatial infinity}
To analyse the conformal constraint eqs. \eqref{Frame-version-conformal-constraint-equations}, introduce the orthonormal frame $\{ \bar{\bme}_{\bmi} \}$ and its dual $\{ \bar{\bmomega}^{\bmi} \}$ related to $\{ \bmpartial /\bmpartial z^{\alpha} \}$ and $\{ \bmd{z^{\alpha}} \}$ by
\begin{equation*}
    \bar{\bme}_{\bmi} = \bar{\bme}^{\alpha}{}_{\bmi} \frac{\bmpartial}{\bmpartial z^{\alpha}}, \qquad \bar{\bmomega}^{\bmi} = \bar{\bmomega}_{\alpha}{}^{\bmi} \bmd{z^{\alpha}}.
\end{equation*}
Given that $\bar{\bmh}(\bar{\bme}_{\bmi},\bar{\bme}_{\bmj})= \delta_{\bmi \bmj}$, one has
\begin{equation*}
    \bar{\bme}^{\alpha}{}_{\bmi}(i) =1, \qquad \bar{\bmomega}_{\alpha}{}^{\bmi} (i)=1.
\end{equation*}
In terms of $\{ \bar{\bme}_{\bmi} \}$, the conformal normal initial data can be written as
\begin{subequations}
    \begin{eqnarray*}
        && \bar{\bmh} = \bar{h}_{\bmi \bmj} \bar{\bmomega}^{\bmi} \otimes \bar{\bmomega}^{\bmj}, \\
        && \bar{\bmK} = \bar{K}_{\bmi \bmj} \bar{\bmomega}^{\bmi} \otimes \bar{\bmomega}^{\bmj},
    \end{eqnarray*}
\end{subequations}
with 
\begin{subequations}
    \begin{eqnarray*}
        && \bar{h}_{\bmi \bmj} = - \delta_{\bmi \bmj} + O(\rho^3), \\
        && \bar{K}_{\bmi \bmj} = - \frac{\zeta}{2} \delta_{\bmi \bmj} - \frac{1}{2} \left( 2 B_{(\bmi} \vartheta_{\bmj)} + \frac{1}{2} (B^{\bmk} \vartheta_{\bmk}) \delta_{\bmi \bmj} \right) + \zeta \vartheta_{\bmi} \vartheta_{\bmj} - 4 (B^{\bmk} \vartheta_{\bmk}) \vartheta_{\bmi} \vartheta_{\bmj} + O(\rho).
    \end{eqnarray*}
\end{subequations}
Substituting in eq. \eqref{Frame-version-conformal-constraint-equations}, one can show that
\begin{equation*}
    \bar{L}_{\bmi \bmj} = O(\rho^{-1}), \qquad \bar{d}_{\bmi \bmj \bmk} = O(\rho^{-3}), \qquad \bar{d}_{\bmi \bmj} = O(\rho^{-3}).
\end{equation*}
Hence, the initial data for the conformal fields $\bar{L}_{\bmi \bmj}, \bar{d}_{\bmi \bmj}, \ldots$ implied by the conformal normal initial data is singular at $\rho=0$. Following Friedrich's formulation \cite{Friedrich98}, the regular initial data for the ECFEs can be introduced by considering the conformal rescaling
\begin{equation}
    \Omega \to \kappa^{-1} \Omega,
    \label{Regular-conformal-rescaling}
\end{equation}
where $\kappa = O(\rho)$ is the arbitrary function introduced in Section \ref{Section:Friedrich formulation of spatial infinity}. Then, introduce the rescaled frame fields $\{ \bme_{\bmi} \}$ and their dual $\{ \bmomega^{\bmi} \}$ as
\begin{equation*}
    \bme_{\bmi} = \kappa \bar{\bme}_{\bmi}, \qquad \bmomega^{\bmi} = \kappa^{-1} \bar{\bmomega}^{\bmi}.
\end{equation*}
Then, the components of the rescaled metric $\bmh = \kappa^{-2} \bar{\bmh}$ with respect to $\{ \bme_{\bmi} \}$ are related to $\bar{h}_{\bmi \bmj}$ by
\begin{subequations}
    \begin{eqnarray*}
        && h_{\bmi \bmj} = \bmh (\bme_{\bmi}, \bme_{\bmj}) \\
        && \phantom{h_{\bmi \bmj}} = \kappa^{-2} \bar{\bmh} (\kappa \bar{\bme}_{\bmi},\kappa \bar{\bme}_{\bmj}) \\
        && \phantom{h_{\bmi \bmj}} = \bar{\bmh} (\bar{\bme}_{\bmi}, \bar{\bme}_{\bmj}) = \bar{h}_{\bmi \bmj}.
    \end{eqnarray*}
\end{subequations}
Moreover, the components of the rescaled extrinsic curvature $\bmK = \kappa^{-1} \bar{\bmK}$ with respect to $\{ \bme_{\bmi} \}$ are related to $\bar{K}_{\bmi \bmj}$ by
\begin{equation*}
    K_{\bmi \bmj}=\bmK(\bme_{\bmi}, \bme_{\bmj}) = \kappa^{-1} \bar{\bmK} (\kappa \bar{\bme}_{\bmi}, \kappa \bar{\bme}_{\bmj}) = \kappa \bar{K}_{\bmi \bmj}.
\end{equation*}
Finally, one can show that the components of the rescaled conformal fields $(L_{\bmi \bmj}, d_{\bmi \bmj \bmk}, d_{\bmi \bmj})$ are related to $(\bar{L}_{\bmi \bmj}, \bar{d}_{\bmi \bmj \bmk}, \bar{d}_{\bmi \bmj})$ by
\begin{equation*}
    L_{\bmi \bmj} = \kappa^{2} \bar{L}_{\bmi \bmj}, \qquad d_{\bmi \bmj \bmk} = \kappa^{3} \bar{d}_{\bmi \bmj \bmk}, \qquad d_{\bmi \bmj} = \kappa^{3} \bar{d}_{\bmi \bmj}.
\end{equation*}
Hence, the conformal rescaling given by eq. \eqref{Regular-conformal-rescaling} introduces regular initial data for the ECFEs since 
\begin{equation*}
    K_{\bmi \bmj} = O(\rho), \qquad L_{\bmi \bmj}= O(\rho), \qquad d_{\bmi \bmj \bmk}= O(1), \qquad d_{\bmi \bmj} = O(1),
\end{equation*}
where the explicit form of these fields is omitted as they are not required for upcoming calculations. 

Note that the conformal freedom in Friedrich's formulation is reflected by the different choices of the conformal factor $\kappa$. In subsequent calculations, assume
\begin{equation}
    \kappa = \omega,
    \label{kappa-omega}
\end{equation}
where $\omega$ is given by eq. \eqref{Definition-omega-function}. This particular choice of $\kappa$ introduces Friedrich's horizontal representation of spatial infinity where $\mathscr{I}^{\pm}_a$ are identified by $0<\rho<a$ and $\tau = \pm 1$ ---see eq \eqref{Sets-null-infinity}. 

Using eq. \eqref{Definition-omega-function} and Taylor expanding $\omega$ around $\rho=0$ gives
\begin{equation}
    \kappa =\rho + \tfrac{1}{2}\Pi_{2}[\omega] \rho^2 + O(\rho^3),
    \label{kappa-expansion}
\end{equation}
with 
\begin{equation*}
    \Pi_{2}[\omega] = - \frac{1}{6} \Pi_{3}[\Omega].
\end{equation*}
Within the framework of the ECFEs, the conformal factor  $\Theta$ relating the spacetime metrics $\bmg$ and $\tilde{\bmg}$ is fixed by the initial data. Given the above choice of $\kappa$ and using eq. \eqref{Conformal-factor-Theta}, the conformal factor $\Theta$ can be expanded near $\rho=0$ as
\begin{equation}
    \Theta = \rho (1-\tau^2) + O(\rho^2).
    \label{Conformal-factor-Theta-expansion}
\end{equation}

\subsubsection{Conformally flat initial data for the space spinor fields}
Given that the final form of the ECFEs is written in terms of spinors, one needs to obtain initial data for the spinor fields appearing in eqs. \eqref{Evolution-system-in-space-spinors} and \eqref{The-Bianchi-evolution-system-in-space-spinors}. This discussion can be carried out for the general initial data discussed in the previous section. However, the calculations in this work only require expressions for the initial data at zero order. Accordingly, the formulae for the conformally flat initial data given in \cite{Kroon04} will be sufficient for this analysis. In the following, assume the conformally flat initial data for eqs. \eqref{Evolution-system-in-space-spinors} and \eqref{The-Bianchi-evolution-system-in-space-spinors}, given by
\begin{subequations}
    \begin{eqnarray*}
        && \bme^{0}{}_{\bmA \bmB} = 0, \\
        && \bme^{1}{}_{\bmA \bmB} = \omega x_{\bmA \bmB}, \qquad \bme^{2}{}_{\bmA \bmB} = \frac{\omega}{\rho} z_{\bmA \bmB}, \qquad \bme^{3}{}_{\bmA \bmB} = \frac{\omega}{\rho} y_{\bmA \bmB}, \\
        && \xi_{\bmA \bmB \bmC \bmD} = \sqrt{2} \left( \frac{\omega}{\rho} (x_{\bmB \bmD} \epsilon_{\bmA \bmC} + x_{\bmA \bmC} \epsilon_{\bmB \bmD}) \right) - \sqrt{2} (\epsilon_{\bmB \bmD} D_{\bmA \bmC} \omega + \epsilon_{\bmA \bmC} D_{\bmB \bmD} \omega), \\
        && f_{\bmA \bmB} = D_{\bmA \bmB} \omega, \\
        && \chi_{((\bmA \bmB) \bmC \bmD)} = 0, \\
        && \Theta_{\bmA \bmB (\bmC \bmD)} = - \frac{\omega^2}{6 \Omega} \left( D_{\bmA \bmB} D_{\bmC \bmD} \Omega + D_{\bmC \bmD} D_{\bmA \bmB} \Omega + 2 D_{\bmA (\bmC} D_{\bmD) \bmB} \Omega + 2 D_{\bmB (\bmC} D_{\bmD) \bmA} \Omega \right), \\
        && \Theta_{\bmA \bmB} = 0, \\
        && \phi_{\bmA \bmB \bmC \bmD} = \frac{\omega^3}{6 \Omega^2} \left( D_{\bmA \bmB} D_{\bmC \bmD} \Omega + D_{\bmC \bmD} D_{\bmA \bmB} \Omega + 2 D_{\bmA (\bmC} D_{\bmD) \bmB} \Omega + 2 D_{\bmB (\bmC} D_{\bmD) \bmA} \Omega \right).
    \end{eqnarray*}
    \label{space-spinor-conformal-flat-initial-data}
\end{subequations}
It is important to note that the initial data obtained from these equations will not be consistent with those obtained from eqs. \eqref{Frame-version-conformal-constraint-equations}. However, the crucial observation is that the zero-order expressions for the initial data are indeed consistent, which is sufficient for the rest of the calculations in this article. Using the xAct package, the non-vanishing conformally flat initial data for the irreducible components of the fields appearing in eqs. \eqref{Evolution-system-in-space-spinors} and \eqref{The-Bianchi-evolution-system-in-space-spinors} can be expressed as
\begin{subequations}
        \begin{eqnarray*}
        && \bme^{1}_{x \star} = \omega, \qquad \bme^{2}_{z \star} = \frac{\omega}{\rho}, \qquad \bme^{3}_{y \star} = \frac{\omega}{\rho}, \\
        && \xi_{x \star} = - \frac{\sqrt{2} \omega + \sqrt{2} \rho \partial_{\rho}(\omega)}{\rho}, \qquad \xi_{y \star} = - \frac{\sqrt{2} X_{-}(\omega)}{\rho}, \qquad \xi_{z \star} = - \frac{\sqrt{2} X_{+}(\omega)}{\rho}, \\
        && f_{x \star} = \partial_{\rho} (\omega), \qquad f_{y \star} = \frac{X_{-}(\omega)}{\rho}, \qquad f_{z \star} = \frac{X_{+}(\omega)}{\rho}, \\
        && \Theta_{0 \star} = - \frac{\omega^2 X^{2}_{+}(\Omega)}{2 \rho^2 \Omega}, \qquad \Theta_{1 \star} = - \frac{2 \rho \omega^2 \partial_{\rho}(X_{+}(\Omega)) - 2 \omega^2 X_{+}(\Omega)}{\rho^2 \Omega}, \\
        && \Theta_{2 \star} = \frac{- 4 \rho \omega^2 \partial_{\rho}(\Omega) + 4 \rho^2 \omega^2 \partial^{2}_{\rho}(\Omega) -  \omega^2 X_{-}(X_{+}(\Omega))-  \omega^2 X_{+}(X_{-}(\Omega))}{2 \rho^2 \Omega}, \\
        && \Theta_{3 \star} = - \frac{- 2 \rho \omega^2 \partial_{\rho}(X_{-}(\Omega)) + 2 \omega^2 X_{-}(\Omega)}{\rho^2 \Omega}, \qquad \Theta_{4 \star} = - \frac{\omega^2 X^{2}_{-}(\Omega)}{2 \rho^2 \Omega}, \\
        && \phi_{0 \star} = \frac{\omega^3 X^{2}_{+}(\Omega)}{2 \rho^2 \Omega^2}, \qquad \phi_{1 \star} = \frac{- 2 \omega^3 X_{+}(\Omega)}{\rho^2 \Omega^2}, \\
        && \phi_{2 \star} = \frac{- 4 \rho \omega^3 \partial_{\rho}(\Omega) + 4 \rho^2 \omega^{3} \partial^{2}_{\rho}(\Omega) -  \omega^{3} X_{-}(X_{+}(\Omega))- \omega^{3} X_{+}(X_{-}(\Omega))}{2 \rho^2 \Omega^2}, \\
        && \phi_{3 \star} = \frac{- 2 \rho \omega^3 \partial_{\rho}(X_{-}(\Omega))+ 2 \omega^{3} X_{-}(\Omega)}{\rho^2 \Omega^2}, \qquad \phi_{4 \star} = \frac{\omega^3 X^{2}_{-}(\Omega)}{2 \rho^2 \Omega^{2}},
    \end{eqnarray*}
\end{subequations} 
where $X_{\pm}^{2}(\Omega) \equiv X_{\pm}(X_{\pm}(\Omega))$ and $\partial^{2}_{\rho}(\Omega) \equiv \partial_{\rho}(\partial_{\rho}(\Omega))$. Using eqs. \eqref{Omega-expansion} and \eqref{kappa-expansion}, the non-vanishing initial data for the zeroth-order equations obtained from eqs. \eqref{Evolution-system-in-space-spinors} and \eqref{The-Bianchi-evolution-system-in-space-spinors} can be written as
\begin{subequations}
    \begin{eqnarray}
        && \bme^{2 (0)}_{z \star} = 1, \qquad \bme^{3 (0)}_{y \star} = 1, \qquad f^{(0)}_{x \star} = 1, \\
        && \phi^{(0)}_{0 \star} = \frac{1}{12} X^{2}_{+}(\Pi_{3}[\Omega]) , \qquad \phi^{(0)}_{1 \star} = \frac{2}{3} X_{+}(\Pi_{3}[\Omega]) , \\
        && \phi^{(0)}_{2 \star} = \frac{1}{12}\left( 12 \Pi_{3}[\Omega] - 2 X_{-}(X_{+}(\Pi_{3}[\Omega])) \right), \\
        && \phi^{(0)}_{3 \star} = - \frac{2}{3} X_{-}(\Pi_{3}[\Omega]), \qquad \phi^{(0)}_{4 \star} = \frac{1}{12} X^{2}_{-}(\Pi_{3}[\Omega]),
    \end{eqnarray}
    \label{Zero-order-initial-data}
\end{subequations}
where the superscript ${}^{(0)}$ indicates that the initial data is evaluated at $\rho=0$. Given that 
\begin{equation}
    \bmd = \Theta \bmf + \bmd{\Theta},
    \label{Definition-d-1-form}
\end{equation}
one can show that the zero-order initial data for the irreducible components of $d_{\bmA \bmB}$ is given by
\begin{equation*}
    d_{x \star}^{(0)}=0, \qquad d_{y \star}^{(0)}=0, \qquad d_{z \star}^{(0)}=0.
\end{equation*}

\begin{remark}
    {\em The explicit calculation of the conformal factor $\theta$ and the transformation matrices $\Lambda^{\bmA}{}_{\bmB}$ will require the first-order solutions of $\bme^{1}_{x}, \bme^{1}_{y}$ and $\bme^{1}_{z}$. For this purpose, we list the first-order initial data for these components ---namely
    \begin{equation}
        \bme^{1 (1)}_{x \star} =1, \qquad \bme^{1 (1)}_{y \star}=0, \qquad \bme^{1 (1)}_{z \star} =0.
        \label{First-order-initial-data}
    \end{equation}}
\end{remark}
Given the initial data for the spinor fields appearing in the ECFEs, it is now possible to obtain solutions for the background evolution eqs. \eqref{Evolution-system-in-space-spinors} and the boundary-adapted evolution and constraint eqs. \eqref{The-Bianchi-evolution-system-in-space-spinors}-\eqref{The-Bianchi-constraint-system-in-space-spinors}. 

\section{Evaluating the BMS-supertranslation charges at the critical sets}
\label{Section:Evaluating-BMS-charges}
This section aims to bring together this analysis's various elements and obtain an expression of BMS-supertranslation charges $\mathcal{Q}$ at the critical sets $\mathcal{I}^{\pm}$. Recall that $\mathcal{Q}$ depends on the solutions of eqs. \eqref{Evolution-system-in-space-spinors} and \eqref{The-Bianchi-evolution-system-in-space-spinors}-\eqref{The-Bianchi-constraint-system-in-space-spinors}, the conformal factor $\theta$ and the transformation matrices $\Lambda^{\bmA}{}_{\bmB}$ ---see eq. \eqref{Asymptotic-BMS-charges-F-gauge}. In the following, the solutions of eqs. \eqref{Evolution-system-in-space-spinors} and \eqref{The-Bianchi-evolution-system-in-space-spinors}-\eqref{The-Bianchi-constraint-system-in-space-spinors} will be used to obtain an explicit transformation from the NP-gauge to the F-gauge, allowing us to assess the contribution from the background term $\sigma^{\bullet ab} N^{\bullet}_{ab}$ and the components of $\phi_{ABCD}$ to $\mathcal{Q}$ at zero-order. As will become evident, this analysis will reveal that the generic initial data provided in \cite{Huang10} does not give rise to well-defined BMS-asymptotic charges $\mathcal{Q}$ at $\mathcal{I}^{\pm}$. 

\subsection{Asymptotic solution of the extended conformal field equations}
In subsequent calculations, the $\Pi_{n}$ notation introduced in Section \ref{Section:Normal coordinates and conformal normal initial data} will be used to express the solutions of eqs. \eqref{Evolution-system-in-space-spinors} and \eqref{The-Bianchi-evolution-system-in-space-spinors}-\eqref{The-Bianchi-constraint-system-in-space-spinors}. Using the $\Pi_{n}$ notation, the expansion of the conformal factor $\Theta$ near $\rho=0$ can be written as
\begin{equation*}
    \Theta = \Pi_{0}[\Theta] + \Pi_{1}[\Theta] \rho + \frac{1}{2} \Pi_{2}[\Theta] \rho^2 + O(\rho^3).
\end{equation*}
Comparing with eq. \eqref{Conformal-factor-Theta-expansion}, one has
\begin{equation}
    \Pi_{0}[\Theta] =0, \qquad \Pi_{1}[\Theta] = (1-\tau^2).
    \label{Pi-Theta}
\end{equation}
From eq. \eqref{Gaussian-Gauge-d-conformalfactor}, the components $d_{x}, d_{y}$ and $d_{z}$ of the 1-form $d_{AB}$ can be shown to satisfy
\begin{subequations}
    \begin{eqnarray}
        && \Pi_{0}[d_{x}] =0, \qquad \Pi_{0}[d_{y}]=0, \qquad \Pi_{0}[d_{z}] =0, \\
        && \Pi_{1}[d_{x}] =2, \qquad \Pi_{1}[d_{y}]=0, \qquad \Pi_{1}[d_{z}] =0.
    \end{eqnarray}
    \label{Pi-1-form-d}
\end{subequations}
Taylor expanding the spinor fields appearing in eqs. \eqref{Evolution-system-in-space-spinors} and using the xAct package, it is possible to show that eqs. \eqref{Evolution-system-in-space-spinors} decouple from the boundary-adapted evolution and constraint eqs. \eqref{The-Bianchi-evolution-system-in-space-spinors}-\eqref{The-Bianchi-constraint-system-in-space-spinors} at zero-order. In other words, the zero-order system for the background fields does not depend on  $\phi_{0}, \phi_{1}, \phi_{2}, \phi_{3}$ and $\phi_{4}$. Integrating, one can write zero-order solutions of eqs. \eqref{Evolution-system-in-space-spinors} as
\begin{eqnarray}
    && \Pi_{0}[\bme^{0}_{x}] = - \tau , \qquad \Pi_{0}[\bme^{0}_{y}] = \Pi_{0}[\bme^{0}_{z}] = \Pi_{0}[\bme^{1}_{x}] = \Pi_{0}[\bme^{1}_{y}] = \Pi_{0}[\bme^{1}_{z}] = 0, \nonumber \\
    && \Pi_{0}[\bme^{2}_{x}] = \Pi_{0}[\bme^{2}_{y}] = 0, \qquad \Pi_{0}[\bme^{2}_{z}] = 1, \nonumber \\
    && \Pi_{0}[\bme^{3}_{x}] = 0, \qquad \Pi_{0}[\bme^{3}_{y}] = 1, \qquad \Pi_{0}[\bme^{3}_{z}] = 0, \nonumber \\
    && \Pi_{0}[\xi_{0}] = \Pi_{0}[\xi_{1}] = \Pi_{0}[\xi_{2}] = \Pi_{0}[\xi_{3}] =  \Pi_{0}[\xi_{4}] =  \Pi_{0}[\xi_{h}] = 0, \nonumber \\
    && \Pi_{0}[\xi_{x}] = \Pi_{0}[\xi_{y}] = \Pi_{0}[\xi_{z}] = 0, \nonumber \\
    && \Pi_{0}[f_{x}] = 1, \qquad \Pi_{0}[f_{y}] = \Pi_{0}[f_{z}] = 0, \label{Zero-order-solution-background-fields} \\
    && \Pi_{0}[\chi_{0}] =  \Pi_{0}[\chi_{1}] =  \Pi_{0}[\chi_{2}] = \Pi_{0}[\chi_{3}] =  \Pi_{0}[\chi_{4}] = \Pi_{0}[\chi_{h}] = 0, \nonumber \\
    && \Pi_{0}[\chi_{x}] =   \Pi_{0}[\chi_{y}] =   \Pi_{0}[\chi_{z}] = 0, \nonumber \\
    && \Pi_{0}[\Theta_{0}] =   \Pi_{0}[\Theta_{1}] =   \Pi_{0}[\Theta_{2}] = \Pi_{0}[\Theta_{3}] =   \Pi_{0}[\Theta_{4}] =   \Pi_{0}[\Theta_{h}] = 0, \nonumber \\
    && \Pi_{0}[\Theta_{x}] =   \Pi_{0}[\Theta_{y}] =   \Pi_{0}[\Theta_{z}] = 0, \nonumber \\
    && \Pi_{0}[\theta_{x}] =   \Pi_{0}[\theta_{y}] =   \Pi_{0}[\theta_{z}] = 0. \nonumber
\end{eqnarray}
For the first-order system, only the equations for $\bme^{1}_{x}, \bme^{1}_{y}$ and $\bme^{1}_{z}$ will decouple from the boundary-adapted evolution and constraint equations. Integrating, it can be shown that $\Pi_{1}[\bme^{1}_{x}], \Pi_{1}[\bme^{1}_{y}]$ and $\Pi_{1}[\bme^{1}_{z}]$ are constants and fixed by the initial data given in eq. \eqref{First-order-initial-data}. Then, the first-order solution for $\bme^{1}_{x}, \bme^{1}_{y}$ and $\bme^{1}_{z}$ is given by
\begin{equation*}
    \Pi_{1}[\bme^{1}_{x}]=1, \qquad \Pi_{1}[\bme^{1}_{y}]=0, \qquad \Pi_{1}[\bme^{1}_{z}]=0.
\end{equation*}
From the above discussion, the solutions of eqs. \eqref{Evolution-system-in-space-spinors} can be written as 
\begin{subequations}
    \begin{eqnarray}
    && \bme^{0}_{x} = - \tau + O(\rho), \qquad \bme^{1}_{x} = \rho + O(\rho^2) , \\
    && \bme^{2}_{z} = 1 + O(\rho), \qquad \bme^{3}_{y} = 1 + O(\rho), \\
    && f_{x} = 1 + O(\rho),
    \label{Transport-system-solution}
    \end{eqnarray}
\end{subequations}
where all other components are $O(\rho)$ or higher order. 

The next step in this analysis is to examine the zero-order boundary-adapted evolution and constraint eqs. \eqref{The-Bianchi-evolution-system-in-space-spinors}-\eqref{The-Bianchi-constraint-system-in-space-spinors}. By Taylor-expanding all the components of the spinor fields in eqs. \eqref{The-Bianchi-evolution-system-in-space-spinors}-\eqref{The-Bianchi-constraint-system-in-space-spinors} and substituting the zero-order solution of the background fields eqs. \eqref{Zero-order-solution-background-fields} and using eqs. \eqref{Pi-Theta}-\eqref{Pi-1-form-d}, the zero-order boundary-adapted evolution system can be written as
\begin{subequations}
    \begin{eqnarray}
        && \sqrt{2}(1+\tau) \partial_{\tau}(\Pi_{0}[\phi_{0}]) + \frac{1}{2 \sqrt{2}} X_{+}(\Pi_{0}[\phi_{1}]) = - 2 \sqrt{2} \Pi_{0}[\phi_{0}], \label{Evolution-phi-Eq1} \\
        && \frac{1}{2 \sqrt{2}} \partial_{\tau}(\Pi_{0}[\phi_{1}]) + \frac{1}{\sqrt{2}} X_{-}(\Pi_{0}[\phi_{0}]) + \frac{1}{6\sqrt{2}} X_{+}(\Pi_{0}[\phi_{2}]) = - \frac{1}{2 \sqrt{2}} \Pi_{0}[\phi_{1}], \quad \label{Evolution-phi-Eq2} \\
        && \frac{1}{3 \sqrt{2}} \partial_{\tau}(\Pi_{0}[\phi_{2}]) + \frac{1}{4 \sqrt{2}} X_{-}(\Pi_{0}[\phi_{1}]) + \frac{1}{4 \sqrt{2}} X_{+}(\Pi_{0}[\phi_{3}])=0, \label{Evolution-phi-Eq3} \\
        && \frac{1}{2 \sqrt{2}} \partial_{\tau}(\Pi_{0}[\phi_{3}]) + \frac{1}{6 \sqrt{2}} X_{-}(\Pi_{0}[\phi_{2}]) + \frac{1}{\sqrt{2}} X_{+}(\Pi_{0}[\phi_{4}]) = \frac{1}{2 \sqrt{2}} \Pi_{0}[\phi_{3}], \label{Evolution-phi-Eq4} \\
        && \sqrt{2} (1-\tau) \partial_{\tau}(\Pi_{0}[\phi_{4}]) + \frac{1}{2 \sqrt{2}} X_{-}(\Pi_{0}[\phi_{3}]) = 2 \sqrt{2} \Pi_{0}[\phi_{4}], \label{Evolution-phi-Eq5}
    \end{eqnarray}
    \label{zero-order-Bianchi-evolution}
\end{subequations}
while the zero-order boundary-adapted constraint system is given by
\begin{subequations}
    \begin{eqnarray}
        && \frac{1}{2 \sqrt{2}} \tau \partial_{\tau}(\Pi_{0}[\phi_{1}]) - \frac{1}{\sqrt{2}} X_{-}(\Pi_{0}[\phi_{0}]) + \frac{1}{6\sqrt{2}} X_{+}(\Pi_{0}[\phi_{2}])=0, \label{Constraint-phi-Eq1} \\
        && \frac{1}{3 \sqrt{2}} \tau \partial_{\tau}(\Pi_{0}[\phi_{2}]) - \frac{1}{4 \sqrt{2}} X_{-}(\Pi_{0}[\phi_{1}]) + \frac{1}{4 \sqrt{2}} X_{+}(\Pi_{0}[\phi_{3}]) =0, \label{Constraint-phi-Eq2} \\
        && \frac{1}{2 \sqrt{2}} \tau \partial_{\tau}(\Pi_{0}[\phi_{3}]) - \frac{1}{6 \sqrt{2}} X_{-}(\Pi_{0}[\phi_{2}]) + \frac{1}{\sqrt{2}} X_{+}(\Pi_{0}[\phi_{4}]) =0. \label{Constraint-phi-Eq3}
    \end{eqnarray}
    \label{zero-order-Bianchi-constraint}
\end{subequations}
To simplify subsequent calculations, define the complex-valued functions $T_{m}{}^{j}{}_{k}$ as follows
\begin{align*}
    T_{m}{}^{j}{}_{k}: SU(2,\mathbb{C}) &\to \mathbb{C}, \\
    \phantom{T_{m}{}^{j}{}_{k}:} \bmt  &\mapsto T_{m}{}^{j}{}_{k}(\bmt),
\end{align*}
where
\begin{equation*}
    T_{m}{}^{j}{}_{k}(\bmt)= \begin{pmatrix}
        m \\
        j
    \end{pmatrix}^{1/2} \begin{pmatrix}
        m \\
        k
    \end{pmatrix}^{1/2} \bmt^{(b_{1}}{}_{(a_{1}} \ldots \bmt^{b_{m})}{}_{a_{m})_{k}}, \qquad T_{0}{}^{0}{}_{0}(\bmt)=1.
\end{equation*}
Here, $m=0,1,2,\ldots$, and $j,k=0,\ldots,m$ and 
\begin{equation*}
    \begin{pmatrix}
        m \\
        j
    \end{pmatrix} = \frac{m!}{j! (m-j)!}.
\end{equation*}
Note that the complex conjugate of $T_{m}{}^{j}{}_{k}(\bmt)$ is defined by
\begin{equation*}
    \overline{T_{m}{}^{j}{}_{k}(\bmt)} = (-1)^{j+k} T_{m}{}^{m-j}{}_{m-k}(\bmt),
\end{equation*}
while the action of $X, X_{+}$ and $X_{-}$ on $T_{m}{}^{j}{}_{k}$ is given by
\begin{eqnarray*}
    &&  X(T_{m}{}^{j}{}_{k}) = (m-2k) T_{m}{}^{j}{}_{k}, \qquad X_{+}(T_{m}{}^{j}{}_{k}) = \sqrt{k (1-k+m)} T_{m}{}^{j}{}_{k-1}, \\
    && X_{-}(T_{m}{}^{j}{}_{k}) = - \sqrt{(1+k)(m-k)} T_{m}{}^{j}{}_{k+1}.
\end{eqnarray*}
\begin{definition}[spin-weight]
    A function $f$ is said to be of spin-weight $s$ if it satisfies $X(f)= 2 s f$, where $s$ is an integer or half-integer.
\end{definition}

\begin{remark}
{\em The functions $T_{m}{}^{j}{}_{k}(\bmt)$ on $SU(2,\mathbb{C})$ are closely related to the standard spin-weighted harmonics on $\mathbb{S}^2$ ---see e.g. \cite{FriedrichKannar00}   }
\end{remark}

\begin{remark}
    {\em By construction, functions on $\mathcal{M}_{a,\kappa}$ will have a well-defined spin-weight.}
\end{remark}
The components $\Pi_{0}[\phi_{n}]$ are functions on $\mathbb{R} \times SU(2,\mathbb{C})$. Hence, they admit a decomposition in terms of $T_{m}{}^{j}{}_{k}$. In the following, assume that:
\begin{assumption}
    The components $\Pi_{0}[\phi_{n}]$ admit an expansion of the form
    \begin{equation}
        \Pi_{0}[\phi_{n}] = \sum_{l=|2-n|}^{\infty} \sum_{m=0}^{2l} a_{n;2l,m}(\tau) T_{2l}{}^{m}{}_{l-2+n}, \qquad \tau \in [-1,1],
        \label{phi-n-decomposition}
    \end{equation}
    where $a_{n;2l,m}(\tau): \mathbb{R} \to \mathbb{C}$.
\end{assumption}
Furthermore, note that $\Pi_{3}[\Omega]$ at $\rho=0$ can be decomposed as
\begin{equation}
    \Pi_{3}[\Omega]|_{\rho=0} = \sum_{l=0}^{\infty} \sum_{m=0}^{2l} \Pi_{3}[\Omega]_{2l,m} T_{2l}{}^{m}{}_{l}.
    \label{Expansion-Pi-3-Omega}
\end{equation}
\begin{remark}
    {\em The argument of $a_{n;2l,m}(\tau)$ will be omitted in upcoming calculations. i.e., write $a_{n;2l,m}$ instead of $a_{n;2l,m}(\tau)$. Moreover, we use $a_{n;2l,m \star}$ to refer to $a_{n;2l,m}(0)$.}
\end{remark}

For $l=0, m=0$, one can use eq. \eqref{phi-n-decomposition} in eqs. \eqref{zero-order-Bianchi-evolution}-\eqref{zero-order-Bianchi-constraint} to show that $a_{2;0,0}$ is constant and is given by
\begin{equation}
    a_{2;0,0} = \Pi_{3}[\Omega]_{0,0}.
    \label{Solution-a2-00-mode}
\end{equation}
For $l=1, m=0,1,2$, the evolution eqs. \eqref{zero-order-Bianchi-evolution} imply
\begin{subequations}
    \begin{eqnarray}
        && \frac{1}{2 \sqrt{2}} \dot{a}_{1;2,m} + \frac{1}{6} a_{2;2,m} = - \frac{1}{2 \sqrt{2}} a_{1;2,m}, \\
        && \frac{1}{3 \sqrt{2}} \dot{a}_{2;2,m} - \frac{1}{4} a_{1;2,m} + \frac{1}{4} a_{3;2,m}=0, \label{l=1-zero-order-Bianchi-evolution-Eq2} \\
        && \frac{1}{ 2 \sqrt{2}} \dot{a}_{3;2,m} - \frac{1}{6} a_{2;2,m} = \frac{1}{2 \sqrt{2}} a_{3;2,m},
    \end{eqnarray}
    \label{l=1-zero-order-Bianchi-evolution}
\end{subequations}
while the constraint eqs. \eqref{zero-order-Bianchi-constraint} give 
\begin{subequations}
    \begin{eqnarray}
        && \frac{\tau}{ 2 \sqrt{2}} \dot{a}_{1;2,m} + \frac{1}{6} a_{2;2,m} =0, \label{l=1-zero-order-Bianchi-constraint-Eq1} \\
        && \frac{\tau}{3 \sqrt{2}} \dot{a}_{2;2,m} + \frac{1}{4} a_{1;2,m} + \frac{1}{4} a_{3;2,m} =0, \label{l=1-zero-order-Bianchi-constraint-Eq2}\\
        && \frac{\tau}{2 \sqrt{2}} \dot{a}_{3;2,m} + \frac{1}{6} a_{2;2,m}=0,
    \end{eqnarray}
    \label{l=1-zero-order-Bianchi-constraint}
\end{subequations}
where $\dot{a}_{n;2l,m} \equiv \partial_{\tau}(a_{n;2l,m})$. By multiplying eqs. \eqref{l=1-zero-order-Bianchi-evolution} with $\tau$ and substracting from eqs. \eqref{l=1-zero-order-Bianchi-constraint}, one obtain linear equations to be solved for $a_{1;2,m}, a_{2;2,m}$ and $a_{3;2,m}$. Substituting back into the evolution eqs. \eqref{l=1-zero-order-Bianchi-evolution} and simplifying, one gets
\begin{subequations}
    \begin{eqnarray}
        && (-1+\tau) \dot{a}_{1;2,m} - a_{1;2,m} =0. \label{l=1-a1-equation} \\
        && \tau \dot{a}_{2;2,m} - a_{2;2,m}=0, \\
        && (1+\tau) \dot{a}_{3;2,m} - a_{3;2,m} =0.
    \end{eqnarray}
\end{subequations}
Then, the solution for $a_{1;2,m},a_{2;2,m}$ and $a_{3;2,m}$ is given by
\begin{equation*}
    a_{1;2,m} = - \mathcal{C}_{1,m} (1-\tau), \qquad a_{2;2,m} = \mathcal{C}_{2,m} \tau, \qquad a_{3;2,m} = \mathcal{C}_{3,m} (1+\tau),
\end{equation*}
where $\mathcal{C}_{1,m}, \mathcal{C}_{2,m}$ and $\mathcal{C}_{3,m}$ are some constants that depend on $m$ with $m \in \{0,1,2 \}$. The initial data for $\Pi_{0}[\phi_{1}]$, given by eq. \eqref{Zero-order-initial-data}, indicates that
\begin{equation*}
    \mathcal{C}_{1,m} = - \frac{2 \sqrt{2}}{3} \Pi_{3}[\Omega]_{2,m}, \qquad m=0,1,2.
\end{equation*}
Using eq. \eqref{l=1-zero-order-Bianchi-constraint-Eq2}, we have
\begin{equation*}
    \mathcal{C}_{3,m} = \mathcal{C}_{1,m}, \qquad m=0,1,2.
\end{equation*}
Finally, eq. \eqref{l=1-zero-order-Bianchi-evolution-Eq2} gives
\begin{equation*}
    \mathcal{C}_{2,m} = - \frac{3}{\sqrt{2}} \mathcal{C}_{1,m}, \qquad m=0,1,2.
\end{equation*}
Therefore, $a_{1;2,m},a_{2;2,m}$ and $a_{3;2,m}$ can be written as
\begin{subequations}
    \begin{eqnarray}
    && a_{1;2,m} = \frac{2 \sqrt{2}}{3} \Pi_{3}[\Omega]_{2,m} (1-\tau), \qquad a_{2;2,m} = 2 \Pi_{3}[\Omega]_{2,m} \tau, \label{Solution-a2-1m-mode} \\
    && a_{3;2,m} = - \frac{2 \sqrt{2}}{3} \Pi_{3}[\Omega]_{2,m} (1+\tau).
\end{eqnarray}
\end{subequations}
where $m \in \{ 0,1,2 \}$. 

\medskip
For $l\geq 2, 0 \leq m \leq 2l$, the evolution eqs. \eqref{zero-order-Bianchi-evolution} imply the ODEs
\begin{subequations}
    \begin{eqnarray}
        && (1+\tau) \dot{a}_{0} + 2 a_{0} + \frac{1}{4} \sqrt{(2+l)(l-1)} a_{1} =0, \label{Evolution-a0} \\
        && \dot{a}_{1} +  a_{1} + \frac{1}{3} \sqrt{l (l+1)} a_{2} - 2 \sqrt{(2+l)(l-1)} a_{0}=0, \label{Evolution-a1} \\
        && \dot{a}_{2} + \frac{3}{4} (a_{3}-a_{1}) =0, \label{Evolution-a2} \\
        && \dot{a}_{3} - a_{3} - \frac{1}{3} \sqrt{l (l+1)} a_{2} + 2 \sqrt{(2+l)(l-1)} a_{4} =0, \label{Evolution-a3} \\
        && (1-\tau) \dot{a}_{4} - 2 a_{4} - \frac{1}{4} \sqrt{(2+l)(l-1)} a_{3} =0, \label{Evolution-a4}
    \end{eqnarray}
\end{subequations}
where $a_{n}$ and $\dot{a}_{n}$ refer to $a_{n;2l,m}$ and $\dot{a}_{n;2l,m}$, respectively. The constraint eqs. \eqref{zero-order-Bianchi-constraint} give
\begin{subequations}
    \begin{eqnarray}
        && \tau \dot{a}_{1} + \frac{1}{3} \sqrt{l (l+1)} a_{2} + 2 \sqrt{(2+l)(l-1)} a_{0} =0, \label{Constraint-a1} \\
        && \tau \dot{a}_{2} + \frac{3}{4} \sqrt{l(l+1)} (a_{3}+a_{1})=0, \label{Constraint-a2} \\
        && \tau \dot{a}_{3} + \frac{1}{3} \sqrt{l (l+1)} a_{2} + 2 \sqrt{(2+l)(l-1)} a_{4} =0. \label{Constraint-a3}
    \end{eqnarray}
    \label{Constraint-an}
\end{subequations}
After some manipulations, the equations for $a_{1},a_{2}$ and $a_{3}$ can be decoupled. In particular, one obtains the following second-order ODEs for $a_{1},a_{2}$ and $a_{3}$:
\begin{subequations}
    \begin{eqnarray}
        && (1-\tau^2) \ddot{a}_{1} + 2 (1-\tau) \dot{a}_{1} + l(l+1) a_{1}=0, \\
        && (1-\tau^2) \ddot{a}_{2} - 2 \tau \dot{a}_{2} + l (l+1) a_{2} =0, \label{Wave-equations-a2} \\
        && (1-\tau^2) \ddot{a}_{3} - 2 (1+\tau) \dot{a}_{3} + l(l+1) a_{3} =0.
    \end{eqnarray}
    \label{Wave-equations-a1-a2-a3}
\end{subequations}
Note that the equations for $a_{0}$ and $a_{4}$ will not decouple. This can be confirmed by solving eqs. \eqref{Evolution-a1}, \eqref{Evolution-a2} and \eqref{Evolution-a3} for $\dot{a}_{1}, \dot{a}_{2}$ and $\dot{a}_{3}$, respectively, and substituting in eq. \eqref{Constraint-an} to obtain an algebraic system that can be written as 
\begin{equation*}
    A \begin{pmatrix}
        a_{1} \\ 
        a_{2} \\
        a_{3}
    \end{pmatrix} = 6 \sqrt{(2+l)(l-1)} \begin{pmatrix}
        (1+\tau) a_{0} \\
        0 \\
        - (1-\tau) a_{4}
    \end{pmatrix},
\end{equation*}
where $A$ is a $3 \times 3$ matrix given by
\begin{equation*}
    A = \begin{pmatrix}
        3 \tau & (-1+\tau) \sqrt{l (l+1)} & 0 \\
        (1+\tau) \sqrt{l (l+1)} & 0 & - (-1 + \tau) \sqrt{l (l+1)} \\
        0 & (1+\tau) \sqrt{l (l+1)} & 3 \tau
    \end{pmatrix}.
\end{equation*}
One can confirm that $\text{det}(A) =0$. Hence, the equations for $a_{0}$ and $a_{4}$ will not decouple. A solution for $a_{0}$ and $a_{4}$ can be determined by treating the terms involving $a_{1}$ and $a_{3}$ as source terms. Remarkably, as it will be seen in the next subsection, the value of the coefficients $a_{0}$ and $a_{4}$ are not required in the computation of the BMS asymptotic charges at the critical sets.

%However, the question remains whether $\Pi_{0}[\phi_{0}]$ and $\Pi_{4}[\phi_{4}]$ contribute to $\mathcal{Q}$ at $\mathcal{I}^{\pm}$. 

%To determine whether a solution for $a_{0}$ and $a_{4}$ is required, we need to explore the explicit transformation from the NP-gauge to the F-gauge.

\subsection{The explicit transformation from the NP to the F-gauge}
\label{Subsection: Explicit-transformation-NP-F-gauge}
In this section, the solution of the background fields given by eqs. \eqref{Zero-order-solution-background-fields} will be used to obtain asymptotic expansions of the conformal factor $\theta$ and the transformation matrices $\Lambda^{\bmA}{}_{\bmB}$ relating the NP-gauge and the F-gauge. The analysis in this section will be concerned with the relation between the F-gauge frame $\{ \bme_{\bmA \bmA'} \}$ and the NP-gauge frame $\{ \bme^{\bullet}_{\bmA \bmA'} \}$ at future null infinity $\mathscr{I}^{+}$. An analogous calculation can be carried out for the transformation at $\mathscr{I}^{-}$. The discussion in this section follows that of the transformation from the NP-gauge to the F-gauge originally presented in \cite{FriedrichKannar00}. However, our initial data differ from those considered in \cite{FriedrichKannar00}, which results in differences in the asymptotic expansions of some of the fields in this section. 

\medskip
Consider the adapted frame $\{ \bme^{\circ}_{\bmA \bmA'}  \}$ at $\mathscr{I}^{+}$ satisfying the conditions introduced in Section \ref{Section: The NP-gauge}, then the vector field $ \bme^{\circ}_{\bmone \bmone'}$ must satisfy 
\begin{equation}
    \bme^{\circ \mu}{}_{\bmone \bmone'} = f \nabla^{\mu} \Theta, 
    \label{Definition-e-circ-11'}
\end{equation}
where $\nabla$ is the Levi-Civita connection associated with $\bmg$ and $\Theta$ is the conformal factor defining $\mathscr{I}^{+}$ ---see eqs. \eqref{Conformal-factor-Theta-expansion} and \eqref{Sets-null-infinity}. From eq. \eqref{kappa-omega}, the set $\mathscr{I}^{+}$ can be identified by $\tau=1$. To determine the function $f$ in eq. \eqref{Definition-e-circ-11'}, consider the parallel propagation condition $ \bme^{\circ \mu}{}_{\bmone \bmone'} \nabla_{\mu} \bme^{\circ \nu}{}_{\bmone \bmone'} =0$ satisfied by $ \bme^{\circ}_{\bmone \bmone'} $ which can be written as
\begin{equation*}
    f \nabla^{\mu} \Theta \nabla_{\mu} f \nabla^{\nu} \Theta + f^{2} \nabla^{\nu} (\frac{1}{2} \nabla_{\mu}\Theta \nabla^{\mu} \Theta) =0.
\end{equation*}
A contraction with a vector field ${\bm Z}$ transverse to $\mathscr{I}^{+}$ yields
\begin{equation}
    \nabla^{\mu} \Theta \nabla_{\mu} (\log{f}) = - \frac{\bmZ\left( \frac{1}{2} \nabla_{\mu} \Theta \nabla^{\mu} \Theta \right)}{{ \bmZ}(\Theta)}.
    \label{ODE-f}
\end{equation}
By setting ${\bm Z}= \partial_{\tau}$, one can show that 
\begin{equation}
    \frac{\bmZ\left( \frac{1}{2} \nabla_{\mu} \Theta \nabla^{\mu} \Theta \right)}{\bmZ(\Theta)} = 2 \Pi_{1}[f_{x}] \rho^2 + O(\rho^3).
    \label{ODE-f-RHS-expanded}
\end{equation}
Moreover, the LHS of eq. \eqref{ODE-f} can be written as 
\begin{eqnarray}
    && \nabla^{\mu} \Theta \nabla_{\mu} (\log{f}) = \left( - 2 \rho^{2} + O(\rho^3) \right) \partial_{\rho}(\log{f}) + \label{ODE-f-LHS-expanded} \\
    && \phantom{\nabla^{\mu} \Theta \nabla_{\mu} (\log{f}) =} \left( -\frac{1}{2} \left( 4 \Pi_{1}[\bme^{2}_{x}] + \frac{1}{6} X_{-}(\Pi_{3}[\Omega]) \right)\rho^{2} + O(\rho^3) \right) X_{-}(\log{f}) + \nonumber \\
    && \phantom{\nabla^{\mu} \Theta \nabla_{\mu} (\log{f}) =} \left( -\frac{1}{2} \left( 4 \Pi_{1}[\bme^{3}_{x}] + \frac{1}{6} X_{+}(\Pi_{3}[\Omega]) \right)\rho^{2} + O(\rho^3) \right) X_{+}(\log{f}). \nonumber
\end{eqnarray}
To solve for $f$, assume $\log{f}$ can be expanded as
\begin{equation*}
    \log{f} = F_{0} + F_{1} \rho + O(\rho^2),
\end{equation*}
with $X_{-}(F_{0})=X_{+}(F_{0})=0$, $X_{-}(F_{1}) \neq 0$ and $X_{+}(F_{1}) \neq 0$. Then, $\partial_{\rho}(\log{f}), X_{-}(\log{f})$ and $X_{+}(\log{f})$ can be written as
\begin{eqnarray*}
    && \partial_{\rho}(\log{f}) = F_{1} + O(\rho), \\
    && X_{-}(\log{f}) = X_{-}(F_{1}) \rho + O(\rho^2), \\
    && X_{+}(\log{f}) = X_{+}(F_{1}) \rho + O(\rho^2).
\end{eqnarray*}
Using the above and eqs. \eqref{ODE-f}, \eqref{ODE-f-RHS-expanded} and \eqref{ODE-f-LHS-expanded}, one has
\begin{equation*}
    F_{1} = - \Pi_{1}[f_{x}].
\end{equation*}
Next, consider the expansion 
\begin{equation*}
    f = f_{0} + f_{1} \rho + O(\rho^2).
\end{equation*}
But $f = e^{\log{f}}$, so 
\begin{equation*}
    f_{0} = e^{F_{0}}, \qquad f_{1} = e^{F_{0}} F_{1}. 
\end{equation*}
In the following, the coefficient $F_{0}$ is chosen such that 
\begin{equation*}
    f_{0} = - \frac{1}{2 \sqrt{2}}, \qquad f_{1} = - \frac{F_{1}}{2 \sqrt{2}} = \frac{\Pi_{1}[f_{x}]}{2 \sqrt{2}}.
\end{equation*}
Thus, the function $f$ can be written as
\begin{equation*}
    f = - \frac{1}{2 \sqrt{2}} - \frac{\Pi_{1}[f_{x}]}{2 \sqrt{2}} \rho  + O(\rho^2).
\end{equation*}
\begin{remark}
    { \em While the function $f$ depends on the first-order solution of $f_{x}$, it will become evident that the explicit solution will not be required for the following calculations. In the rest of this analysis, first-order solutions for the background fields will be treated as unknowns, and it will be shown that the final expression of $\mathcal{Q}$ will only depend on the zero-order solution. }
\end{remark}
In the following, assume that the adapted frame $\{ \bme^{\circ}_{\bmA \bmA'} \}$ and the F-gauge frame $\{ \bme_{\bmA \bmA'} \}$ are related by 
\begin{equation}
    \bme^{\circ}_{\bmA \bmA'} = \lambda^{\bmB}{}_{\bmA} \bar{\lambda}^{\bmB'}{}_{\bmA'} \bme_{\bmB \bmB'},
    \label{Adapted-frame-transformation}
\end{equation}
where $\lambda^{\bmA}{}_{\bmB}$ denote an $SL(2,\mathbb{C})$ transformation matrix. To determine $\lambda^{\bmB}{}_{\bmA}$, note that eq. \eqref{Definition-e-circ-11'} can be written as 
\begin{eqnarray*}
    && \bme^{\circ \mu}{}_{\bmone \bmone'} = f \bme^{\circ \mu}{}_{\bmA \bmA'} \bme^{\circ \bmA \bmA'} (\Theta) \\
    && \phantom{\bme^{\circ \mu}{}_{\bmone \bmone'} } = f \bme^{\circ \mu}{}_{\bmA \bmA'} \epsilon^{\bmA \bmB} \epsilon^{\bmA' \bmB'} \bme^{\circ}_{\bmB \bmB'} (\Theta).
\end{eqnarray*}
Using eq. \eqref{Adapted-frame-transformation} and substituting in the above, one obtains 
\begin{subequations}
    \begin{eqnarray}
        && \lambda^{\bmzero}{}_{\bmone} \bar{\lambda}^{\bmzero'}{}_{\bmone'} = f \bme_{\bmone \bmone'} (\Theta), \label{lambda01-equation} \\
        && \lambda^{\bmzero}{}_{\bmone} \bar{\lambda}^{\bmone'}{}_{\bmone'} = - f \bme_{\bmone \bmzero'}(\Theta), \label{lambda01-11-equation} \\
        && \lambda^{\bmone}{}_{\bmone} \bar{\lambda}^{\bmone}{}_{\bmone'} = f \bme_{\bmzero \bmzero'}(\Theta). \label{lambda11-equation}
    \end{eqnarray}
\end{subequations}
But,
\begin{subequations}
    \begin{eqnarray*}
        && \bme_{\bmzero \bmzero'}(\Theta) = \frac{\Pi_{4}[\Omega]}{6 \sqrt{2}} \rho^3 + O(\rho^4), \\
        && \bme_{\bmzero \bmone'}(\Theta) = - \frac{X_{+}(\Pi_{3}[\Omega])}{6 \sqrt{2}} \rho^2 + O(\rho^{3}), \\
        && \bme_{\bmone \bmzero'}(\Theta) = - \frac{X_{-}(\Pi_{3}[\Omega])}{6 \sqrt{2}} \rho^2 + O(\rho^{3}), \\
        && \bme_{\bmone \bmone'}(\Theta) = - 2 \sqrt{2} \rho - \frac{\Pi_{3}[\Omega]}{\sqrt{2}} \rho^2 + O(\rho^{3}),
    \end{eqnarray*}
\end{subequations}
Hence, eq. \eqref{lambda01-equation} yields
\begin{equation*}
    |\lambda^{\bmzero}{}_{\bmone}|^2 = \rho + \frac{1}{2} \left( \frac{\Pi_{3}[\Omega]}{2} - 2 \Pi_{1}[f_{x}] \right) \rho^2 + O(\rho^3).
\end{equation*}
By writing $\lambda^{\bmzero}{}_{\bmone}$ as $\lambda^{\bmzero}{}_{\bmone} = |\lambda^{\bmzero}{}_{\bmone}| e^{i \omega_{1}}$, where $\omega_{1}$ is the phase of $\lambda^{\bmzero}{}_{\bmone}$, and choosing $\omega_{1}$ such that $e^{i \omega_{1}}=1$, one shows that 
\begin{equation*}
    \lambda^{\bmzero}{}_{\bmone} =  \sqrt{\rho} + \frac{1}{8} \left( \Pi_{3}[\Omega] - 4 \Pi_{1}[f_{x}] \right) \rho^{3/2} + O(\rho^{5/2}),
\end{equation*}
From eq. \eqref{lambda11-equation}, one has
\begin{equation*}
    |\lambda^{\bmone}{}_{\bmone}|^2 = - \frac{1}{24} \Pi_{4}[\Omega] \rho^{3} + O(\rho^4).
\end{equation*}
Assume $\lambda^{\bmone}{}_{\bmone} = |\lambda^{\bmone}{}_{\bmone}| e^{i \omega_{2}}$, and make use of eq. \eqref{lambda01-11-equation} to fix $e^{i \omega_{2}}$. Then, $\lambda^{\bmone}{}_{\bmone}$ can be written as
\begin{equation*}
    \lambda^{\bmone}{}_{\bmone} = \frac{\Pi_{4}[\Omega]}{X_{-}(\Pi_{3}[\Omega])} \rho^{3/2} + O(\rho^{5/2}).
\end{equation*}
Note that eq. \eqref{Adapted-frame-transformation} implies
\begin{equation*}
    \lambda^{\bmzero}{}_{\bmone} \bme^{\circ}_{\bmzero \bmone'} - \lambda^{\bmzero}{}_{\bmzero} \bme^{\circ}_{\bmone \bmone'} = - \bar{\lambda}^{\bmzero'}{}_{\bmone'} \bme_{\bmone \bmzero'} - \bar{\lambda}^{\bmone'}{}_{\bmone'} \bme_{\bmone' \bmone'}. 
\end{equation*}
By applying the above to the affine parameter $u^{\circ}$ and by making use of the conditions $\bme^{\circ}_{\bmzero \bmone'}(u^{\circ}) =0$ and $\bme^{\circ}_{\bmone \bmone'}(u^{\circ}) =1$ on $\mathscr{I}^{+}$, one can show that
\begin{equation}
    \lambda^{\bmzero}{}_{\bmzero} = \bar{\lambda}^{\bmzero'}{}_{\bmone'} \bme_{\bmone \bmzero'}(u^{\circ}) + \bar{\lambda}^{\bmone'}{}_{\bmone'} \bme_{\bmone \bmone'}(u^{\circ}).
    \label{lambda-zero-zero-condition}
\end{equation}
The above equation allow us to determine $\lambda^{\bmzero}{}_{\bmzero}$ given $\bar{\lambda}^{\bmzero'}{}_{\bmone'}, \bar{\lambda}^{\bmone'}{}_{\bmone'}, \bme_{\bmone \bmzero'}(u^{\circ})$ and $\bme_{\bmone \bmone'}(u^{\circ})$. Then, the condition $\text{det}(\lambda^{\bmA}{}_{\bmB}) =1$ can be used to determine $\lambda^{\bmone}{}_{\bmzero}$. 

To obtain an asymptotic expression for the affine parameter $u^{\circ}$, note that eq. \eqref{Adapted-frame-transformation} implies that
\begin{eqnarray*}
    && \bme^{\circ}_{\bmone \bmone'} = \left( \frac{\Pi_{1}[\bme^{0}_{x}]}{\sqrt{2}} \rho^2 + O(\rho^{3}) \right) \partial_{\tau} + \left( \frac{1}{\sqrt{2}} \rho^2 + O(\rho^{3}) \right) \partial_{\rho} + \\
    && \phantom{\bme^{\circ}_{\bmone \bmone'} =} \left( \frac{1}{24 \sqrt{2}} \left( 24 \Pi_{1}[\bme^{2}_{x}] + X_{-}(\Pi_{3}[\Omega])\right) \rho^2 + O(\rho^{3}) \right) X_{+} + \\
    && \phantom{\bme^{\circ}_{\bmone \bmone'} =} \left( \frac{1}{24 \sqrt{2}} \left( 24 \Pi_{1}[\bme^{3}_{x}] + X_{+}(\Pi_{3}[\Omega])\right) \rho^2 + O(\rho^{3} ) \right) X_{-}.
\end{eqnarray*}
Consider the ansatz 
\begin{equation*}
    u^{\circ} = \frac{A^{\circ}}{\rho} + \mathcal{W}^{\circ} + B^{\circ} \log{\rho},
\end{equation*}
where $\partial_{\rho}(A^{\circ}) = \partial_{\rho}(B^{\circ})=0$ and $\mathcal{W}^{\circ}$ can be written as
\begin{equation*}
    \mathcal{W}^{\circ} = u^{\circ}_{\star} + O(\rho).
\end{equation*}
Then, one has 
\begin{subequations}
    \begin{eqnarray*}
        && \partial_{\tau} (u^{\circ}) = \frac{\partial_{\tau}(A^{\circ})}{\rho} + \partial_{\tau}(\mathcal{W}^{\circ}) + \partial_{\tau}(B^{\circ}) \log{\rho}, \\
        && \partial_{\rho} (u^{\circ}) = - \frac{A^{\circ}}{\rho^2} + \frac{B^{\circ}}{\rho} + \partial_{\rho}(\mathcal{W}^{\circ}), \\
        && X_{+}(u^{\circ}) = \frac{X_{+}(A^{\circ})}{\rho} + X_{+}(\mathcal{W}^{\circ}) + X_{+}(B^{\circ}) \log{\rho}, \\
        && X_{-}(u^{\circ}) = \frac{X_{-}(A^{\circ})}{\rho} + X_{-}(\mathcal{W}^{\circ}) + X_{-}(B^{\circ}) \log{\rho}.
    \end{eqnarray*}
\end{subequations}
Moreover, the condition $\bme^{\circ}_{\bmone \bmone'}(u^{\circ}) =1$ on $\mathscr{I}^{+}$ yields
\begin{equation*}
    A^{\circ} = - \frac{1}{\sqrt{2}}, \qquad B^{\circ}=0.
\end{equation*}
From the above, one has
\begin{equation}
    u^{\circ} = - \frac{1}{\sqrt{2} \rho} + \mathcal{W}^{\circ} = - \frac{1}{\sqrt{2} \rho} + u^{\circ}_{\star} + O(\rho).
    \label{Affine-parameter-circ}
\end{equation}
\begin{remark}
{\em In the following, we use $\Pi_{0}[\mathcal{W}^{\circ}]$ to refer to $u^{\circ}_{\star}$.}
\end{remark}
Using eq. \eqref{Affine-parameter-circ}, we have
\begin{eqnarray*}
    && \bme_{\bmone \bmzero'}(u^{\circ}) = \left( - \frac{1}{4} \Pi_{2}[\bme^{1}_{y}] - \frac{1}{\sqrt{2}} X_{-}(\Pi_{0}[\mathcal{W}^{\circ}]) \right) + O(\rho), \\
    && \bme_{\bmone \bmone'}(u^{\circ}) = - \frac{1}{2 \rho} + \left( \frac{2 \partial_{\tau}(\Pi_{0}[\mathcal{W}^{\circ}])}{\sqrt{2}} - \frac{1}{4} \Pi_{2}[\bme^{2}_{x}] \right) + O(\rho).
\end{eqnarray*}
Substituting in eq. \eqref{lambda-zero-zero-condition} and using $\text{det}(\lambda^{\bmA}{}_{\bmB}) =1$ yields
\begin{equation*}
    \lambda^{\bmzero}{}_{\bmzero} = O(\rho^{1/2}), \qquad \lambda^{\bmone}{}_{\bmzero} = \frac{1}{\sqrt{\rho}} + \left( \frac{\Pi_{3}[\Omega]}{8} - \frac{\Pi_{1}[f_{x}]}{2} \right) + O(\rho^{3/2}).
\end{equation*}
From eq. \eqref{F-NP-gauge-Tensor-Frame-relation}, the relation between the NP frame $\{ \bme^{\bullet}_{\bmA \bmA'} \}$ and the F-gauge frame $\{ \bme^{\bullet}_{\bmA \bmA'} \}$ is given by
\begin{equation}
    \bme^{\bullet}_{\bmA \bmA'} = \Lambda^{\bmB}{}_{\bmA} \bar{\Lambda}^{\bmB'}{}_{\bmA'} \bme_{\bmB \bmB'}. 
    \label{NP-frame-transformation}
\end{equation}
Comparing eq. \eqref{NP-frame-transformation} with eq. \eqref{Adapted-frame-transformation}, one can show that 
\begin{equation}
    \Lambda^{\bmzero}{}_{\bmone} = \theta^{-1/2} \lambda^{\bmzero}{}_{\bmone} e^{i c}, \qquad \Lambda^{\bmone}{}_{\bmone} = \theta^{-1/2} \lambda^{\bmone}{}_{\bmone} e^{i c},
    \label{Lambda-01-Lambda-11}
\end{equation}
where $c$ is a function encoding the phase freedom. From eq. \eqref{Definition-Phi-22}, one can obtain an asymptotic expansion for $\Phi^{\circ}_{22}$. Then, eq. \eqref{theta-equation} with initial data 
\begin{equation*}
    \lim_{\rho \to 0} \theta = 1,
\end{equation*}
can be used to obtain an asymptotic expansion for $\theta$. However, the strategy in this analysis is to use eq. \eqref{theta-equation} to confirm the expansion of $\theta$ obtained in \cite{FriedrichKannar00} with the assumptions that $\Phi^{\circ}_{22}$ can be written as
\begin{equation*}
    \Phi^{\circ}_{22} = \Pi_{0}[\Phi^{\circ}_{22}] + \Pi_{1}[\Phi^{\circ}_{22}] \rho + \frac{1}{2} \Pi_{2}[\Phi^{\circ}_{22}] \rho^2 + \frac{1}{6} \Pi_{3}[\Phi^{\circ}_{22}] \rho^3 + O(\rho^4).
\end{equation*}
From eq. \eqref{theta-equation}, one can confirm that
\begin{equation*}
    \Pi_{0}[\Phi^{\circ}_{22}] = \Pi_{1}[\Phi^{\circ}_{22}] = \Pi_{2}[\Phi^{\circ}_{22}] =0, \qquad \Pi_{1}[\theta] = \frac{1}{6} \Pi_{3}[\Phi^{\circ}_{22}].
\end{equation*}
Then, the conformal factor $\theta$ can be written as
\begin{equation*}
    \theta = 1 + \frac{1}{6} \Pi_{3}[\Phi^{\circ}_{22}] \rho + O(\rho^2).
    \label{expansion-theta}
\end{equation*}
Using the above and substituting in eqs. \eqref{Lambda-01-Lambda-11}, one can show that $\Lambda^{\bmzero}{}_{\bmone}$ and $\Lambda^{\bmone}{}_{\bmone}$ are given by
\begin{subequations}
    \begin{eqnarray}
        && \Lambda^{\bmzero}{}_{\bmone} = \left( \sqrt{\rho} - \frac{1}{24} \left( - 3 \Pi_{3}[\Omega] + 2 \Pi_{3}[\Phi^{\circ}_{22}] + 12 \Pi_{1}[f_{x}] \right) \rho^{3/2} + O(\rho^{5/2}) \right) e^{ic}, \qquad \\
        && \Lambda^{\bmone}{}_{\bmone} = \left( \frac{\Pi_{4}[\Omega]}{X_{-}(\Pi_{3}[\Omega])} \rho^{3/2} + O(\rho^{5/2})  \right) e^{ic}.
    \end{eqnarray}
    \label{Expansion-Lambda-01-Lambda-11}
\end{subequations}
Then, eq. \eqref{NP-frame-transformation} show that the frame $\bme^{\bullet}_{\bmone \bmone'}$ can be written as
\begin{eqnarray*}
    && \bme^{\bullet}_{\bmone \bmone'} = \left( \frac{\Pi_{1}[\bme^{0}_{x}]}{\sqrt{2}} \rho^2 + O(\rho^3) \right) \partial_{\tau} + \left( \frac{1}{\sqrt{2}} \rho^{2} + O(\rho^3) \right) \partial_{\rho} + \\
    && \phantom{\bme^{\bullet}_{\bmone \bmone'} =} \left( \left( \frac{\Pi_{1}[\bme^{2}_{x}]}{\sqrt{2}} - \frac{\overline{\Pi_{4}[\Omega]}}{\sqrt{2} X_{+}(\overline{\Pi_{3}[\Omega]})} \right) \rho^2 + O(\rho^3) \right) X_{+} + \\
    && \phantom{\bme^{\bullet}_{\bmone \bmone'} =} \left( \left( \frac{\Pi_{1}[\bme^{3}_{x}]}{\sqrt{2}} - \frac{\Pi_{4}[\Omega]}{\sqrt{2} X_{-}(\Pi_{3}[\Omega])} \right) \rho^2 + O(\rho^3) \right) X_{-} .
\end{eqnarray*}
Applying the above to the affine parameter $u^{\bullet}$ and using the conditions $\bme^{\bullet}_{\bmzero \bmone'}(u^{\bullet}) =0$ and $\bme^{\bullet}_{\bmone \bmone'}(u^{\bullet}) =1$ on $\mathscr{I}^{+}$, one can show that
\begin{equation}
    \Lambda^{\bmzero}{}_{\bmzero} = \bar{\Lambda}^{\bmzero'}{}_{\bmone'} \bme_{\bmone \bmzero'}(u^{\bullet}) + \bar{\Lambda
    }^{\bmone'}{}_{\bmone'} \bme_{\bmone \bmone'}(u^{\bullet}).
    \label{Lambda-zero-zero-condition}
\end{equation}
Then, the condition $\text{det}(\Lambda^{\bmA}{}_{\bmB})=1$ can be used to solve for $\Lambda^{\bmone}{}_{\bmzero}$. To determine $u^{\bullet}$, consider the ansatz 
\begin{equation*}
    u^{\bullet} = \frac{A^{\bullet}}{\rho} + \mathcal{W}^{\bullet} + B^{\bullet} \log{\rho},
\end{equation*}
with $\partial_{\rho}(A^{\bullet}) = \partial_{\rho}(B^{\bullet})=0$. Then, the condition $\bme^{\bullet}_{\bmone \bmone'}(u^{\bullet})=1$ on $\mathscr{I}^{+}$ can be used to show that 
\begin{equation*}
    A^{\bullet} = - \sqrt{2}, \qquad B^{\bullet} = 0.
\end{equation*}
Given the above, the affine parameter $u^{\bullet}$ can be written as
\begin{equation*}
    u^{\bullet} = - \frac{\sqrt{2}}{\rho} + \mathcal{W}^{\bullet}.
\end{equation*}
Assume $\mathcal{W}^{\bullet}$ can be written as $\mathcal{W}^{\bullet} = u^{\bullet}_{\star} + O(\rho)$, then
\begin{equation*}
     u^{\bullet} = - \frac{\sqrt{2}}{\rho} + u^{\bullet}_{\star} + O(\rho).
\end{equation*}
Similar to earlier discussion, one has $\Pi_{0}[\mathcal{W}^{\bullet}]=u^{\bullet}_{\star}$. Using eq. \eqref{Lambda-zero-zero-condition} and the condition $\text{det}(\Lambda^{\bmA}{}_{\bmB})=1$ yields
\begin{subequations}
    \begin{eqnarray}
        && \Lambda^{\bmzero}{}_{\bmzero} = \left( - \frac{1}{2} \left( \Pi_{2}[\bme^{1}_{y}] + \sqrt{2} X_{-}(\Pi_{0}[\mathcal{W}^{\bullet}]) \right) \sqrt{\rho} + O(\rho^{3/2}) \right) e^{-ic}, \\
        && \Lambda^{\bmone}{}_{\bmzero} = \left( - \frac{1}{\sqrt{\rho}} + O(\rho^{1/2}) \right)  e^{-ic}.
    \end{eqnarray}
    \label{Expansion-Lambda-00-Lambda-10}
\end{subequations}
\begin{remark}
    {\em The phase parameter $c$ can be determined using eq. \eqref{phase-parameter-equation} and the initial data $c=0$ on $\mathcal{C}$. For our purpose, it will be sufficient to note that 
    \begin{equation*}
        e^{ic} = 1 + O(\rho).
    \end{equation*}}
\end{remark}
Given the expansions of the components of the matrix $\Lambda^{\bmA}{}_{\bmB}$, one can show that the components of the inverse matrix $\Lambda_{\bmA}{}^{\bmB}$ can be written as
\begin{subequations}
    \begin{eqnarray*}
        && \Lambda_{\bmzero}{}^{\bmzero} = \left( - \frac{\Pi_{4}[\Omega]}{X_{-}(\Pi_{3}[\Omega])} \rho^{3/2} + O(\rho^{5/2}) \right) e^{ic}, \\
        && \Lambda_{\bmzero}{}^{\bmone} = \left( - \frac{1}{\sqrt{\rho}} - \left( \frac{\Pi_{3}[\Omega]}{8} + \frac{\Pi_{3}[\Phi^{\circ}_{22}]}{12} + \frac{\Pi_{1}[f_{x}]}{2} \right) \sqrt{2} + O(\rho^{3/2}) \right) e^{-ic}, \\
        && \Lambda_{\bmone}{}^{\bmzero} = \left( \sqrt{\rho} - \frac{1}{24} \left( -3 \Pi_{3}[\Omega] + 2 \Pi_{3}[\Phi^{\circ}_{22}] + 12 \Pi_{1}[f_x] \right) \rho^{3/2} + O(\rho^{5/2}) \right) e^{ic}, \\
        && \Lambda_{\bmone}{}^{\bmone} = \left( \frac{1}{2} \left( \Pi_{2}[\bme^{1}_{y}] + \sqrt{2} X_{-}(\Pi_{0}[\mathcal{W}^{\bullet}]) \right) + O(\rho^{3/2}) \right) e^{-ic}.
    \end{eqnarray*}
\end{subequations}
Given eqs. \eqref{expansion-theta}, \eqref{Expansion-Lambda-01-Lambda-11}-\eqref{Expansion-Lambda-00-Lambda-10} and eqs. \eqref{NP-connection-coefficient-transformation}, it is possible to obtain asymptotic expansions for the NP-connection coefficients $\sigma^{\bullet}, \mu^{\bullet}$ and $\gamma^{\bullet}$. Then, eq. \eqref{The-background-term-scri+} can be used to confirm whether $\sigma^{\bullet ab} N^{\bullet}_{ab}$ will contribute to $\mathcal{Q}$ at $\mathcal{I}^{+}$. Using eq. \eqref{phi-bullet-transformation}, one can also determine which components of $\phi_{ABCD}$ will appear in the expression of $\mathcal{Q}$ at $\mathcal{I}^{+}$. From eqs. \eqref{NP-connection-coefficient-transformation}, one can show that
\begin{subequations}
    \begin{eqnarray*}
        && \mu^{\bullet} = O(\rho^{2}), \\
        && \gamma^{\bullet}= - \frac{\rho}{2 \sqrt{2}} + O(\rho^{2}), \\
        && \sigma^{\bullet} =  - \frac{1}{2 \sqrt{2}} \left( X_{-}(\Pi_{2}[\bme^{1}_{y}]) + \sqrt{2} X_{-}(X_{-}(\Pi_{0}[\mathcal{W}^{\bullet}])) \right) + O(\rho).
    \end{eqnarray*}
\end{subequations}
Then, the term $\Delta|\sigma^{\bullet}|^{2}$ satisfies
\begin{equation*}
    \Delta|\sigma^{\bullet}|^{2} = O(\rho^2). 
\end{equation*}
This readily implies that
\begin{equation*}
    \lim_{\rho \to 0} \sigma^{\bullet ab} N^{\bullet}_{ab} \simeq 0. 
\end{equation*}
From eq. \eqref{phi-bullet-transformation}, one can show that
\begin{equation*}
    \bar{\phi}^{\bullet}_2 = - \Pi_{0}[\bar{\phi}_{2}] + O(\rho).
\end{equation*}
Hence, the charges $\mathcal{Q}$ at $\mathcal{I}^{+}$ can be written as
\begin{equation}
    \mathcal{Q}(f,\mathcal{C})|_{\mathcal{I}^{+}} = \oint_{\mathcal{C}} -2 \bm{\varepsilon}_2 f \Pi_{0}[\bar{\phi}_2],
    \label{Q-Iplus-F-gauge}
\end{equation}
where $\Pi_{0}[\bar{\phi}_2]$ is evaluated at $\tau=1$. A similar analysis for the transformation between the NP-gauge frame $\{ \bme^{\bullet}_{\bmA \bmA'} \}$ and the F-gauge frame $\{ \bme_{\bmA \bmA'} \}$ at $\mathscr{I}^{-}$ reveals that the background term $\sigma^{\bullet ab} N^{\bullet}_{ab}$ given by eq. \eqref{The-background-term-scri-} will not contribute to $\mathcal{Q}$ at $\mathcal{I}^{-}$ and that $\bar{\phi}^{\bullet}_2 = - \Pi_{0}[\bar{\phi}_{2}]$ at $\rho=0$. Thus, 
\begin{equation}
    \mathcal{Q}(f,\mathcal{C})|_{\mathcal{I}^{-}} = \oint_{\mathcal{C}} -2 \bm{\varepsilon}_2 f \Pi_{0}[\bar{\phi}_2],
    \label{Q-Iminus-F-gauge}
\end{equation}
where $\Pi_{0}[\bar{\phi}_2]$ is evaluated at $\tau=-1$.

\begin{remark}
    {\em It should be highlighted the limited amount of explicit information about the asymptotic expansions, which is required for the evaluation of the BMS asymptotic charges at $\mathcal{I}^{\pm}$. In particular, expressions in eqs. \eqref{Q-Iplus-F-gauge} and \eqref{Q-Iminus-F-gauge} are formally identical to their spin-2 equivalents given in \cite{MohamedKroon22}.}
\end{remark}

\subsection{BMS-supertranslation charges at the critical sets}
\label{Subsection:MainResult}

The discussion in the previous section indicates that in order to evaluate $\mathcal{Q}$ at $\mathcal{I}^{\pm}$, we require a solution for $\Pi_{0}[\bar{\phi}_{2}]$ at $\tau = \pm 1$. Given the second-order ODE for $a_{2}$ in eq. \eqref{Wave-equations-a1-a2-a3}, one can show that
\begin{proposition}
For $ l \geq 2$ and $0 \leq m \leq 2l$, the solution to eq. \eqref{Wave-equations-a2} is given by:
\begin{equation}
    a_{2}(\tau) = \mathcal{A}_{l,m} P_{l}(\tau) + \mathcal{B}_{l,m} Q_{l}(\tau) 
    \label{Solution-a20l-geq-2}
\end{equation}
where $P_{l}(\tau)$ is the Legendre polynomial of order $l$ and
$Q_{l}(\tau)$ is the Legendre function of the second kind of order
$l$, $\mathcal{A}_{l,m}$ and $\mathcal{B}_{l,m}$ are constants that can be expressed in terms of the coefficients $\Pi_{3}[\Omega]_{2l,m}$ appearing in eq. \eqref{Expansion-Pi-3-Omega}. In particular, one has
\begin{subequations}
    \begin{eqnarray}
        && \mathcal{A}_{l,m} = -\frac{(6+l+l^2) Q_{l+1}(0) \Pi_{3}[\Omega]_{2l,m}}{6 P_{l+1}(0) Q_{l}(0) - 6 P_{l}(0) Q_{l+1}(0)}, \\
        && \mathcal{B}_{l,m} = \frac{(6+l+l^2) P_{l+1}(0) \Pi_{3}[\Omega]_{2l,m}}{6 P_{l+1}(0) Q_{l}(0) - 6 P_{l}(0) Q_{l+1}(0)}. \label{Blm-GR}
    \end{eqnarray}
\end{subequations}
\label{Solution-GR-a2}
\end{proposition}
Note that the recurrence relation of $Q_{l}(\tau)$ implies that $Q_{l}(\tau)$ diverges logarithmically near $\tau = \pm 1$, i.e., 
\begin{equation*}
    Q_l(\tau) = \mathcal{C}_l \ln(1 \pm \tau) + O(1), \mbox{for some constant
$\mathcal{C}_l$.}
\end{equation*}
Thus, it is straightforward to see that the solution for $a_{2}$ given in eq. \eqref{Solution-a20l-geq-2} will diverge at the critical sets unless $\mathcal{B}_{l,m} =0$.
From eq. \eqref{Blm-GR}, we have that $\mathcal{B}_{l,m}=0$ for even $l$. For odd $l$, one must restrict the initial data set to ensure that $\mathcal{B}_{l,m} =0$.

\begin{lemma}
    The solution in Proposition \ref{Solution-GR-a2} is regular at $\tau=\pm 1$ if and only if the coefficients $\Pi_{3}[\Omega]_{2l,m}$ satisfy
    \begin{equation*}
        \Pi_{3}[\Omega]_{2l,m} =0, \qquad \text{for odd } l \geq 2, \quad 0 \leq m \leq 2l.
    \end{equation*}
    \label{Regularity-conditions-GR}
\end{lemma}

To simplify the integrand in eqs. \eqref{Q-Iplus-F-gauge}-\eqref{Q-Iminus-F-gauge}, we rewrite the expansion of $\Pi_{0}[\phi_{2}]$ as 
\begin{equation}
    \Pi_{0}[\phi_{2}] = \sum_{l=0}^{\infty} \sum_{m=-l}^{l} a_{2;2l,m+l}(\tau) T_{2l}{}^{m+l}{}_{l},
    \label{Phi-2-alt-expansion}
\end{equation}
and the expansion for $\Pi_{3}[\Omega]$ at $\rho=0$ as
\begin{equation}
    \Pi_{3}[\Omega]|_{\rho=0} = \sum_{l=0}^{\infty} \sum_{m=-l}^{l} \Pi_{3}[\Omega]_{2l,m+l} T_{2l}{}^{m+l}{}_{l}.
    \label{Expansion-Pi_3_Omega_Alt}
\end{equation}
Then, the regularity condition in Lemma \ref{Regularity-conditions-GR} can be written as:
\begin{lemma}
    The solution in Proposition \ref{Solution-GR-a2} is regular at $\tau=\pm 1$ if and only if the coefficients $\Pi_{3}[\Omega]_{2l,m+l}$ satisfy
    \begin{equation*}
        \Pi_{3}[\Omega]_{2l,m+l} =0, \qquad \text{for odd } l \geq 2 \quad -l \leq m \leq l.
    \end{equation*}
    \label{Regularity-conditions-GR-2}
\end{lemma}

\begin{remark}
    {\em Recalling eq. \eqref{Pi-3-Omega_ALT}, one readily sees that the regularity condition in Lemma \ref{Regularity-conditions-GR-2} is, in fact, a statement about the multipolar structure of the freely specifiable function $\xi$ on $\mathbb{S}^2$. More precisely, the condition excludes from $\xi$ the modes with odd parity. This condition is, to the best of our knowledge, new.}
\end{remark}

Given initial data that satisfy the above regularity condition, the solution for $a_{2;2l,m+l}$ for $l\geq 2$ and $-l \leq m \leq l$ can be written as
\begin{equation}
    a_{2;2l,m+l}(\tau) = \mathcal{A}_{l,m+l} P_{l}(\tau).
    \label{Solution-a2-lm-mode}
\end{equation}
Note that $\mathcal{A}_{l,m+l}=0$ for odd $l$. Therefore, for $l \geq 2$, the solution for $a_{2;2l,m+l}$ is only non-vanishing for even $l$. Substituting in eqs. \eqref{Q-Iplus-F-gauge}-\eqref{Q-Iminus-F-gauge} using eqs. \eqref{Phi-2-alt-expansion}, we have
\begin{equation}
    \mathcal{Q}(f,\mathcal{C})|_{\mathcal{I}^{\pm}} =-2 \sum_{l=0}^{\infty} \sum_{m=-l}^{l} \bar{a}_{2;2l,m+l}(\pm 1) \oint_{\mathcal{C}} \bm{\varepsilon}_2 f \bar{T}_{2l}{}^{m+l}{}_{l}.
    \label{Charges-I-plus-minus}
\end{equation}
But, the functions $\bar{T}_{2l}{}^{m+l}{}_{l}$ are related to the complex conjugate of the spherical harmonics $\bar{Y}_{l,m}$. In particular, we have the correspondence 
\begin{equation*}
    \bar{T}_{2l}{}^{m+l}{}_{l} \mapsto C_{l,m} \bar{Y}_{l,m},
\end{equation*}
where $C_{l,m}$ is a constant that depends on $l$ and $m$ and whose specific form is not required for our discussion ---see \cite{Kroon04}. Given that $f$ is a function on $\mathbb{S}^{2}$, one can always write $f$ as
\begin{equation}
    f = \frac{1}{C_{l',m'}} Y_{l'm'}.
    \label{BMS-supertranslation-definition-function-f}
\end{equation}
Substituting in eq. \eqref{Charges-I-plus-minus}, we get
\begin{eqnarray*}
    && \mathcal{Q}(f,\mathcal{C})|_{\mathcal{I}^{\pm}} = -2 \sum_{l=0}^{\infty} \sum_{m=-l}^{l} \frac{C_{l,m}}{C_{l',m'}} \bar{a}_{2;2l,m+l}(\pm 1) \oint_{\mathcal{C}} \bm{\varepsilon}_2 Y_{l'm'} \bar{Y}_{l,m} \\
    && \phantom{\mathcal{Q}(f,\mathcal{C})|_{\mathcal{I}^{\pm}}} = -2 \bar{a}_{2;2l',m'+l'}(\pm 1)
\end{eqnarray*}
Then, using eqs. \eqref{Solution-a2-00-mode}, \eqref{Solution-a2-1m-mode} and \eqref{Solution-a2-lm-mode}, one has
\begin{equation*}
\mathcal{Q}|_{\mathcal{I}^{\pm}} = 
    \begin{cases}
         - 2 \Pi_{3}[\Omega]_{0,0}, \qquad & \text{for } l =0, m=0, \\
         \mp 4 \Pi_{3}[\Omega]_{2,m+1}, \qquad 
 & \text{for } l=1, -1\leq m \leq 1, \\
         0, \qquad & \text{for odd } l \geq 2, -l \leq m \leq l, \\
         - 2 \mathcal{A}_{l,m+l}, \qquad & \text{for even } l \geq 2, -l \leq m \leq l,
    \end{cases}
\end{equation*}
where $\mathcal{A}_{l,m+l}$ is given by
\begin{equation*}
    \mathcal{A}_{l,m+l} = -\frac{(6+l+l^2) Q_{l+1}(0) \Pi_{3}[\Omega]_{2l,m+l}}{6 P_{l+1}(0) Q_{l}(0) - 6 P_{l}(0) Q_{l+1}(0)}.
\end{equation*}
Using eqs. \eqref{Pi-3-Omega_ALT} and \eqref{Expansion-Pi_3_Omega_Alt}, and expanding the freely specifiable data $\xi$ as 
\begin{equation*}
    \xi = \sum_{l=0}^{\infty} \sum_{m=-l}^{l} \xi_{2l,m+l} T_{2l}{}^{m+l}{}_{l},
\end{equation*}
one can see that
\begin{subequations}
    \begin{eqnarray*}
        && \Pi_{3}[\Omega]_{0,0} = \frac{21}{2} \xi_{0,0} - 3 A, \\
        && \Pi_{3}[\Omega]_{2l,m+l} = \frac{21}{2} \xi_{2l,m+l}.
    \end{eqnarray*}
\end{subequations}
Then, the final result of our analysis can be summarised in the following:
\begin{theorem}
\label{Theorem:ChargesGR}
Given the generic initial data in Proposition \ref{Proposition:physical-initial-data}, the asymptotic BMS-supertranslation charges $\mathcal{Q}$ are not well-defined at $\mathcal{I}^{\pm}$ unless the freely specifiable data satisfy the regularity condition:
\begin{equation}
    \xi_{2l,m+l} =0 \qquad \text{for odd } l \geq 2 \quad -l \leq m \leq l.
    \label{Free-specificable-data-regularity-condition}
\end{equation}
Given initial data that satisfy the extra regularity condition, the charges $\mathcal{Q}$ at $\mathcal{I}^{\pm}$ can be expressed in terms of the freely specifiable data $A$ and $\xi$:
\begin{equation*}
\mathcal{Q}|_{\mathcal{I}^{\pm}} = 
    \begin{cases}
         - 21 \xi_{0,0} + 6 A, \qquad & \text{for } l =0, m=0, \\
         \mp 42 \xi_{2,m+1}, \qquad 
 & \text{for } l=1, -1\leq m \leq 1, \\
         0, \qquad & \text{for odd } l \geq 2, -l \leq m \leq l, \\
         - 2 \mathcal{A}'_{l,m+l}, \qquad & \text{for even } l \geq 2, -l \leq m \leq l,
    \end{cases}
\end{equation*}
where $\mathcal{A}'_{l,m+l}$ is given by
\begin{equation*}
    \mathcal{A}'_{l,m+l} = - \frac{ 21 (6+l+l^2) Q_{l+1}(0) \xi_{2l,m+l}}{  12 P_{l+1}(0) Q_{l}(0) - 12 P_{l}(0) Q_{l+1}(0)}.
\end{equation*}
Moreover, there is a natural correspondence between the charges at $\mathcal{I}^{+}$ and $\mathcal{I}^{-}$ expressed as follows:
\begin{equation*}
    \mathcal{Q}|_{\mathcal{I}^{+}} = (-1)^{l} \mathcal{Q}|_{\mathcal{I}^{-}}.
\end{equation*}
\end{theorem}

\begin{remark}
    {\em It should be stressed that the above results express the BMS asymptotic charges at $\mathcal{I}^{\pm}$ in terms of freely specifiable data coming from the initial metric. Remarkably, there is no contribution to the charges coming from the extrinsic curvature. }
\end{remark}

\begin{remark}
   {\em As a consequence of the regularity condition in Lemma \ref{Regularity-conditions-GR}, one finds that for $l \geq 2$, only the even parity charges have a non-trivial content. Observe that the regularity condition given in Lemma \ref{Regularity-conditions-GR} are necessary conditions for the BMS charges to be well-defined at the critical sets $\mathcal{I}^\pm$. The regularity condition eliminates the odd parity modes for $l \geq 2$ in the function $\xi$ over $\mathbb{S}^2$. Thus, compared with the equivalent conditions arising from analysing the spin-2 field on the Minkowski spacetime in \cite{MohamedKroon22}, one finds that the full non-linear GR situation is much more restrictive. This should not be surprising as the Einstein constraints are known to be more rigid than their linearised counterpart.}
\end{remark}

\begin{remark}
\label{Remark:C-K}   
{\em The initial data used in the analysis of the non-linear stability of the Minkowski spacetime by Christodoulou and Klainerman \cite{Christodoulou91} is of the form
\begin{eqnarray*}
&&\tilde{h}_{\alpha\beta} = \left( 1+ \frac{2m}{r}\right)\delta_{\alpha\beta} + o_4(r^{-3/2}),\\
&& \tilde{K}_{\alpha\beta} = o_3(r^{-5/2}).
\end{eqnarray*}
Comparing with the data in Proposition \ref{Proposition:DataLan}, it follows that the function $\xi$ from which the asymptotic charges are computed necessarily vanishes. Thus, for the Christodoulou-Klainerman spacetimes the asymptotic charges at the critical sets vanish. This observation is, to the best of our knowledge new. A similar statement can be made of the spacetimes constructed by Rodianski-Lindblad \cite{RodnianskiLindblad05}. The spacetimes constructed by Klainerman-Nicolo \cite{KlainermanNicolo} arise from data with a 3-metric having an even stronger decay ---thus, their asymptotic BMS charges vanish as well. Notice, however, that the more general analysis of the stability of the Minkowski spacetime by Bieri \cite{Bie10} requires 
\begin{eqnarray*}
&&\tilde{h}_{\alpha\beta} = \delta_{\alpha\beta} + O_{H^3}(r^{-1/2}),\\
&& \tilde{K}_{\alpha\beta} = O_{H^2}(r^{-3/2}).
\end{eqnarray*}
This is consistent with non-vanishing asymptotic charges at the critical sets.}
\end{remark}

\section{Conclusions}
In this article, we presented an analysis of BMS asymptotic charges using Friedrich's formulation of spatial infinity. Our analysis indicates that the initial data given in \cite{Huang10} does not give rise to well-defined asymptotic charges at $\mathcal{I}^{\pm}$ unless the initial data satisfies an extra regularity condition. The regularity condition eliminates the odd $l \geq 2$ modes on the freely specifiable leading term of the initial metric. If the initial data are chosen to satisfy the regularity condition in eq. \eqref{Free-specificable-data-regularity-condition}, BMS asymptotic charges are well-defined at $\mathcal{I}^{\pm}$. Then, if the function $f \in C^{\infty}(\mathbb{S}^{2})$ is written as eq. \eqref{BMS-supertranslation-definition-function-f}, one can show that BMS asymptotic charges at $\mathcal{I}^{\pm}$ can be fully expressed in terms of the freely specifiable data and that there exists a natural correspondence between the charges at $\mathcal{I}^{+}$ and $\mathcal{I}^{-}$. As a consequence of the regularity condition on the initial data, only even parity BMS asymptotic charges for $l \geq 2$ have a non-trivial content at the critical sets $\mathcal{I}^\pm$. Observe that the vanishing of the odd parity BMS asymptotic charges for $l \geq 2$ does not necessarily mean that the corresponding BMS asymptotic charges vanish everywhere on $\mathscr{I}^\pm$ and generically they will not be conserved.

Our result expands the discussion in \cite{HerberthsonLudvigsen92} that showed that the component of the Weyl tensor appearing in the Bondi mass at $\mathscr{I}^{+}$ match antipodally with that appearing at $\mathscr{I}^{-}$. In \cite{HerberthsonLudvigsen92}, the antipodal matching of these components also follows from the regularity conditions on $\mathscr{I}^{\pm}$ and at $i^{0}$. This provides further evidence that the antipodal matching is a regularity statement, i.e., it follows directly from the regularity condition on the initial data. It would be very interesting to prove a converse of the latter ---namely, that the antipodal matching implies regularity conditions on initial data similar to the ones obtained in the present analysis. This question goes beyond the scope of this article and will be considered elsewhere. 

\section{Acknowledgements}
M.M.A.Mohamed would like to thank Mahdi Godazgar, Grigalius Taujanskas and Marica Minucci for the useful discussions while working on this project. This work was supported in part by the NSF grant PHY-2107939 to the University of California, Santa Barbara. Calculations in this project used the computer algebra system Mathematica with the package xAct ---see \cite{xAct}.

\appendix

\section{The boundary adapted evolution and constraint equations}
\label{Appendix: The boundary adapted evolution and constraint equations}
In the main text of this article, we present the analysis of the zero-order boundary adapted evolution and constraint eqs. \eqref{The-Bianchi-evolution-system-in-space-spinors}-\eqref{The-Bianchi-evolution-system-in-space-spinors}. The zero-order system in given in eqs. \eqref{zero-order-Bianchi-evolution}-\eqref{zero-order-Bianchi-constraint} while the full system is presented in this section. 

The boundary-adapted evolution system eq. \eqref{The-Bianchi-evolution-system-in-space-spinors} can be written as 
\begin{subequations}
    \begin{align}
    & \frac{1}{8}\left(4 \sqrt{2} \xi_2 \phi_0-24 \sqrt{2} \xi_h \phi_0-16 \xi_x \phi_0-3 \sqrt{2} \xi_1 \phi_1+\right. \nonumber \\
    & 4 \xi_z \phi_1+4 \sqrt{2} \xi_0 \phi_2+16 \sqrt{2} \phi_0 f_x-4 \sqrt{2} \phi_1 f_z-4 \sqrt{2} \phi_2 \chi_0+ \nonumber \\
    & 3 \sqrt{2} \phi_1 \chi_1-4 \sqrt{2} \phi_0 \chi_2+24 \sqrt{2} \phi_0 \chi_h+16 \phi_0 \chi_x-4 \phi_1 \chi_z+  \nonumber \\
    & 2 \sqrt{2}\left(-4 \partial_\tau\left[\phi_0\right]\left(-1+\bme_{x}^0\right)-4 \partial_\rho\left[\phi_0\right] \bme_{x}^1+\partial_\tau\left[\phi_1\right] \bme_{z}^0+\partial_\rho\left[\phi_1\right] \bme_{z}^1-\right. \nonumber \\
    & \left.\left.4 \bme_{x}^3 X_{-}\left[\phi_0\right]+\bme_{z}^3 X_{-}\left[\phi_1\right]-4 \bme_{x}^2 X_{+}\left[\phi_0\right]+\bme_{z}^2 X_{+}\left[\phi_1\right]\right)\right)=0 \tag{A1} \\
    & \frac{1}{6 \sqrt{2}}\left(3 \partial_\tau\left[\phi_1\right]+6 \partial_\tau\left[\phi_0\right] \bme_{y}^0+6 \partial_\rho\left[\phi_0\right] \bme_{y}^1+\partial_\tau\left[\phi_2\right] \bm\bme_{z}^0+\right. \nonumber \\
    & \left.\partial_\rho\left[\phi_2\right] \bme_{z}^1+6 \bme_{y}^3 X_{-}\left[\phi_0\right]+\bme_{z}^3 X_{-}\left[\phi_2\right]+6 \bme_{y}^2 X_{+}\left[\phi_0\right]+\bme_{z}^2 X_{+}\left[\phi_2\right]\right)= \nonumber \\
    & \frac{1}{8}\left(-4 \sqrt{2} \xi_3 \phi_0-16 \xi_y \phi_0+\sqrt{2} \xi_2 \phi_1+6 \sqrt{2} \xi_h \phi_1-4 \xi_x \phi_1-\right. \nonumber \\
    & 2 \sqrt{2} \xi_0 \phi_3-2 \sqrt{2} \phi_1 f_x+4 \sqrt{2} \phi_0 f_y+2 \sqrt{2} \phi_2 f_z+2 \sqrt{2} \phi_3 \chi_0- \nonumber \\
    & \left.\sqrt{2} \phi_1 \chi_2+4 \sqrt{2} \phi_0 \chi_3-6 \sqrt{2} \phi_1 \chi_{\mathrm{h}}+4 \phi_1 \chi_{x}+16 \phi_0 \chi_{y}\right) \tag{A2} \\
    & \frac{1}{24}\left(\frac { 1 } { 2 } \left(24 \sqrt{2} \phi_4\left(\xi_0-\chi_0\right)+24 \sqrt{2} \phi_0\left(\xi_4-\chi_4\right)-\right.\right. \nonumber \\
    & 4 \sqrt{2} \phi_2\left(\xi_2+6 \xi_h-\chi_2-6 \chi_h\right)+3 \phi_1\left(\sqrt{2} \xi_3+4 \xi_y-4 \sqrt{2} f_y-\sqrt{2} \chi_3-4 \chi_y\right)+ \nonumber \\
    & \left.3 \phi_3\left(\sqrt{2} \xi_1+4 \xi_z-4 \sqrt{2} f_z-\sqrt{2} \chi_1-4 \chi_z\right)\right)+ \nonumber \\
    & \sqrt{2}\left(4 \partial_\tau\left[\phi_2\right]+3\left(\partial_\tau\left[\phi_1\right] \bme_{y}^0+\partial_\rho\left[\phi_1\right] \bme_{y}^1+\partial_\tau\left[\phi_3\right] \bme_{z}^0+\partial_\rho\left[\phi_3\right] \bme_{z}^1+\right.\right. \nonumber \\
    & \left.\left.\left.\bme_{y}^3 X_{-}\left[\phi_1\right]+\bme_{z}^3 X_{-}\left[\phi_3\right]+\bme_{y^2} X_{+}\left[\phi_1\right]+\bme_{z}^2 X_{+}\left[\phi_3\right]\right)\right)\right)=0 \tag{A3} \\
    & \frac{1}{6 \sqrt{2}}\left(3 \partial_\tau\left[\phi_3\right]+\partial_\tau\left[\phi_2\right] \bme_{y}^0+\partial_\rho\left[\phi_2\right] \quad \bme_{y}^1+6 \partial_\tau\left[\phi_4\right] \bm\bme_{z}^0+\right. \nonumber \\
    & \left.6 \partial_\rho\left[\phi_4\right] \bm\bme_{z}^1+\bme_{y}^3 X_{-}\left[\phi_2\right]+6 \bme_{z}^3 X_{-}\left[\phi_4\right]+\bme_{y}^2 X_{+}\left[\phi_2\right]+6 \bme_{z}^2 X_{+}\left[\phi_4\right]\right)= \nonumber \\
    & \frac{1}{8}\left(-2 \sqrt{2} \xi_4 \phi_1+\sqrt{2} \xi_2 \phi_3+6 \sqrt{2} \xi_h \phi_3+4 \xi_x \phi_3-4 \sqrt{2} \xi_1 \phi_4-\right. \nonumber \\
    & 16 \xi_z \phi_4+2 \sqrt{2} \phi_3 f_x+2 \sqrt{2} \phi_2 f_y+4 \sqrt{2} \phi_4 f_z+4 \sqrt{2} \phi_4 \chi_1- \nonumber \\
    & \left.\sqrt{2} \phi_3 \chi_2+2 \sqrt{2} \phi_1 \chi_4-6 \sqrt{2} \phi_3 \chi_{\mathrm{h}}-4 \phi_3 \chi_{x}+16 \phi_4 \chi_z\right) \tag{A4} \\
    & \frac{1}{2 \sqrt{2}}\left(4 \partial_\tau\left[\phi_4\right]\left(1+\bme_{x}^0\right)+4 \partial_\rho\left[\phi_4\right] \bme_{x}^1+\partial_\tau\left[\phi_3\right] \bme_{y}^0+\right. \nonumber \\
    & \left.\partial_\rho\left[\phi_3\right] \bme_{y}^1+\bme_{y}^3 X_{-}\left[\phi_3\right]+4 \bme_{x}^3 X_{-}\left[\phi_4\right]+\bme_{y}^2 X_{+}\left[\phi_3\right]+4 \bme_{x}^2 X_{+}\left[\phi_4\right]\right)= \nonumber \\
    & \frac{1}{8}\left(-4 \sqrt{2} \xi_4 \phi_2+3 \sqrt{2} \xi_3 \phi_3-4 \xi_y \phi_3-4 \sqrt{2} \xi_2 \phi_4+24 \sqrt{2} \xi_h \phi_4-\right. \nonumber \\
    & 16 \xi_x \phi_4+16 \sqrt{2} \phi_4 f_x+4 \sqrt{2} \phi_3 f_y+4 \sqrt{2} \phi_4 \chi_2- \nonumber \\
    & \left.3 \sqrt{2} \phi_3 \chi_3+4 \sqrt{2} \phi_2 \chi_4-24 \sqrt{2} \phi_4 \chi_h+16 \phi_4 \chi_x+4 \phi_3 \chi_y\right) \tag{A5}
    \end{align}
\end{subequations}
The constraint equations eq. \eqref{The-Bianchi-constraint-system-in-space-spinors} can be written as
\begin{subequations}
    \begin{align}
    & \frac{1}{6 \sqrt{2}}\left(-3 \partial_\tau\left[\phi_1\right] \bme_{x}^0-3 \partial_\rho\left[\phi_1\right] \bme_{x}^1-6 \partial_\tau\left[\phi_0\right]\bme_{y}^0-\right. \nonumber \\
    & 6 \partial_\rho\left[\phi_0\right]\bme_{y}^1+\partial_\tau\left[\phi_2\right] \bme_{z}^0+\partial_\rho\left[\phi_2\right] \bme_{z}^1-6 \bme_{y}^3 X_{-}\left[\phi_0\right]-3 \bme_{x}^3 X_{-}\left[\phi_1\right]+ \nonumber \\
    & \left.\bme_z^3 X_{-}\left[\phi_2\right]-6 \bme_{y}^2 X_{+}\left[\phi_0\right]-3 \bme_{x}^2 X_{+}\left[\phi_1\right]+\bme_z^2 X_{+}\left[\phi_2\right]\right)= \nonumber \\
    & \frac{1}{8}\left(2 \sqrt{2} \xi_3 \phi_0+24 \xi_y \phi_0-\sqrt{2} \xi_2 \phi_1+12 \xi_x \phi_1+\sqrt{2} \xi_1 \phi_2-4 \xi_z \phi_2-2 \sqrt{2} \xi_0 \phi_3+\right. \nonumber \\
    & \left.2 \sqrt{2} \phi_3 \chi_0-\sqrt{2} \phi_2 \chi_1+\sqrt{2} \phi_1 \chi_2-2 \sqrt{2} \phi_0 \chi_3-12 \phi_1 \chi_x-24 \phi_0 \chi_y+4 \phi_2 \chi_z\right) \tag{A6} \\
    & -\frac{1}{12 \sqrt{2}}\left(4 \partial_\tau\left[\phi_2\right] \bme_x^0+4 \partial_\rho\left[\phi_2\right] \bme_{x}^1+3 \partial_\tau\left[\phi_1\right]\bme_{y}^0+3 \partial_\rho\left[\phi_1\right]\bme_{y}^1-\right. \nonumber \\
    & 3 \partial_\tau\left[\phi_3\right] \bme_{z}^0-3 \partial_\rho\left[\phi_3\right] \bme_{z}^1+3 \bme_{y}^3 X_{-}\left[\phi_1\right]+4 \bme_{x}^3 X_{-}\left[\phi_2\right]- \nonumber \\
    & \left.3 \bme_z^3 X_{-}\left[\phi_3\right]+3 \bme_{y}^2 X_{+}\left[\phi_1\right]+4 \bme_{x}^2 X_{+}\left[\phi_2\right]-3 \bme_{z}^2 X_{+}\left[\phi_3\right]\right)= \nonumber \\
    & \frac{1}{16}\left(8 \sqrt{2} \xi_4 \phi_0-\sqrt{2} \xi_3 \phi_1+12 \xi_y \phi_1+16 \xi_x \phi_2+\sqrt{2} \xi_1 \phi_3-12 \xi_z \phi_3-8 \sqrt{2} \xi_0 \phi_4+\right. \nonumber \\
    & \left.8 \sqrt{2} \phi_4 \chi_0-\sqrt{2} \phi_3 \chi_1+\sqrt{2} \phi_1 \chi_3-8 \sqrt{2} \phi_0 \chi_4-16 \phi_2 \chi_x-12 \phi_1 \chi_y+12 \phi_3 \chi_z\right) \tag{A7}  \\
    & -\frac{1}{6 \sqrt{2}}\left(3 \partial_\tau\left[\phi_3\right] \bme_{x}^0+3 \partial_\rho\left[\phi_3\right] \bme_x^1+\partial_\tau\left[\phi_2\right] \bme_y^0+\partial_\rho\left[\phi_2\right] \bme_{y}^1 \right. \nonumber \\
    & 6 \partial_\tau\left[\phi_4\right] \bme_z^0-6 \partial_\rho\left[\phi_4\right] \bme_z^1+\bme_{y}^3 X_{-}\left[\phi_2\right]+3 \bme_{x}^3 X_{-}\left[\phi_3\right]- \nonumber \\
    & \left.6 \bme_{z}^3 X_{-}\left[\phi_4\right]+\bme_{y}^2 X_{+}\left[\phi_2\right]+3 \bme_{x}^2 X_{+}\left[\phi_3\right]-6 \bme_{z}^2 X_{+}\left[\phi_4\right]\right)= \nonumber \\
    & \frac{1}{8}\left(2 \sqrt{2} \xi_4 \phi_1-\sqrt{2} \xi_3 \phi_2+4 \xi_y \phi_2+\sqrt{2} \xi_2 \phi_3+12 \xi_x \phi_3-2 \sqrt{2} \xi_1 \phi_4-24 \xi_z \phi_4+\right. \nonumber \\
    & \left.2 \sqrt{2} \phi_4 \chi_1-\sqrt{2} \phi_3 \chi_2+\sqrt{2} \phi_2 \chi_3-2 \sqrt{2} \phi_1 \chi_4-12 \phi_3 \chi_x-4 \phi_2 \chi_y+24 \phi_4 \chi_z\right) \tag{A8}
    \end{align}
\end{subequations}

\bibliographystyle{unsrt} 
\bibliography{refs} 

\end{document}